 \documentclass[prd,a4paper,
twocolumn,showpacs, floatfix, nofootinbib,wrapfloat byrevtex]{revtex4}
 \usepackage{amsmath, amssymb,mathrsfs,upgreek}
 \usepackage{graphicx,epstopdf,placeins,float}
 \usepackage{color}
 \usepackage{booktabs}
 \usepackage{multirow}

 \newcommand{\beq}[1]{\begin{equation}\label{#1}}
 \newcommand{\eeq}{\end{equation}}
 \newcommand{\bea}[1]{\begin{eqnarray}\label{#1}}
 \newcommand{\eea}{\end{eqnarray}}
 \newcommand{\m}{{\rule[1.5pt]{2pt}{0.5pt}}}
 \newcommand{\mm}{{\rule[1.5pt]{2pt}{0.5pt}}}

\begin{document}

 \title{Spontaneous Radiation of Black Holes}
 \author{Ding-fang Zeng}
 \email{dfzeng@bjut.edu.cn}
 \affiliation{Institute of Theoretical Physics, Beijing University of Technology, China, Bejing 100124}
 \begin{abstract}
We provide an explicitly hermitian hamiltonian description for the spontaneous radiation of black holes, which is a many-level, multiple-degeneracy generalization of the usual Janeys-Cummings model for two-level atoms.  By standard Wigner-Wiesskopf approximation, we show that for the first one or few particles' radiation our model yields completely the same power spectrum as hawking radiation requires. While in the many-particle radiation cases, numeric methods allow us to follow the evolution of microscopic state of a black hole exactly, from which we can get the firstly increasing then decreasing entropy variation trend for the radiation particles just as the Page-curve exhibited. Basing on this model analysis, we claim that two ingredients are necessary for resolutions of the information missing puzzle, a spontaneous radiation like mechanism for the production of hawking particles and proper account of the macroscopic superposition happening in the full quantum description of a black hole radiation evolution and, the working logic of replica wormholes is an effect account of this latter ingredient.
 
As the basis for our interpretation of black hole Hawking radiation as their spontaneous radiation, we also provide a fully atomic like inner structure models for their microscopic states definition and origins of their Bekenstein-Hawking entropy, that is, exact solution families to the Einstein equation sourced by matter constituents oscillating across the central point and their quantization. Such a first quantization model for black holes' microscopic state is non necessary for our spontaneous radiation description, but has advantages comparing with other alternatives, such as string theory fuzzball or brick wall models.

 \end{abstract}
 \maketitle
 
 \tableofcontents
 
 \section{Motivation}
 
Although the replica wormhole \cite{qes1408,qes1705,penington1905,Almheiri1905,Almheiri1908,islandsoutside1910,replicaWH1911westcoast,replicaWH1911eastcoast,AImheiriReview2006,BartlReview2108} or quantum extremal surface method, or island formula is considered a great progress towards final solutions to the black hole information missing puzzle \cite{fireworksAMPS2012,fireworksAMPS2013,fireworksCOP2016,hawking1509,hawking1975cmp,Preskill1992,Giddings1995,LectureNotesPolchinski}, it still leaves many questions to answer. To us, a very difficult point to understand is, although the function form of a physical quantity such as the $n-$th order Rene entropy is continuatible to general complex $n$-field, its diagram representation should not be so. Because the latter is a topological object in essence, whose variation with $n$ is jumping-featured and usually non-continuatable. This brings us a very natural suspicion that the picture structure of
\bea{}
&&\hspace{5mm}\mathrm{tr}\rho^2_{\scriptscriptstyle R}\sim({}_{\scriptscriptstyle C\!}\langle\psi_i|\psi_j\rangle_{\scriptscriptstyle C})({}_{\scriptscriptstyle C\!}(\langle\psi_i|\psi_j\rangle)_{\scriptscriptstyle C})^*=
\label{replicaWormholeA}\\
&&\hspace{-5mm}\parbox{80mm}{\includegraphics[totalheight=35mm]{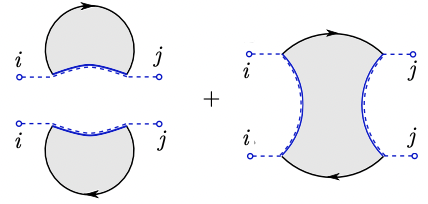}}
\label{replicaWormholeB}
\eea
could be analyically continuated to the calculation of
\beq{}
\mathrm{tr}\rho^1_{\scriptscriptstyle R}\sim({}_{\scriptscriptstyle C\!}\langle\psi_i|\psi_j\rangle_{\scriptscriptstyle C})=
\parbox{40mm}{\includegraphics[totalheight=15mm]{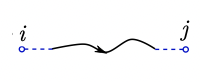}}
\label{replicaWormholeC}
\eeq
and be used further to justify declarations of the formula
\beq{}
S(R){=}\lim_{n\rightarrow1^+}\frac{1}{1{-}n}\log\mathrm{tr}(\rho^n_{\scriptscriptstyle R}){\approx}\min\{\mathrm{conn},\mathrm{disconn}\},
\label{replicaWormholeD}
\eeq
Obviously, in quantum field theory, it is totally non-acceptable to derive the math structure of 1$\rightarrow$1 propagators from the 2$\rightarrow$2 scattering amplitudes. However, this logics of reasoning play key roles in the path integral derivation of Page curves in the replica worm hole method.

Essentially, the starting point of replica wormhole method to the information missing puzzle is path integrals for quantum gravitation. However, researchers do not know how to calculate, in fact even not know how to define such integrals exactly. The so called derivation or proof is nothing but saddle point approximation something like those in Gibbons-Hawking's derivation of Bekenstein-Hawking entropy through Euclideanized free energy contributed by the known geometry of Schwarzschld black hole. It is unimaginable that such saddles' contribution like that by replica wormholes could be found if no one tells the researcher that proper trend of hawking radiation's entropy variation is of page curve type.

On the other hand, basing on string theory fuzzy ball pictures,  S. Mathur proves a very powerful small correction theorem \cite{fuzzball2009mathur,fuzzball2021mathur} which states that, any small correction to low energy dynamics around the horizon cannot solve the puzzle 
\beq{}
S_{N+1}>S_N+\log2-2\epsilon,
\eeq
that is, any particles arising from pair production around the classic horizon and escaping from there always cause increases of the entropy of hawking radiation, $S_{N+1}$ and  $S_N$ exemplifies just two such events. Just as we point out above, the derivation for hawking radiation's entropy in the replica wormhole is in the sense of Gibbons-Hawking's derivation of Bekenstein-Hawking entropy through Euclidean path integration and saddle point approximation, it says nothing on the microscopic mechanism of hawking radiation. If the mechanism of hawking radiation is still pair production from vacuum around the classic horizon and escaping from there, then Mathur's small correction theorem applies, we could not resolve the information missing puzzle such a way. The fact that replica wormhole method works has only two possibility, (i) it is circular proof, the researchers know the variation trend of hawking radiation's entropy so they add extra terms to $\mathrm{tr}\rho^1_{\scriptscriptstyle\!R}$ to get the desired results; (ii) it indeed involves new mechanism for hawking radiation through extra saddle point contribution to $\mathrm{tr}\rho^1_{\scriptscriptstyle\!R}$ and calls them island or replica wormholes. For more criticizes on working logics of the replica wormhole, we recommend readers to \cite{fuzzball2021mathur}, \cite{stojkovic2020prd} and \cite{giddings2202}

Different from the replica wormhole method, we pursued in references\cite{dfzeng2017,dfzeng2018a,dfzeng2018b,dfzeng2020} an alternative way to the information missing puzzle. Our basic idea is to interpret the hawking radiation as an approximation of spontaneous radiation of black holes, instead of pair productions around the classic horizon and escaping from there. Just like the spontaneous radiation of usual atoms, in such an interpretation, the black hole is no longer classic objects with clear-cut horizon surface, but becomes inner structured quantum objects with highly blurred horizons. This is very similar with the string theory fuzzy balls with the difference that, we build this fuzzy ball pictures through quantizations of classic solutions family to the Einstein equation sourced by matters periodically oscillating across the central point. From this fuzzy ball picture and quantum description of hawking radiation, we observed that a black hole's radiation evolution is not progressing in such a way that each intermedia state could be considered as entanglement of given size classic hole with corresponding radiation particles, 
\beq{}
|\raisebox{-3pt}{\includegraphics[totalheight=4mm]{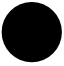}}\rangle
\rightarrow
|\raisebox{-4pt}{\includegraphics[totalheight=5mm]{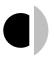}}\!\rangle
\rightarrow
|\raisebox{-5pt}{\includegraphics[totalheight=5mm]{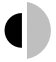}}\!\rangle
\rightarrow
|\raisebox{-5pt}{\includegraphics[totalheight=5mm]{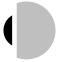}}\rangle
\rightarrow
|\raisebox{-5pt}{\includegraphics[totalheight=5mm]{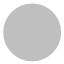}}\rangle
\Rightarrow
\raisebox{-5pt}{\includegraphics[totalheight=7mm]{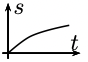}}
\label{SponRadiationChainB}
\eeq
where dark pie and its light-gray counterpart or segments of them represent the black hole and corresponding radiation products respectively, but the way
\bea{}
|\raisebox{-3pt}{\includegraphics[totalheight=4mm]{fig0cDecay1.png}}\rangle
&&\hspace{-3mm}\rightarrow
|\raisebox{-3pt}{\includegraphics[totalheight=4mm]{fig0cDecay1.png}}\rangle
+c_b^{t_1}|\raisebox{-4pt}{\includegraphics[totalheight=5mm]{fig0dDecay2whl.png}}\!\rangle
+c_m^{t_1}|\raisebox{-5pt}{\includegraphics[totalheight=5mm]{fig0eDecay3whl.png}}\!\rangle
+c_s^{t_1}|\raisebox{-5pt}{\includegraphics[totalheight=5mm]{fig0fDecay4whl.png}}\rangle
+|\raisebox{-5pt}{\includegraphics[totalheight=5mm]{fig0gDecay5whl.png}}\rangle
\label{SponRadiationChainC}\\
&&\hspace{-3mm}\rightarrow
|\raisebox{-3pt}{\includegraphics[totalheight=4mm]{fig0cDecay1.png}}\rangle
+c_b^{t_2}|\raisebox{-4pt}{\includegraphics[totalheight=5mm]{fig0dDecay2whl.png}}\!\rangle
+c_m^{t_2}|\raisebox{-5pt}{\includegraphics[totalheight=5mm]{fig0eDecay3whl.png}}\!\rangle
+c_s^{t_2}|\raisebox{-5pt}{\includegraphics[totalheight=5mm]{fig0fDecay4whl.png}}\rangle
+|\raisebox{-5pt}{\includegraphics[totalheight=5mm]{fig0gDecay5whl.png}}\rangle
\nonumber\\
&&\hspace{-3mm}\rightarrow\cdots
\nonumber
\eea
That is, each intermediate state of an evaporating black hole is a superposition of all possible mass black holes and their radiation products, at the same time all equal mass black holes are superposition of various inner structured microscopic state. A two levels of superposition exists here. 

This two levels of superposition makes physic quantities such as the entropy of radiation products a quantum expectation of quantum expectations({\bf no typos here}),
\beq{}
s(t)=|c^t_b|^2s_{b\;\!\!^{b}r\;\!\!^{b}}+|c^t_m|^2s_{b\;\!\!^{m}r\;\!\!^{m}}+|c^t_s|^2s_{b\;\!\!^{s}r\;\!\!^{s}}
\label{entropySuperposition}
\eeq
where $s_{b\;\!\!^{b}r\;\!\!^{b}}$, $s_{b\;\!\!^{m}r\;\!\!^{m}}$, $s_{b\;\!\!^{s}r\;\!\!^{s}}$ are entanglement entropies of the big, median, small holes and their radiation, their calculation involves quantum expectations already. Big, median and small here are only three representation of intermediate states. They are not exhaustive. At early times, $c_b^{t_1}\gg c^{t_1}_m\gg c^{t_1}_s$,  so $s(t_1)$ is dominated by $s_{b\;\!\!^{b}r\;\!\!^{b}}$ and increases with time. As time passes by, it will be dominated by $s_{b\;\!\!^{m}r\;\!\!^{m}}$ and reaches maximum at some intermediate time just as the Page-curve requires, then decreases due to the domination of $s_{b\;\!\!^{s}r\;\!\!^{s}}$. Comparing with the replica wormhole method, our derivation contains no argument chain like those going from equations \eqref{replicaWormholeA} to \eqref{replicaWormholeD}, but it requires us to have a comprehensive understanding of Hawking radiation as a quantum mechanic process such as the spontaneous radiation of usual atoms. 

\begin{figure}[h]
\includegraphics[totalheight=50mm]{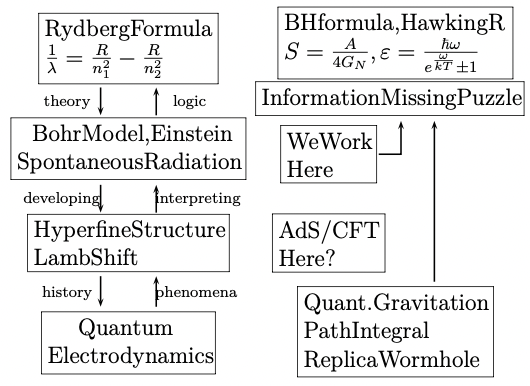}
\caption{Comparisons between the history and current status of quantum electrodynamics and those of quantum gravitation. We do not assume that these two theories are on same footings. What we adopt is just a practical as well as enterprising attitude to shrink back to a phenomenological model level when we are faced with chaoses brought about by the information missing puzzle.}
\label{figWorkPosition}
\end{figure}

Some people may argue that such an understanding is possible only in some final theory of quantum gravitation, which we obviously have none at hand currently. However, if we recall and compare developments of quantum gravitation with the history of quantum electrodynamics, we will easily see that some kind of toy model or phenomenological theory for black holes' radiation on the same level as those of spontaneous radiation of usual atoms is not only possible, but also necessary to some degree, see FIG.\ref{figWorkPosition} and captions there. Since resolutions of the information missing puzzle are a conclusion-known question, without guidance from phenomenological model, explorations basing on fundamental assumptions for quantum gravitation or its speculated feature of structures is very easy to walking into traps of cyclic arguments. String theory fuzzy ball provides such models. However, in the first-quantization frame, the quantum wave function of such fuzzy balls is non-available. For this reason we resort to Schwarzschild fuzzy balls of our own \cite{dfzeng2017,dfzeng2018a,dfzeng2018b,dfzeng2020}, whose wave functions can be written down explicitly and are mandatory in a full quantum description of spontaneous radiation of black holes. Any wide-visioned readers will find that, our explorations here are not replacements of other works with the same goal such as string theory fuzzy balls or replica worm holes, but are their complementarity indeed.

This section is the stimulation and key ideas for our proposition of black hole spontaneous radiation concepts. We will provide in \S II our explicitly hermitian hamiltonian description for spontaneous radiation of black holes, its approximations to the hawking radiation and resolutions to the information missing puzzle and relations to the replica worm holes. As the basis for our spontaneous radiation theory, we discuss the atomic like structure model for black holes in \S III and IV. \S III is classic geometric descriptions for our inner structured black holes, exact solutions to the Einstein equation and Penrose-Carter diagram representation for these objects. \S IV discusses these inner structure's quantization and the relevant microscopic state definition and degeneracy counting. \S V is a concrete example for the radiation process and inner structure quantization discussed in \S II and IV. \S VI is a discussion on physics meaning of inside horizon oscillation and its relevance with gravitational wave physics. We conclude and make prospects for future researches in the last section.
 
\section{Spontaneous Radiation of Black Holes}
As long as we accept the fact that the microscopic state of black hole corresponds to some kind of inner structure, just as the string theory fuzzy ball model indicates, we can build a theory for their radiation just the same way we do for spontaneous radiation of conventional atoms or molecules. The most big difference between radiations of this two types of objects may be, the latter is induced by dipole coupling between the atom/molecule and their radiation products - electromagnetic wave or photons, while the former could be induced by the mass-energy monopole coupling between the black hole and their radiation products - any wave or particles happens in hawking radiations.

\subsection{Explicitly Hermitian Hamiltonian}

Similar with Janyes-Cummings model describing interactions between the usual 2-levles-atom photon interaction \cite{ScullyQOtextbook}, we will write down an explicitly hermitian hamiltonian to describe gravitational monopole interaction between black holes and their radiation products
\beq{}
H=H_\mathrm{BH}+H_\mathrm{vac}+H_\mathrm{int}
\label{HamiltonianA}
\eeq
\beq{}
=\begin{pmatrix}w^i\\\!&\!\!w_{\mm}^{j}\!\\&\!&\!\!\ddots\!\\\!&\!&\!&\!{\scriptstyle\it0}^{\scriptscriptstyle\it1}\end{pmatrix}\!+\!\sum_q\hbar\omega_qa^\dagger_qa_q
{+}\!\!\sum_{u\;\!\!^n\mm v^\ell}^{\hbar\omega_q=}g_{u\;\!\!^n v^\ell}b^\dagger_{u\;\!\!^n v^\ell}a_q
\label{HamiltonianB}
\eeq
we will call this a Z-model description for spontaneous radiation of black holes. In this model, $w$, $w_\mm{=}w{\mm}{\scriptstyle1}$ and et al denote eigen-energy or degeneracies of the black hole quantum state regarding contexts; $i$, $j$ and et al distinguish microscopic states of equal mass black holes, i.e., $i=1,2\cdots,w$, $w=\exp[\frac{4\pi r_{\!h}^2}{4G\!_N}]$, $j=1,2\cdots,w{\mm}{\scriptstyle1}, w{\mm}1=\exp[\frac{4\pi r_{\!h}^{\prime2}}{4G\!_N}]$; the symbol ${\scriptstyle\it0}^{\scriptscriptstyle\it1}$ is introduced to denote the quantum state of ``nothing'' when the black hole evaporate completely; $a_q^\dagger$ \& $a_q$ are operators describing the vacuum fluctuation, $b^\dagger_{u\;\!\!^nv^\ell}$ \& $a_q$ take responsibilities for the black hole energy level's lowering or encreasing and the vacuum fluctuation mode $\omega^n_q$'s realization or inverse. Hermitian of the hamiltonian implies that $b_{u<v}^\dagger=b_{vu}$, $a_{q<0}=a_{-q}^\dagger$. For symbol conciseness, $\omega^n_q$ will be written as $q^n$ in the following, with $q=u{-}v$, $n$ inheriting from the emission body $u^n$'s superscript. Obviously, in our model, spontaneous radiation and absorption of the black hole are equally probable on hamiltonian levels.  

For simplicity, we will focus on spherically symmetric, that is S-wave radiations only so that spatial-momentum of the radiation particles will be ignored and their quantum state will be characterized by the energy completely. In this case, the basis of Hilbert space for an evaporating black hole and its radiation particles can be written as
\bea{}
\{w^i\otimes{\it\phi},w^j_{\mm}\otimes\omega_1^i,w^k_{{\mm\,\mm}}\otimes\omega_1^j\omega_1^i,w^j_{{\mm\,\mm}}\otimes\omega_2^i,\cdots
\nonumber\\
,u\;\!\!^n\otimes q^k{p}^j\cdots{o}^i(u{+}q{+}p\cdots{+}o=w),\cdots\;
\\
{\scriptstyle\it0}^{\scriptscriptstyle\it1}\otimes\omega_1^z\cdots\omega_1^j\omega_1^i,{\it\scriptstyle0}^{\it\scriptscriptstyle1}\otimes\omega_1^y\cdots\omega_2^i,\cdots,{\it\scriptstyle0}^{\it\scriptscriptstyle1}\otimes\omega_w^i
\}\nonumber
\eea
$w^i\otimes{\it\phi}$ denotes the state of the system before any radiations leave away, $w^j_{\mm}\otimes\omega^i_1$ denotes the state of the system after one particle is radiated away from the black hole and becomes free so that the degeneracy of the black hole decreases by 1. $w^k_{\mm\,\mm}\otimes\omega^j_1\omega^i_1$ denotes the state of the system when two particles are radiated away from the black hole one by one and become free so that the degeneracy of the black hole decreases by 1+1, $w^j_{\mm\,\mm}\otimes\omega^i_2$ denotes the state of the system when a little bigger particle is radiated away and the degeneracy of the black hole decreases by 2 in one time. Other symbols can be understood similarly, with $w$, $v$, $u$, ${\it\scriptstyle0}$ representing state of the black hole while $\phi$, $\omega$, $o$, $p$ \& $q$, those of the radiation particles.

Common sense basing on Newton mechanics tells us that gravitational interaction between particles happens through monopole couplings, $H_\mathrm{int}=-\frac{GMm}{r}$. Applying this point on the transition of black holes, we will also consider interactions between the black hole and their radiation particles induced by monopole type couplings only. This can be written schematically as follows
\beq{}
g_{u\;\!\!^nv^\ell}\propto \frac{1}{\sqrt{G}}\mathrm{Siml}\{\Psi[M_{u\;\!\!^n}\!(r)],\Psi[M_{v\;\!\!^\ell}\!(r)]\}
\label{monopoleCouplingStrength}
\eeq
where $u^n$ and $v^\ell$ denote initial and final states of the black hole respectively, $\mathrm{Siml}\{\cdot,\cdot\}$ measures their similarity of inner structures characterized by some initial time mass distribution $M_{u^n}(r)$ and $M_{v^\ell}(r)$, the proportionality here will be set to 1 for simplicity. In our inner structure models which we call Schwarzschild Fuzzy ball, each mass distribution has their own quantum wave function and can be written out explicitly. This question will be expanded discussing in latter sections. Let us focus on dynamics of the radiation at this point.

The qualitative reason for expressions \eqref{monopoleCouplingStrength} is, more similar the initial and final black hole's inner structures are, more easily the spontaneous radiation will happen as results. In our inner structure models, the mass distribution is characterized by many concentric shells periodically oscillating across the central point, thus being dynamical instead of static. Each of those shells has its own mass the radius expectation. We will define similarities between two mass distributions as
\bea{}
&&\hspace{-7mm}\mathrm{Siml}\{\Psi_{u^n}[M(r)],\Psi_{v^\ell}[M(r)]\}\equiv(\prod_{\langle i,j\rangle} \frac{2r^i_{un}\cdot r^j_{v\ell}}{r^{i2}_{un}+r^{j2}_{v\ell}})^\frac{1}{q}
\label{similarityA}
\\
&&\hspace{-7mm}q=i_\mathrm{max}j_\mathrm{max};
\nonumber
\eea
\bea{}
&&\hspace{3mm}\mathrm{Siml}\{\Psi_{u^n}[M(r)],\Psi_\mathrm{vacuum}[M]\}
\label{similarityB}
\\
&&\hspace{-5mm}\equiv\mathrm{Min}_{as.v^\ell}^{\mathrm{varies}}\{\mathrm{Siml}\{\Psi_{u^n}[M(r)],\Psi_{v^\ell}[M(r)]\}\}
\nonumber
\eea
where $r^i_{un}$ or $r^j_{v\ell}$ denote shell radiuses in inner structure wave functions $\Psi_{u^n}(r)$ and  $\Psi_{v^\ell}(r)$ respectively, see section \ref{secSimilarityDefinition} for concrete examples.

\begin{figure}[h]
\includegraphics[totalheight=25mm]{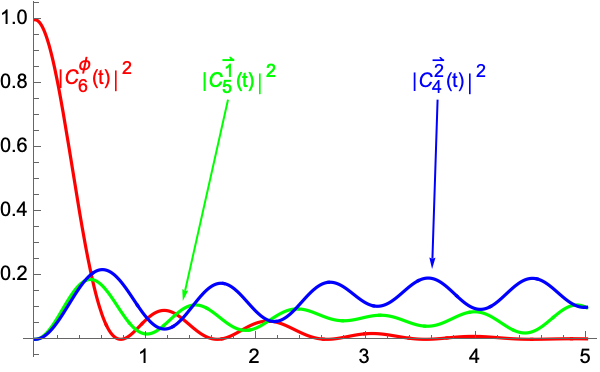}
\includegraphics[totalheight=25mm]{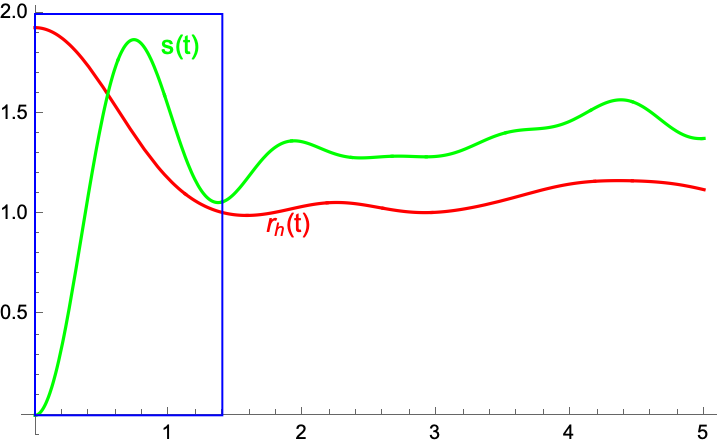}
\caption{Left panel is the coefficient of \eqref{WaveFunctionRadiation} describing an initially $w^i=6^1$ black hole's radiation evolution. The result follows from exact numeric integration of equation \eqref{SchrodingerEq}. The right panel is the horizon size and entropy curve of corresponding radiation products. The latter's inside blue box part exhibits manifestly first increasing then decreasing feature. }
\label{figWsixRadiationExample}
\end{figure}

\begin{figure}[h]
\includegraphics[totalheight=25mm]{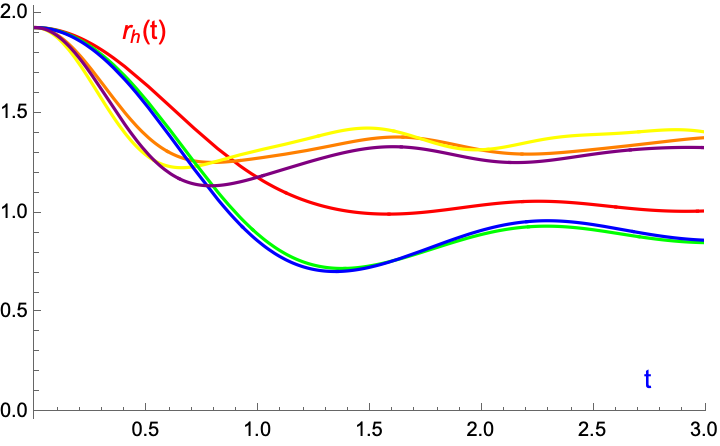}
\includegraphics[totalheight=25mm]{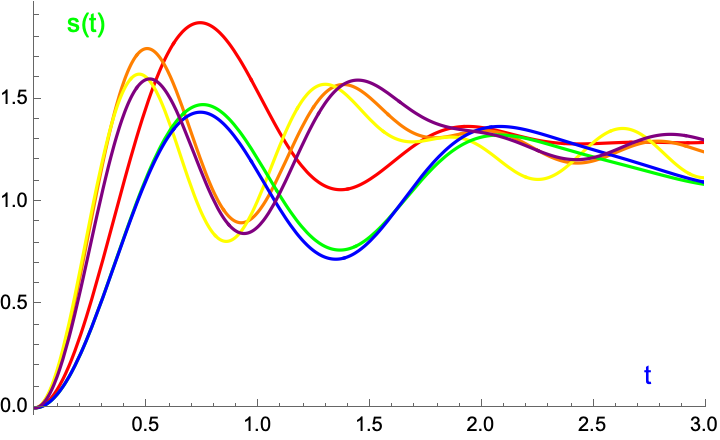}
\caption{The variation of remaining hole's radius and the radiation particles' entropy for 6 initial black holes with equal mass but at different microscopic states initially. The variation of entropies in all six cases exhibits manifestly first increasing then decreasing trends. The latter non-monotone behavior is due to Rabi-oscillation.}
\label{figWsixRadiationCurve}
\end{figure}

Now consider evolutions of a black hole due to spontaneous radiation as well as absorption. On an arbitrary middle time of such radiation/absorption processes, the microscopic state of the black hole and its radiation products around can be written as
\beq{}
|\psi(t)\rangle=\sum_{u=w}^0\sum_{n=1}^{u}\sum_{{\scriptscriptstyle\sum}o^i}^{w{-}u}
e^{-iut-i\omega t}c_{u\;\!\!^n}^{\vec{\omega}}(t)|u\;\!\!^n\otimes\vec{\omega}\rangle
\label{WaveFunctionRadiation}
\eeq
where ${\vec{\omega}}\equiv\{o^i,\cdots p^jq^k\}$ is an abbreviation for the radiation particles' quantum state, with the total energy given by $\omega=w-u$. 

By the standard Schr\"odinger equation $i\hbar\partial_t|\psi(t)\rangle=(H_\mathrm{BH}+H_\mathrm{vac}+H_\mathrm{int})|\psi(t)\rangle$ and the explicit form of Z-model hamiltonian \eqref{HamiltonianB}, we can get evolutions of this wave function as follows
\beq{}
i{\partial}_tc^{\vec{\omega}}_{u\;\!\!^n}(t){=}
\sum_{v\neq u}^{v+{\scriptscriptstyle'\!}\omega=w}\sum_{\ell=1}^{v}g_{u\;\!\!^n v^\ell}c^{{\scriptscriptstyle'\!}{\vec{\omega}}}_{v^\ell}(t)
\label{SchrodingerEq}
\eeq
where $\hbar$ has been set to $1$ and ${{\scriptscriptstyle'\!}{\vec{\omega}}}$ differs from $\vec{\omega}$ only by the last emitted or absorbed particles. Without loss of generality, we will set the initial condition 
\beq{}
c_{w^1}^{\scriptscriptstyle\it\phi}=1,c_{w^{i\neq1}}^{\scriptscriptstyle\it\phi}=0,c_{u\;\!\!^n}^{\vec{\omega}\scriptscriptstyle\neq\phi}=0
\label{iniConSchrodingerEq}
\eeq
That is, we let our black hole lies on eigenstate $w^1$ at initial time $t=0$. With this initial condition as well as appropriate exact value assigning prescription for $g_{u^nv^\ell}$, see subsection \ref{secSimilarityDefinition} for referrings, we can integrate equation \eqref{SchrodingerEq} numerically and get evolutions of the system routinely. We provide in FIG.\ref{figWsixRadiationExample} an example of an initially 6-times degenerating black hole's evolution. While FIG.\ref{figWsixRadiationCurve} compares the horizon size and entanglement entropy of radiation particles among the 6 initially different microscopic black holes. The left hand side of FIG.\ref{figWsixRadiationExample} is the wave function coefficients, the right is the size and entropy variation as time passes by. The non-monotonics of these functions in the very later stage of the evolution is due to Rabbi oscillations of the system. This arises from the hermitian characters of the hamiltonian \eqref{HamiltonianA}-\eqref{HamiltonianB}, due to which emission and absorption terms appear equally well thus enable particles radiated away by the black hole nonzero probabilities be absorbed back and cause renaissance of the black hole as results. This phenomena would disappear or almost invisible if we allow more general instead of only S-wave particles to appear in the final state of the radiation. 

The most noteworthy point of FIG.\ref{figWsixRadiationExample} is not the Rabi oscillation but the first increasing then decreasing feature of entropies of the blackhole/radiation products as the radiation progresses, before the Rabi oscillation happens. This feature was firstly pointed out by Don Page in reference \cite{page1,page2,page3} basing on general quantum mechanic analysis of entanglements between a black hole and its radiation products. Reproducing features of this curve is considered a great progress of the replica worm hole method to the information missing puzzle. However, just as we provide above, this is a completely routinely work in the spontaneous radiation of black holes which is described by the explicitly hermitian hamiltonian. For us, the truly challenging work is, is it possible to get thermal type power spectrum of Hawking radiations from such a spontaneous radiation mechanism controlled by the so called Z-model hamiltonian through a natural enough, but recognizing not so easy approximation? The answer is yes.

\subsection{Hawking Radiation as an Approximation}
Facing equation \eqref{SchrodingerEq}, if we do not resort numeric methods but analytical methods such as Wigner-Wiesskopf type approximation well understood  in atomic/molecular physics, we will have chances to get more instructive results. For example, in the first few particles radiation, all $c^{{{\scriptscriptstyle'\!}\vec{\omega}}}_{v^\ell}(t)=0$ except $c^\phi_{w^1}\!(t)$ and $c^{o{\scriptscriptstyle1}}_{u\;\!\!^n}(t)\neq0$. That is, for the first particle's radiation, equation \eqref{SchrodingerEq} solidifies to
\bea{}
i{\partial}_tc^{o{\scriptscriptstyle1}}_{u\;\!\!^n}(t)=g_{u\;\!\!^n\!w\;\!\!^1}^{u{+}\omega=w}c^\phi_{w^1}\!(t){+}
\!\sum_{v\neq u}^{v\neq w}\sum_{\ell=1}^{v}g_{u\;\!\!^n\!v\;\!\!^\ell}c^{{{\scriptscriptstyle'\!}\vec{\omega}}}_{v^\ell}(t)[{\approx}0]
\label{WignerWiesskopfA}
\\
i\partial_tc^\phi_{w^1}(t)=\!\!\!\sum_{u\neq w}^{u+o{\scriptscriptstyle1}=w}\!\!\sum_{n=1}^{u}g_{w\;\!\!^1\!u\;\!\!^n}c^{o{\scriptscriptstyle1}}_{u\;\!\!^n}(t)
\approx\!-i|g_{w\;\!\!^1\!u\;\!\!^n}|^2c^\phi_{w^1}(t)
\label{WignerWiesskopfB}
\eea
Equation \eqref{WignerWiesskopfB} follows from
\bea{}
&&\hspace{-2mm}c^{o{\scriptscriptstyle1}}_{u\;\!\!^n}(t)\approx-i\int_{0}^tg_{u\;\!\!^n\!w\;\!\!^1}^{u{+}\omega=w}c^\phi_{w^1}(t')dt'
\\
&&\hspace{-5mm}\approx-ig_{u\;\!\!^n\!w\;\!\!^1}c^\phi_{w^1}(t)\int_{0}^{t\rightarrow\infty}e^{-i[\omega-(w-u)]t'}dt'
\eea
Removing $c^\phi_{w^1}(t')$ out of the integration operator is the key of Wigner-Wiesskopf approximation and the validity of this doing is based on the fact that $c^\phi_{w^1}(t')$ varies far more slowly than the factor $e^{-i[\omega-(w-u)]t'}$, so contributions from $c^\phi_{w^1}(t')$'s earlier behavior to the integration is strongly suppressed by the oscillation of $e^{-i[\omega-(w-u)]t'}$. This is essence of Wigner-Wiesskopf approximation. Obviously, this approximation throw away history information encoded in $c^\phi_{w^1}(t')_{t'<t}$. For this reason, we will call it forgetting history approximation sometimes. This approximation makes the radiation of the black hole invertible, thus breaking hermitian of the hamiltonian matrix \eqref{HamiltonianB}. So it is the place where information missing happens. However, it is not the key point why we cannot get proper trend of hawking particle's entropy variation.

Under the Wigner-Wiesskopf approximation, we easily see that for first particle's radiation evolving,
\bea{}
&&\hspace{-5mm}c^\phi_{w^1}(t){=}e^{\mm\Gamma t},
c_{u\;\!\!^n}^{o{\scriptscriptstyle1}}(t){=}\frac{ig_{u\;\!\!^n\!w\;\!\!^1}}{\Gamma}(e^{\mm\Gamma t}{-}1)
,c_{v\;\!\!^\ell}^{{\scriptscriptstyle'\!}\vec{\omega}\neq\phi,o{\scriptscriptstyle1}}{=}0
\label{WignerWiesskopfC}
\\
&&\hspace{-5mm}\Gamma{\equiv}\!\sum_{u,n}^{u{\!\neq\!}w}|g_{w\;\!\!^1\!u\;\!\!^n}\!|^2
\label{GammaDefinition}
\eea
The corresponding power spectrum
\beq{}
\langle E\rangle_{t\rightarrow}^{\infty}=\!\!\!\!\!\sum_{o{\scriptscriptstyle1},n}^{u+o{\scriptscriptstyle1}=w}
\!\!\!{o\scriptstyle1}|c_{u\;\!\!^n}^{o{\scriptscriptstyle1}}|^2
=\!\sum_{k}\!\frac{k\omega e^{-\frac{k\omega}{k_{\scriptscriptstyle\!B\!}T}}}{e^{{\mm}\frac{k\omega}{k_{\scriptscriptstyle\!B\!}T}}{+}\cdots1}{=}\frac{\omega}{e^{\frac{\omega}{k_{\scriptscriptstyle\!B\!}T}}{\pm}1}
\label{powerSpectrum}
\eeq
In the second step of \eqref{powerSpectrum} we assumed that the radiation particle is quantized so that $o{\scriptscriptstyle1}$ is always integer $k$ times of the single particle unit $\omega$, $k=0$ and $1$ for fermion, while for bosons $k=0,1,2\cdots$. The exponential weight factor $e^{-\frac{\omega}{k_{\scriptscriptstyle\!B\!}T}}$ with $k_{\scriptscriptstyle\!B\!}T\equiv(8\pi G_{\scriptscriptstyle N}M)^{-1}$ in this step follows from the normalization condition of $c_{u\;\!\!^n}^{o{\scriptscriptstyle1}}$ and the fact that $g_{u\;\!\!^n\!w\;\!\!^1}$ is approximately constant as $\frac{w-u}{w+u}\ll1$ so that
\beq{}
\frac{g^{2}_{u\;\!\!^n\!w\;\!\!^1}}{\Gamma^2}\approx\frac{u[{=}e^{4\pi G_{\scriptscriptstyle N}(M-o{\scriptscriptstyle1})^2}]}{w_\mm({=}e^{4\pi G_{\scriptscriptstyle N}M^2}\mm1){+}\cdots2{+}1}\approx\frac{e^{-8\pi GMo{\scriptscriptstyle1}}}{e^{4\pi G_{\scriptscriptstyle N}M^2}}
\eeq
Obviously, the power spectrum \eqref{powerSpectrum} of the black hole spontaneous radiation is completely the same as those of Hawking radiation. It follows from the Wigner-Weisskopf approximation of the first particle radiation. 

In principle, for the second and more latter particles radiation, Wigner-Wiesskopf approximation could be used similarly, but the operation will become more complicated and the resulting power spectrum of the radiation will deviate from thermal type step by step. The physics reason for this expectation is, the radiation evolution of a black hole is not an equilibrium process. As the radiation proceeds on, the temperature $k_{\scriptscriptstyle\!B}T$ or horizon size $r_h$ of the black hole varies. The function form of $k_{\scriptscriptstyle\!B}T(t)$ or $r_h(t)$ encode information about the initial black holes. Different microscopic state in initials have different temperature variation curve $k_{\scriptscriptstyle\!B}T(t)$ or $r_h(t)$, see FIG.\ref{figWsixRadiationCurve} for examples. Only for the first few particle's radiation, the high degrees of degeneracy of initial black holes' microscopic state yields universal features of radiation, thermal type power spectrum.

In conventional understanding, during evolutions of a black hole caused by hawking radiation, the black hole is considered a classic object. After a time ${\scriptstyle\Delta}t$, the radiation product $o{\scriptstyle1}$ is believed to flied away from the black hole so the black hole will not be considered on superpositions of the initial and final black holes' microscopic state $c_1|w^1\rangle+c_u|u^x\rangle$, but on state of something like $|u^x\rangle=\mathrm{partial.tr}_{o{\scriptscriptstyle1}}|u^x{\otimes}o^{\scriptstyle1}\rangle$, with classical probability $\sum_{o{\scriptscriptstyle1}}|c^{o\scriptscriptstyle1}_{u^{\!n}}({\scriptstyle\Delta}t)|^2$. $|u^x\rangle$ here is in fact a mixed state due to partially tracing out of the radiation product state. Taking $|u^x\rangle$ as new initial state of the black hole and using approximations similar with \eqref{WignerWiesskopfA}-\eqref{WignerWiesskopfB}, we will get evolutions of $|u^x\rangle$ and so on. Compiling all these evolutions together, we will get an evolution chain of the following form
\beq{}
|\raisebox{-3pt}{\includegraphics[totalheight=4mm]{fig0cDecay1.png}}\rangle
\rightarrow
|\raisebox{-4pt}{\includegraphics[totalheight=5mm]{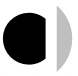}}\rangle
\rightarrow
|\raisebox{-5pt}{\includegraphics[totalheight=5mm]{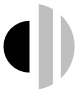}}\rangle
\rightarrow
|\raisebox{-5pt}{\includegraphics[totalheight=5mm]{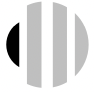}}\rangle
\rightarrow
|\raisebox{-5pt}{\includegraphics[totalheight=5mm]{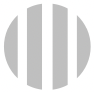}}\rangle
\Rightarrow
\raisebox{-5pt}{\includegraphics[totalheight=7mm]{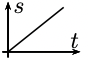}}
\label{SponRadiationChainA}
\eeq
where dark pie or segment of it and their lightgray partners denote the black hole and radiation products at different stages respectively. Since the Wigner-Wiesskopf approximation is used on each stage of the evolution so that history of the system is forgotten all the way till the black hole evaporated completely. Forgetting history implies information missing necessarily so it is very natural that we get a monotonically increasing curve for the entropy of radiation products. Including correlations between the radiation products escaping away from the black hole on different stages so that the evolution chain becomes
\beq{}
|\raisebox{-3pt}{\includegraphics[totalheight=4mm]{fig0cDecay1.png}}\rangle
\rightarrow
|\raisebox{-4pt}{\includegraphics[totalheight=5mm]{fig0dDecay2whl.png}}\!\rangle
\rightarrow
|\raisebox{-5pt}{\includegraphics[totalheight=5mm]{fig0eDecay3whl.png}}\!\rangle
\rightarrow
|\raisebox{-5pt}{\includegraphics[totalheight=5mm]{fig0fDecay4whl.png}}\rangle
\rightarrow
|\raisebox{-5pt}{\includegraphics[totalheight=5mm]{fig0gDecay5whl.png}}\rangle
\Rightarrow
\raisebox{-5pt}{\includegraphics[totalheight=7mm]{fig0hHwkCurve1.png}}
\label{SponRadiationChainB}
\eeq
will not change the qualitative feature that the entropy of radiation products increases monotonically. Because the forgetting history approximation happens on each stage of this evolution process.

Obviously, to retrieve back the information missed during Hawking radiation, the key is not to consider correlations between the radiation products escaping away on different stages of the evolution, but to take the black hole as a quantum object so that during any stage of its evolution, it is not living on mixed state of black holes with precisely specified mass, but on superpositions of mixed state of all possible masses. That is, we need change the evolution chain into forms
\bea{}
&&\hspace{-5mm}|\raisebox{-3pt}{\includegraphics[totalheight=4mm]{fig0cDecay1.png}}\rangle
\rightarrow
|\raisebox{-3pt}{\includegraphics[totalheight=4mm]{fig0cDecay1.png}}\rangle
+|\raisebox{-4pt}{\includegraphics[totalheight=5mm]{fig0dDecay2whl.png}}\!\rangle
+|\raisebox{-5pt}{\includegraphics[totalheight=5mm]{fig0eDecay3whl.png}}\!\rangle
+|\raisebox{-5pt}{\includegraphics[totalheight=5mm]{fig0fDecay4whl.png}}\!\rangle
+|\raisebox{-5pt}{\includegraphics[totalheight=5mm]{fig0gDecay5whl.png}}\!\rangle
\label{SponRadiationChainC}
\\
&&\hspace{-1mm}\rightarrow
|\raisebox{-3pt}{\includegraphics[totalheight=4mm]{fig0cDecay1.png}}\rangle
+|\!\raisebox{-4pt}{\includegraphics[totalheight=5mm]{fig0dDecay2whl.png}}\!\rangle
+|\raisebox{-5pt}{\includegraphics[totalheight=5mm]{fig0eDecay3whl.png}}\!\rangle
+|\raisebox{-5pt}{\includegraphics[totalheight=5mm]{fig0fDecay4whl.png}}\!\rangle
+|\raisebox{-5pt}{\includegraphics[totalheight=5mm]{fig0gDecay5whl.png}}\!\rangle
\nonumber\\
&&\hspace{-3mm}\rightarrow\cdots~\cdots~\cdots
\nonumber\\
&&\hspace{-5mm}\Rightarrow
\raisebox{-3pt}{\includegraphics[totalheight=7mm]{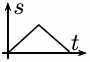}}
\eea
Or mathematically (state of the radiation products such as $r^b$, $r^m$, $r^s$ should be traced out to get mixed state of the black hole. Before that doing, the whole system is pure),
\bea{}
&&\hspace{-5mm}|\psi_0\rangle\rightarrow|\psi_0\rangle{+} c_b^{t_1}|b\;\!\!^b\!r\;\!\!^b\rangle{+} c_m^{t_1}|b\;\!\!^m\!r\;\!\!^m\rangle{+} c_s^{t_1}|b\;\!\!^s\!r\;\!\!^s\rangle{+}|\psi_1\rangle
\label{stateSuperposition}
\\
&&\hspace{2mm}\rightarrow|\psi_0\rangle{+}c_b^{t_2}|b\;\!\!^b\!r\;\!\!^b\rangle{+}c_m^{t_2}|b\;\!\!^m\!r\;\!\!^m\rangle{+}c_s^{t_2}|b\;\!\!^s\!r\;\!\!^s\rangle{+}|\psi_1\rangle
\nonumber
\\
&&\hspace{2mm}\rightarrow\cdots
\nonumber
\eea
with $|\psi_0\rangle$ and $|\psi_1\rangle$ denoting microscopic states of the initially non-radiating black hole and final pure vacuum with all matter contents radiated away respectively, $|{b\;\!\!^{b}r\;\!\!^{b}}\rangle$, $|{b\;\!\!^{m}r\;\!\!^{m}}\rangle$, $|{b\;\!\!^{s}r\;\!\!^{s}}\rangle$ being entanglements of big, median, small black holes with their corresponding radiation products, $t_1$, $t_2$ and et al distinguish time epochs on which the wave function value is taken. 
Reminds here, $|{b\;\!\!^{b}r\;\!\!^{b}}\rangle$, $|{b\;\!\!^{m}r\;\!\!^{m}}\rangle$, $|{b\;\!\!^{s}r\;\!\!^{s}}\rangle$ are only three representative intermediate state of the radiation process, they are not exhaustive.

The entropy of the black hole or radiations at any given time has to be calculated as a quantum expectation, see subsections following closely for more derivation or proof details of this equation,
\beq{}
s(t)=|c^t_b|^2s_{b\;\!\!^{b}r\;\!\!^{b}}+|c^t_m|^2s_{b\;\!\!^{m}r\;\!\!^{m}}+|c^t_s|^2s_{b\;\!\!^{s}r\;\!\!^{s}}
\label{entropySuperposition}
\eeq
where $s_{b\;\!\!^{b}r\;\!\!^{b}}$, $s_{b\;\!\!^{m}r\;\!\!^{m}}$, $s_{b\;\!\!^{s}r\;\!\!^{s}}$ denote the entanglement entropy of the big, median, small hole and their radiations. At early times, $c_b^{t_1}\gg c^{t_1}_m\&c^{t_1}_s$,  so $s(t_1)$ is dominated by $s_{b\;\!\!^{b}r\;\!\!^{b}}$ and increases with time. As time passes by, it will be dominated by $s_{b\;\!\!^{m}r\;\!\!^{m}}$ and reaches maximum at some intermediate epoch just as the Page-curve indicated, then decreases due to the domination of $s_{b\;\!\!^{s}r\;\!\!^{s}}$, and then Rabbi oscillates, as FIG.\ref{figWsixRadiationExample} displays.

At this moment, we need to point out that the reason we get the proper entropy variation curve for hawking particles is not due to our consideration of any small corrections originating from correlations between the radiation particles, but a completely new radiation mechanism and its full quantum description. None of this two factors is negligible. By completely new radiation mechanism, our mean is, hawking radiation happens through monopole type coupling between hawking particles and the inner structured black hole, just as described in the Z-model hamiltonian \eqref{HamiltonianB}. Implications of this coupling for Hawking particles are also discussed in \cite{stojkovic2020unp}. Replacing this mechanism, adopting the viewpoint hat hawking radiation happens through pair production around the purely classic and no hair horizon and escaping away from there, then Mathur's small correction theorem will be an in-escapable barrier to the proper page curve.  Realizing the power of small correction theorem is only one of the two factors to get the page curve, the other one is, we need a fully quantum description of the whole radiation evolving process. Otherwise, the two levels superposition could not be accounted for properly.

\subsection{Two Levels of Superposition}
\label{twoLevelsSuperposition}

\begin{figure}[h]
\includegraphics[totalheight=75mm]{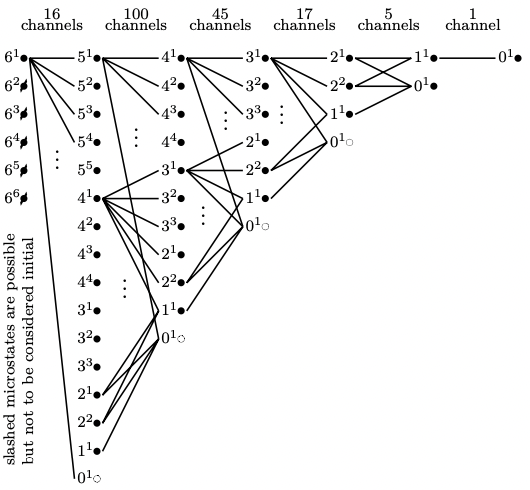}
\caption{All 184 possible transition channels of an initially 6-times degenerating black hole. Each node in the figure is characterized by microscopic state of the black hole, $w^i$, $w\sim e^{A/4G_N}$ is the degeneracy, $i$ distinguishes inner structures in the given degeneracy family.}
\label{figRadiChannel}
\end{figure}

\begin{figure}[h]
\includegraphics[totalheight=115mm]{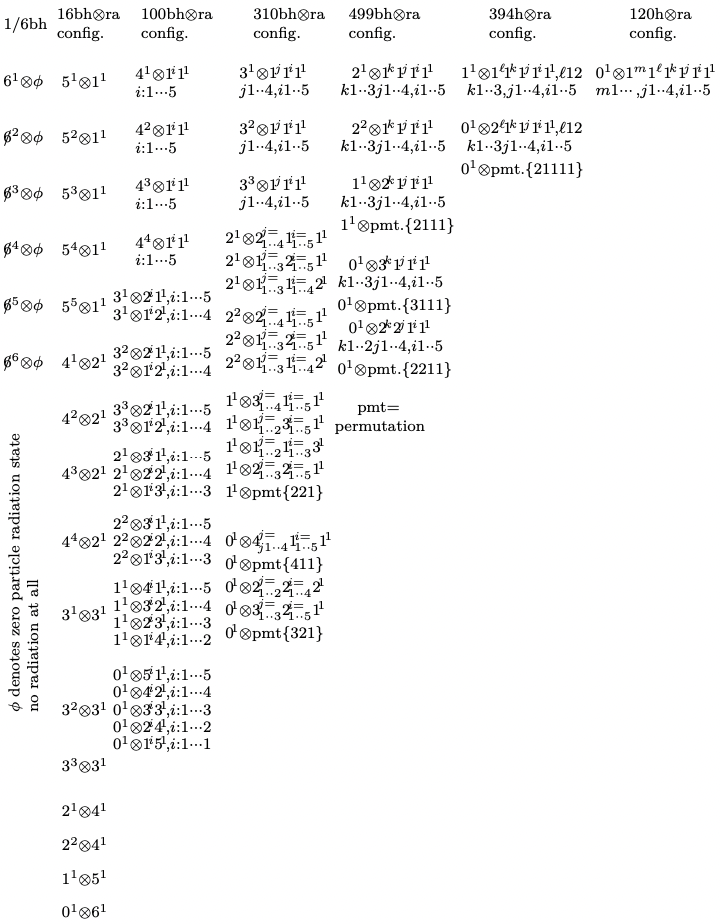}
\caption{All 1440 possible intermediate state produced by the radiation transition of an initially 6-times degenerating black hole.}
\label{figMiddleState}
\end{figure}

There are two levels of quantum state superposition involved in analysis above when deriving features of the page curve. 
\begin{enumerate}
\item[i] superposition of microscopic state of black holes with given mass but different inner structure
\item[ii] superposition of microscopic state of black holes with all possible masses
\end{enumerate} 
Sometimes we will call this latter superposition as macroscopic quantum interference. Just as we pointed above, this is one of the two most important observations in our understanding of information missing puzzle. To help readers more concretely catch this point, we provide in FIG.\ref{figRadiChannel} and \ref{figMiddleState} all possible quantum transition channel and intermediate state involved in the radiation of an inititially 6-times degenerating black hole. Without loss of generality, we assume the black hole lies on microscopic state $w^i=6^1$ at the beginning point.

From the figure, we easily see that, as long as the radiation begins, due to uncertainties of quantum mechanic events, things like a black hole radiates or not, it radiates on what time, it radiates a particle of how much mass, and et al all cannot be known definitely. So, intermediate/final state of the system will become superpositions of rather many basic states, far more than expected in conventional understanding of hawking radiation. For example, in conventionals, when a duration of page time passes by, an initially horizon area $A$ black hole will become an area $\frac{1}{2}A$ one. However, in the full quantum description, this is not the case. Instead, even at that time, the black hole should be considered as superposition of all possible horizon areas ones. Macroscopic interferences happens here. Time coming to the page epoch only means that the original black hole has very large probability to become an area $\frac{1}{2}A$ one, but not that way definitely. If we measure masses of the black hole classically and continuously so that the black hole is observed to transit to smaller and smaller descendants definitely,  then the total entropy of the system will increase monotonically as results of the macroscopic interference's continuously being broken by the classic observation.

This macroscopic quantum interference or two levels superposition character of intermediate state brings about new features for expressions of tracing out microscopic state of the black hole or radiation products. Using $|\psi_i\rangle$ and $|\psi_i\rangle_C$ to denote microstates of the radiation and black hole respectively, their direct product $|\psi_i\rangle\otimes|\psi_i\rangle_C$ will characterize global state of the system and is thus pure manifestly. 
\bea{}
&&\hspace{-6mm}|\psi_i\rangle_{\scriptscriptstyle\!C}\sim\mathrm{tracing.out.states.of.the.radiation.product}
\label{psiiDefinition}
\\
&&\hspace{-2mm}\{|\psi^0\rangle+c_b^t|b^br^b\rangle+c_m^t|b^mr^m\rangle+c_s^t|b^sr^s\rangle+|\psi^1\rangle
\}\nonumber
\\
&&\hspace{-6mm}\sim\sqrt{\rho^0}\oplus c^t_b\sqrt{\rho^b}\oplus c^t_m\sqrt{\rho^m}\oplus c^t_s\sqrt{\rho^s}\oplus\sqrt{\rho^1}
\label{addMixedstate}
\\
&&\hspace{-6mm}\rho^b=\mathrm{tr}_{rb}|b^br^b\rangle\langle{b^br^b}|,~\rho^m=\mathrm{tr}_{rm}|b^mr^m\rangle\langle{b^mr^m}|
\nonumber\\
&&\hspace{-6mm}\rho^s=\mathrm{tr}_{rs}|b^sr^s\rangle\langle{b^sr^s}|
\nonumber
\eea
where $\rho^0$, $\rho^b$, $\rho^m$, $\rho^s$ \& $\rho^1$ are density matrices of initially zero-radiation(no radiated), big(little radiated), median(half radiated), small(largely radiated) \& finally completely-radiated away black holes with microstates of the corresponding radiation products traced out. Since density matrices are not additive quantity, while their square root are wave function level objects thus additive quantity, we use expression \eqref{addMixedstate} to reflect this fact schematically. It is worth to notice that subscript $i$ on $|\psi_i\rangle$ takes value in a very big set of the following form
\bea{}
&&\hspace{-5mm}\cdots\cup\{1_b,2_b,\cdots N_b\}\cup\{1_m,2_m,\cdots N_m\}\cup\cdots
\\
&&\hspace{-3mm}\cdots N_b=e^{\mathrm{Area}_b/4G},N_m=e^{\mathrm{Area}_m/4G},\cdots
\eea
Each factors in the cup product corresponds to the range of indices of the microscopic state supporting the definition of the density matrix appearing in \eqref{addMixedstate}. 

With the help of microscopic state notation \eqref{addMixedstate}, we can express calculation results such as inner product of two quantum state the following way,
\bea{}
&&\hspace{-5mm}{}_{\scriptscriptstyle C\!}\langle\psi_i|\psi_j\rangle_{\scriptscriptstyle C}=
(\sqrt{\rho^0}\oplus\psi^t_b\sqrt{\rho^b}\oplus\cdots\sqrt{\rho^1})(\cdots)^\dagger
\label{psiijDefinition}\\
&&\hspace{-7mm}=\rho^0\oplus|c^{t}_b|^2\rho^b\oplus|c^{t}_m|^2\rho^m\oplus|c^{t}_s|^2\rho^s\oplus\rho^1//\mathrm{normalize}
\nonumber
\eea
Obviously, $\rho^0$ and $\rho^1$ correspond to pure state of the initial black hole and final vacuum when all matter contents are radiated away, they are obviously diagonalized on the black hole microstate basis
\bea{}
&&\hspace{-5mm}\rho^0_{ij}=\delta_{i1}\delta_{j1},i,j\in\{1,2,\cdots e^{\mathrm{Area}/4G}\}
,~S(\rho^0)=0
\label{rho0matrixform}
\\
&&\hspace{-5mm}\rho^1_{ij}=\delta_{i0}\delta_{j0},i,j\in\{0\},S(\rho^1)=0
\label{rho1matrixform}
\eea
However, $\rho^b$, $\rho^m$ and $\rho^s$ are not so simple, see section \ref{secSimilarityDefinition} for concrete working examples.. Due to operations of tracing out states of the radiation product, they are no long diagonal on the black hole microscopic state basis. We have to write generally
\bea{}
&&\hspace{-5mm}\rho^b_{ij}=d_{ij}+c_{ij},~i,j\in\{1,2,\cdots e^{\mathrm{Area}_b/4G}\}
\label{rhobmatrixform}
\\
&&\hspace{-5mm}\rho^m_{ij}=d_{ij}+c_{ij},~i,j\in\{1,2,\cdots e^{\mathrm{Area}_m/4G}\}
\label{rhommatrixform}
\\
&&\hspace{-5mm}\rho^s_{ij}=\cdots,\cdots
\label{rhosmatrixform}
\eea
where $d_{ij}$ and $c_{ij}$ represent the diagonal and off-diagonal elements of $\rho_{ij}$ respectively. 

\subsection{Why Replica Worm Hole Works?}
\label{workingLogicsReplicWormHole}

The replica worm hole method does not provide full description for the whole radiation process, but focuses on calculations of the radiation products' second order Rene entropy related quantities like
\bea{}
&&\hspace{5mm}\mathrm{tr}\rho^2_{\scriptscriptstyle R}\sim({}_{\scriptscriptstyle C\!}\langle\psi_i|\psi_j\rangle_{\scriptscriptstyle C})({}_{\scriptscriptstyle C\!}(\langle\psi_i|\psi_j\rangle)_{\scriptscriptstyle C})^*=
\label{replicaWormholeA2}\\
&&\hspace{-5mm}\parbox{80mm}{\includegraphics[totalheight=35mm]{fig0aReplica.png}}
\label{replicaWormholeB2}
\eea
where $|\psi_i\rangle_{\scriptscriptstyle C}$, $|\psi_j\rangle_{\scriptscriptstyle C\!}$ are complementary state of the radiation product or equivalently mixed state of the black hole following from tracing out of the radiation factors from the global pure state $|\psi_i\rangle_{\scriptscriptstyle\!C}\otimes|i_R\rangle$, $|\psi_j\rangle_{\scriptscriptstyle\!C}\otimes|j_R\rangle$. Tracing of $\rho_R^2$ is to be done over the remaining $ii$, $jj$ index of the radiation state. The picture structure of \eqref{replicaWormholeB2} for $\mathrm{tr}\rho_R^2$ is then argued to appear in mathematical expressions for
\beq{}
\mathrm{tr}\rho^1_{\scriptscriptstyle R}\sim({}_{\scriptscriptstyle C\!}\langle\psi_i|\psi_j\rangle_{\scriptscriptstyle C})=
\parbox{40mm}{\includegraphics[totalheight=15mm]{fig0bReplica.png}}
\label{replicaWormholeC2}
\eeq
so that $\mathrm{tr}\rho_R^1$ includes contributions also from two terms
\beq{}
S(R){=}\lim_{n\rightarrow1^+}\frac{1}{1{-}n}\log\mathrm{tr}(\rho^n_{\scriptscriptstyle R}){\approx}\min\{\mathrm{disconn},\mathrm{conn}\},
\label{replicaWormholeD2}
\eeq
As is well known, the structure of Feynamn diagrams in perturbative quantum field theory is topological. For example, the 2->2 scattering process and the 1->1 propagation have totally different diagram representation and algebraic structure. It is almost impossible to get algebraic structures of the latter by continuation of the former's digram representation. So it should be very difficult to prove the statement that $\mathrm{tr}\rho^1_R$ and $\mathrm{tr}\rho^2_R$ have similar picture structure. We suspect that such provments exist at all. 

However, just as we pointed out above, $|\psi_i\rangle_{\scriptscriptstyle\!C}$ \& $|\psi_j\rangle_{\scriptscriptstyle\!C}$ are not microscopic state of black holes with specified masses, but superposition of mixed state of different mass ones, see equations \eqref{psiiDefinition} and \eqref{addMixedstate} for refferrings. When inner product of $|\psi_i\rangle$ and $|\psi_j\rangle$ are considered, our results are
\bea{}
&&\hspace{-5mm}{}_{\scriptscriptstyle C\!}\langle\psi_i|\psi_j\rangle_{\scriptscriptstyle C}
=\rho^0\oplus|c^{t}_b|^2\rho^b\oplus|c^{t}_m|^2\rho^m\oplus|c^{t}_s|^2\rho^s\oplus\rho^1
\eea
instead of simple $\langle\psi_i|\psi_j\rangle=\delta_{ij}$ of naive orthonormal relation implies. Here, only $\rho^0$ and $\rho^1$ are diagonalized on the basis of black hole inner structure eigenstate. $\rho^b$, $\rho^m$ and $\rho^s$ are not, see equations \eqref{rho0matrixform}-\eqref{rhosmatrixform} for referrings. 

Noticing this fact, we easily understand that in calculating quantities such as $\mathrm{tr}\rho_R^2$, why the method of replica worm hole provides us results diagrammatically interpretable the way like those in \eqref{replicaWormholeB2}. However, this is not evidence that similar diagrammatic representation would also appear in the calculation of $\mathrm{tr}\rho_R^1$, like those in \eqref{replicaWormholeD2}. Instead, from equations \eqref{addMixedstate} and \eqref{psiijDefinition}, we see that
\bea{}
&&\hspace{5mm}S(R)=\lim_{n\rightarrow1^+}\frac{1}{1{-}n}\log\mathrm{tr}(\rho^n_{\scriptscriptstyle R})
\\
&&\hspace{-9mm}=0+|c_b^t|^2S(\rho^b)+|c_m^t|^2S(\rho^m)+|c_s^t|^2S(\rho^s)+0
\eea
During early stages of the radiation, $c^t_b\gg c^t_m\gg c^t_s$, so that the entropy of the black hole is mainly contributed by $S(\rho^b)$. While during latter stages, directions of the wave function value inequality reverse, $c^t_b\ll c^t_m\ll c^t_s$, so the entropy is mainly contributed by $S(\rho^s)$. Both $S(\rho^b)$ and $S(\rho^s)$ are very small, but $S(\rho^m)$ is very large, which dominates contributions to the entropy of the system during middle stages of the radiation. As results, we get first increasing then decreasing trend for the entropy variation of an evaporating black hole. Obviously, this trend follows directly from the fact that, even after the radiation comes to the latter stage, the quantum state of the system is superpositions of mircrostates of all possible mass black holes.

The quantum extremal surface or island formula also have correspondences in our two levels superposition description of the black hole radiation. We just need to go back to \eqref{psiiDefinition}-\eqref{addMixedstate}, but only partially tracing out microscopic state of the radiation products in such a way that each to be added density matrix square root has equal size so that they could be added together by conventional ways instead of the direct summation we adopted previously,
\bea{}
&&\hspace{-8mm}|\psi_i\rangle_{\scriptscriptstyle\!C}\sim\mathrm{partially.tracing.out.state.of.the.radiation}
\label{psiiDefinition2}
\\
&&\hspace{-2mm}\{|\psi^0\rangle+c_b^t|b^br^b\rangle+c_m^t|b^mr^m\rangle+c_s^t|b^sr^s\rangle+|\psi^1\rangle
\}\nonumber
\\
&&\hspace{-8mm}\sim\sqrt{\rho^0}+c^t_b\sqrt{\rho^b{\otimes}\rho_r^b}+c^t_m\sqrt{\rho^m{\otimes}\rho_r^m}+c^t_s\sqrt{\rho^s{\otimes}\rho_r^s}
\label{addMixedstate2}\\
&&\hspace{35mm}+\sqrt{\rho^1{\otimes}\rho_r^1}
\nonumber
\\
&&\hspace{-8mm}(\rho^b{\otimes}\rho_r^b)_{N\times N}=\mathrm{tr}_{rb}^\mathrm{partially}|b^br^b\rangle\langle{b^br^b}|
\eea
\bea{}
&&\hspace{-8mm}\rho^0_{ij}=\delta_{i1}\delta_{j1},i,j\in\{1,2,\cdots N\}
,~N=e^{A_\mathrm{ini}/4G}
\\
&&\hspace{-8mm}(\rho^m{\otimes}\rho_r^m)_{N\times N}=\mathrm{tr}_{rm}^\mathrm{partially}|b^mr^m\rangle\langle{b^mr^m}|
\nonumber
\\
&&\hspace{-8mm}(\rho^s{\otimes}\rho_r^s)_{N\times N}=\mathrm{tr}_{rs}^\mathrm{partially}|b^sr^s\rangle\langle{b^sr^s}|
\nonumber
\\
&&\hspace{-8mm}(\rho^1{\otimes}\rho_r^1)_{N\times N}=1_{1\times1}\otimes\rho^0
\nonumber
\eea
Using this $|\psi_i\rangle_{C}$ definition, the inner product ${}_{\scriptscriptstyle C\!}\langle\psi_i|\psi_j\rangle_{\scriptscriptstyle C}$ has forms
\bea{}
&&\hspace{-6mm}{}_{\scriptscriptstyle C\!}\langle\psi_i|\psi_j\rangle_{\scriptscriptstyle C}=\rho^0+|c_b^t|^2\rho^b{\otimes}\rho_r^b+|c_m^t|^2\rho^m{\otimes}\rho_r^m+\cdots
\\
&&\hspace{-6mm}\nonumber+c_b^t\sqrt{\rho^0.(\rho^b\otimes\rho^b_r)}\mathrm{like~crossing~terms}
\\
&&\hspace{-6mm}\equiv \rho_R
\nonumber
\eea
These accompanying factors $\rho^b_r$, $\rho^m_r$ and $\rho^m_r$ will enter calculations of $S(R)=\lim\limits_{n\rightarrow1^+}{\log\mathrm{tr}\rho_R^n}/(1-n)$ and appear as island, for which quantum extremal surface formula provides effective accounting.

\section{Inside Horizon Structure of Black Holes}
 
To this point, we believe broadminded readers may accept that, our resolutions to the information missing puzzle through spontaneous radiation of black holes, i.e. Z-model and its full quantum description catches key point of the question somehow. The remaining questions become, do physic black holes allow inner structure like those of usual atoms at all, so that spontaneous radiation of them bear meanings at all? And does such inner structure forms contradiction with the classic no-hair theorem and singularity theorem? As is well known, realities of the Bekenstein-Hawking's entropy tell us that black holes have many possible microscopic states indeed. The question for us is, can we find classic solutions to the Einstein equation corresponding such microscopic states?  Can we implement quantization of such classic solutions and get the proper area law entropy formula? Our answers to all these questions are, yes.
 
 \subsection{The Physic Ideas Overview}
 
Early attempts to understand the microscopic state and radiation of black holes on atomic physics levels is J. Bekenstein \& V. Mukhanov's area quantization scheme \cite{Bekenstein1974a,Mukhanov1986,Bekenstein1995,Bekenstein1997a,Bekenstein1997b} and Nambu \& Sasaki et al's inside-horizon wave function model \cite{Allen1987,NambuSasaki1988,Nagai1989,BFZ1994,KMY2013}. Other related works include D. Stojkovic \& collaborator's collapsing domain wall picture \cite{Stojkovic2007,Stojkovic2008a,Stojkovic2008b,Stojkovic2009,Stojkovic2013,Stojkovic2014,Stojkovic2015}. B\&M's starting point is the adiabatic invariance of horizon area. Basing on this invariance, they argue that the mass spectrum of black holes is quantized and uniformly-spaced. As results, their emission and absorption spectrum is of line shape in the single-quanta process approximation. To get the continuous spectrum of Hawking radiation, some unknown principle must be invoked to encourage the multi-quanta process but forbid the single quanta ones. In B\&M's idea, the microscopic state of black holes is a reflection of the fact that their horizon could be constructed by sticking unit patches one by one, thus a feature of the horizon itself, but nothing with the motion of matter constituents of the black hole.

\begin{figure}[h]
\includegraphics[totalheight=35mm]{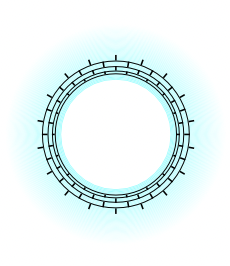}
\includegraphics[totalheight=35mm]{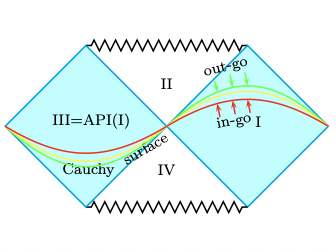}
\caption{Left is the old brick wall model, the microscopic state of black holes is proxied by spherical harmonic modes of an auxiliary fields living outside of the classic horizon. Right is the API-augmented brick wall model, the auxiliary field is thrown away and attentions are put on microscopic state of in-going and out-going particles on an arbitray initial Cauchy surface in the API-folded Penrose diagram of an eternal black hole. The direction of time flow arrow in the API-world is reversed relative the normal world. Particles going toward the horizon with ever-increasing energy will leave imprints on particles escaping away from there with ever-decreasing energy thus letting information being carried out, with no information missing at all.}
\label{figBrickWallModel}
\end{figure}

Different from B\&M, G. 't Hooft's brick wall model \cite{tHooft1984,tHooft1985,Aichelburg1971,Bonnor1969,tHooft1985b} and its API (antipodal point identifcation) augmented version and FWT (firewall transformation)\cite{tHooft2019,tHooft1996,tHooftISSP2016,tHooft1601,tHooft1605,tHooft1612,tHooft1804,tHooft1809,tHooftFP2018,tHooftISSP2018} - could be considered a pure classic prescription for the black hole's microscopic state interpretation and information missing puzzle. In the old model, the microscopic state of black holes is associated with motion modes - spherical harmonics - of an auxiliary field living in a finitely thick shell-region outside the horizon. This auxiliary field is essentially a proxy for matters consisting of the black hole. In its API augmented version, the auxiliary field is thrown away and the focus becomes an initial Cauchy surface wrapping around the horizon of the black hole. Observers far away from that surface see endless streams of in-going and out-going particles there, evolutions of the black hole are determined by purely gravitational interaction between these two streams. By API, space-time points in region III of the Penrose Diagram are identified as antipodal partners in region I, see FIG.\ref{figBrickWallModel} for intuition. While FWT transcripts information carried by the in-going particles with ever-increasing energy to those with ever-decreasing one through Shapiro time delaying. With this two ingredients, 't Hooft successfully avoids all ultra high energy degrees of freedom around the horizon and proves unitarity for general ingoing-outing processes happening on the background of eternal black holes through pure general relativity idea instead of any unknown quantum gravitation arguments. In both the old brick wall model and its API-augmented version, 't Hooft only resort physics occurring outside the horizon of black holes. The microscopic state of black holes is related with all possible spherical harmonic modes of matters consisting of the black hole, even for the spherically symmetric Schwarzschild one.

\begin{figure}[h]
\includegraphics[scale=0.32]{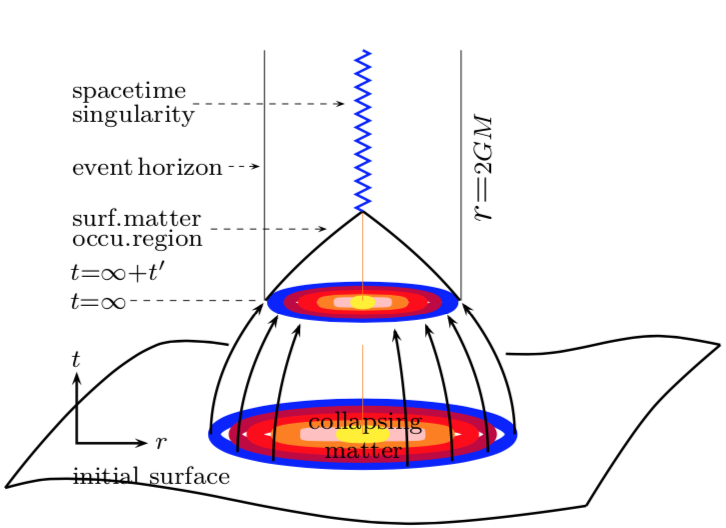}
\includegraphics[scale=0.32]{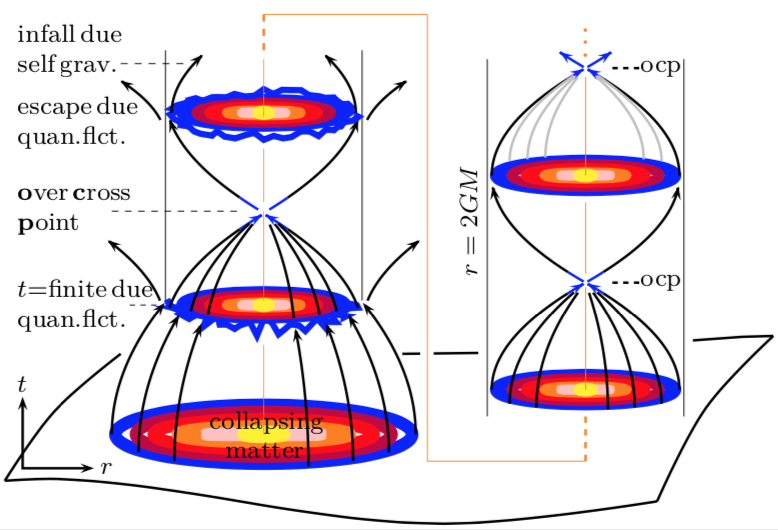}
\caption{Left panel is the picture of conventional gravitation collapses, infinite coordinate time is needed for horizons to form. After that matters aggregate to the central point in finite proper time and form singularities there. Singularity symbolizes classic physics especially the time concepts' terminal. The right panel considers the uncertainty principle, only finite coordinate time is needed for horizons to form. As matters aggregate to the central point, they will over cross each other due to forward amplitude domination of quantum collision and then periodically oscillate around there. In the early few rounds, every time matters getting close to the horizon, due to randomness of quantum fluctuation, part of the matters will fall across and into, while others escape away from the horizon. But in latter duration, they will form stable oscillation pattern inside the horizon, except the very weak spontaneous radiation effects. From pure classic aspects, the inside horizon oscillation happens beyond the domain of time definition of outside observers. It can be considered physics in those observers' parallel universe or statistic ensemble world.}
\label{figCollapsingProcess}
\end{figure}

\begin{figure}[h]
\includegraphics[totalheight=30mm]{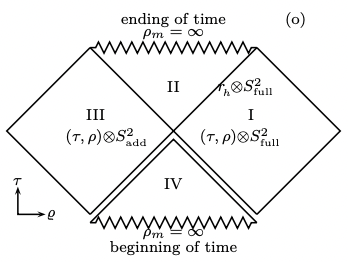}
\includegraphics[totalheight=32mm]{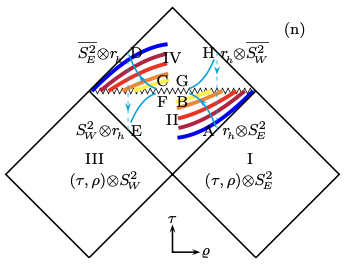}
\caption{The left hand side is the conventional Penrose-Carter diagram of a maximally extended Schwarzschild spacetime. The diagram space maps to the 3+1 spacetime through $(\tau\!,\varrho){\otimes}S^2_\mathrm{full}$ or $(\tau\!,\varrho){\otimes}S^2_\mathrm{add}$. The spacetime has two asymptotically flat region I and III, which are connected by an Einstein Rosen bridge. The time has a beginning as well as an ending point, a white hole exists before the black hole forms. Singularities on the future-most of II and past-most of IV is non-overcross-able. The right hand side is our Penrose-Carter diagram representation of a Schwarzschild blackhole with its consisting matters' motion inside the horizon, which happens beyond the domain of time definition of outside observers. We consider it as physics in those observers' parallel universe or statistic ensemble world. Only half of the matter core's evolving is plotted and denoted with a family of colorful curves for good looking. The diagram space maps to the 3+1 spacetime through $(\tau\!,\varrho){\otimes}S^2_{\scriptscriptstyle E}\cup(\tau\!,\varrho){\otimes}S^2_{\scriptscriptstyle W}$, with $S^2_{\scriptscriptstyle E/W}$ denoting the east/west panel of a full sphere and $\overline{S}_{\scriptscriptstyle E/W}=S{\scriptscriptstyle W/E}$. Regions III and I are  simply connected both physically and mathematically. Region IV lies in the future of region II, with the border overcross-able. A representation point co-moving with matters consisting of the blackhole follow periodical tracks like A-B-C-D-E-F-G-H-A$\cdots$, C is the antipodal of B, E is the identified with D, and et al.}
\label{figPCdiagramSchwz}
\end{figure}

What we have been pursuing in references \cite{dfzeng2017,dfzeng2018a,dfzeng2018b,dfzeng2020} and what we will report in the following as the basis for our spontaneous radiation theory of black holes is a road to the microscopic state interpretation and is an idea hybridizing Bekenstein's atomic level quantization and 't Hooft's API-augmented brick wall model, rather similar with the string theory fuzzy balls, but uses no hyper-physics idea such as extra-dimension and super-symmetry et al. Sometimes we will call such an inner structure model as the Z-structure of black holes for convenience. Different from Mukhanov and Bekenstein, we do not attribute the microscopic state of black holes directly to the horizon area degrees of freedom, but to oscillation modes - especially the spherically symmetric ones - of matter contents of the black hole. We will build a more close analogy between the black hole and usual atoms, deriving out its matter contents' equation of motion in details and writing down the quantum wave function explicitly. Different from 't Hooft, Z-structure does not refuse to discuss physics related with matter contents of the black hole fall across the horizon and oscillate across the central point. Because such motion happens outside the domain of time definition of outside observers, it can be interpreted as physics in the parallel universe or statistic ensembles of the outside observer. As is well known, all details of the microscopic state related motion is sensible only to those Maxwell demon like observers. The difference between black hole and other statistic mechanic systems is, horizons of the black hole set a definite watershed between the outside observer and Maxwell demon. Similar with 't Hooft, we will also use antipodal identification to help description of matters oscillation across the central point of the system. In 't Hooft, this happens on the two sides of horizon surface. While in ours, this happens on the two sides of central point of the system.  Out pictures are illustrated in Figure.\ref{figCollapsingProcess} and \ref{figPCdiagramSchwz}.

FIG.\ref{figCollapsingProcess} compares physics of the conventional gravitation collapse and ours. In conventional one, an observer sitting far away need to wait infinite time to see a matter cluster's contraction into its to-be-horizon surface under self gravitation. But this is a pure classic picture because uncertainty principle is not considered. When the uncertainty principle is included, the cluster has always probabilities be seen by distant observers to collapse into the horizon in finite duration. Of course due to energy conservation, there must be some matters sputtering outside as the collapsing events occur due to uncertainties.  In conventional picture, as long as matters falling across the horizon, they will aggregate to the central point and form singularities there. Singularity means terminal of all physics, especially the  time concept itself.  But this is in fact an artificial doctrine without physical basis. In our pictures, matter particles aggregating to the central point is allowed to over-cross and pass each other without any harmless. This is nothing but a counterpart of the classic wave superposition principle or the simple forward amplitude domination in high energy scatterings. We will give explicit metric description for this inside horizon oscillation across the central point in the next section. The singularity theorem of Penrose and Hawking is well respected here because matters falling across the horizon indeed get on the central point in finite proper time. Different from the popular saying, we do not consider such an event as the terminal of anything, it is a just normal intermediate point of many cycles of inside horizon oscillation. In this picture, API happens across the central point of the matter contents, it is a physic mechanism - over cross oscillation instead any work assumption. While in 't Hooft's idea of fire wall transformation, API is a working assumption imposed on the horizon surface because in there, all inside horizon matters' motion bears no meaning for outside observers.

FIG.\ref{figPCdiagramSchwz} is the Penrose Carter diagram representation of our inside horizon oscillation, i.e. Z-structure in details. Parallel with 't Hooft's API-folding operation, we introduce tricks that right panel of the diagram space represents only east semi-sphere of real world 3 dimensional space, while left panel of the diagram represents west semi-sphere of the real world. But different from 't Hooft's constraining discussion only in regions I and III, when defining microscopic state of black holes, we will use physics in regions II and IV and shift IV to the future of II and introduce antipodal identification so that points like D and E, or A and H are identified as the same point physically. Under this identification, white hole concepts disappear from our physics and we no longer have the necessity to distinguish event horizons of the future and past type, only one kind exists here. Also different from 't Hooft's definition of the microscopic state of black holes as all possible spherical harmonic modes, our microscopic state comprises of the matter contents' radial modes only. It is very difficult to imagine that superposition with modes of many nonzero angular momentum could lead to ideally spherical symmetric black holes such as a Schwarzschild one. Similar difficulty also happens to those ideas directly using horizon surface to save information of black holes \cite{Bekenstein1974a,Mukhanov1986,Bekenstein1995,Bekenstein1997a,Bekenstein1997b}, but not to those \cite{Carlip1995,Strominger1996,Strominger1997,Banados1997,Carlip1997,Banados1998,Carlip2005,Carlip1998} working through holographic principles. However, in a black hole of exactly spheric symmetric, non-trivial inside horizon radial motion of matter contents is possible. We will consider the whole matter core of a black hole as the nest of many concentric shells\footnote{Arbitrariness of this doing will be reduced by quantization. The number of shell decomposition and the way of assigning masses to each shell will be determined by quantization.} and quantize them one by one, taking their direct product as the wave functional of the whole system, thus getting spectrums of all possible microscopic states, which is finite and countable, with the result consistent with the area law entropy of Bekenstein Hawking formula. However, different from the horizon area quantization scheme of Bekenstein and Mukhanov, our mass spectrum of the black hole is continuous. This provides us basis to get desired continuous shape of radiation spectrum as hawking formula requires.

As long as we adopt quantum viewpoints with the motion of matters consisting of the black hole, we will get a highly blurred horizon region instead a cut-clear surface, all matter shells consisting of the black hole will have nonzero probabilities be observed outside the horizon. This provides us a picture very similar with the string theory fuzzy balls, which we call Schwarzschild fuzzy ball in \cite{dfzeng2018a,dfzeng2018b}.  With this quantum description for Z-structure of black holes, we master the most important ingredients to describe spontaneous radiation of black holes the same way we are told in atomic physics. Obviously, to implement such a description, an intuitive picture such as string theory fuzzy ball is not enough, explicit wave functions for each microscopic state of the black holes are necessary. Nevertheless, our Z-structure model for black holes is not irreplaceable. For readers which hold strongly negative attitude with Z-structures can neglect relevant contents in the following and jump to the concluding section directly. They can replace such an inner structure model with any other alternatives they prefer, such as the string theory fuzz ball or other things \cite{strominger1996,MT0103,LuninMathur0202,LuninMaldacenaMaoz0212,GMathur0412,mathur0502,mathur0510,Jejjala0504,KST0704,Mathur0706,lqgEntropy1996,lqgEntropy1997,gravMemoryChristodoulou1991,HPS201601,HPS201611,emMemorySabrina2017}. Of course, before such doing, they need to make their models more perfect so that similarities of the initial-final microscopic state could be calculated quantitatively. The bottom line is, they need to accept one thing, microscopic states embodied in the Bekenstein-Hawking entropy correspond truly-existing inner structure of black holes \cite{Bekenstein1973,Bardeen1973,Bekenstein1974b,Bekenstein1981,Wald1999} and the Hawking radiation is a result of quantum transition between those microscopic states. Nevertheless, Z-structure model has many attractive features, both its classic metric and quantum wave function can be written down explicitly. Just as we will explore in the following, interpretation of black hole entropy and Bekenstein-Hawking formula basing on this model leads to a new theorem on real number's partition which is provable from pure mathematic aspects. At the same time, merging of binary black holes with this inner structure will yield characteristic shape of gravitational waves thus disprovable observationally \cite{cardoso2017}. Let us begin from their classic geometric description firstly.

\subsection{3+1 Dimensional Schwarzschild Internal}  
\label{subsec31exactsolution}

Early works on exact solutions related with collapsing matters include \cite{oppenheimer1939,MTWbook,Yodzis1973,Yodzis1974,Vaidya1950,Vaidya1970}. But as long as we know, we may be the first to consider the possibility that collapsing matters may execute oscillations inside the horizon instead the once and for all collapses and causing of eternal singularity \cite{dfzeng2017,dfzeng2018a,dfzeng2018b,dfzeng2020}. Reference \cite{dfzeng2020} provides two exact solution families to the Einstein equation with oscillatory dust matter contents. One is in 3+1 dimensional asymptotically Minkowskian background, while the other is in asymptotically 2+1 asymptotically Anti de Sitter spacetime. Both solutions are spherical symmetric and characterized by a radial mass profile function $M[\varrho]$ and total mass $M[\varrho]_{\varrho>\varrho_\mathrm{min}}=M_\mathrm{tot}=\mathrm{const}$. Although we choose the no-pressure dust as the proxy for matter sources, in the case when gravitation dominates all other interaction, this is a precise enough approximation for general matter sources. 

Using Lemaitre style coordinate system, the inside horizon metric of our first solution family can be written as follows,
\bea{}
&&\hspace{-5mm}ds^2_\mathrm{in}=-d\tau^2\!+\!\frac{\big[1\!-\!\big(\frac{2GM}{\varrho^3}\big)\!^\frac{1}{2}\!\frac{M'\!\varrho}{2M}\tau\big]^2\!d\varrho^2}{a[\tau,\varrho]}{+}a[\tau,\varrho]^2\varrho^2d\Omega^2_2
\label{genOSmetric}
\\
&&\hspace{-5mm}a[\tau,\varrho]\!=\!\big[1\!-\!\frac{3}{2}\big(\frac{2GM[\varrho]}{\varrho^3}\big)\!^\frac{1}{2}\tau\big]^{\!\frac{2}{3}}
,~M[\varrho\!\geqslant\!\varrho_\mathrm{max}]=M_\mathrm{tot}
\label{genOSscalefactor}
\\
&&\hspace{-5mm}a[\tau\!\in\!|\!_{\frac{p\;\!\!^\varrho}{4}}^{\frac{p\;\!\!^\varrho}{2}},\varrho]=-a[\frac{p\;\!\!^\varrho}{2}{-}\tau,\varrho]
,a[\tau|\!_{\frac{p\;\!\!^\varrho}{2}}^\frac{p\;\!\!^\varrho}{1},\varrho]=-a[p^\varrho{-}\tau,\varrho]
\label{aExtension}
\\
&&\hspace{-5mm}a[\tau,\varrho]=a[\tau{+}p^{\varrho},\varrho]
,p^\varrho\equiv\frac{8}{3}\big(\frac{\varrho^3}{2GM[\varrho]}\big)\!^\frac{1}{2}
\label{aperiodic}
\eea
\beq{}
T_{\mu\nu}\!=\!\mathrm{diagonal}\{\frac{M'[\varrho]/8\pi\varrho^2}{a^{\!\frac{3}{2}}
\!\!+\!\!\frac{3GM'\tau^2}{4\varrho^2}
\!-\!\big(\frac{GM}{\varrho^3}\big)\!^\frac{1}{2}\frac{M'\varrho\tau}{2M}},0,0,0\}
\label{dustEMTensor}
\eeq 
where $\tau$ and $\varrho$ is the proper time and radial coordinate of co-moving observers, $M[\varrho]$ is the characterizing function of the family, whose concrete form at classic levels is completely arbitrary except increases monotonically and gets maximal at some pre-specified $\varrho=\varrho_\mathrm{max}$. $\varrho$ is very like the oscillatory x-coordinate of a test particle in a linear-inverse potential well, while the exact meaning of $M[\varrho]$ is the mass-energy of matter-contents inside the co-moving sphere of radius $\varrho$.

It can be easily proven \cite{GRweinberg} that the inside horizon metric \eqref{genOSmetric} smoothly joins to the outside Schwarzschild one in the same Lemaitre style coordinate
\beq{}
ds^2_\mathrm{out}=-d\tau^2{+}\frac{d\varrho^2}{(1{-}\frac{3\tau}{2r_g})^\frac{2}{3}}{+}(1{-}\frac{3\tau}{2r_g})^\frac{4}{3}r_g^2d\Omega_2^2
\label{SchwarzschildOutsideLemaitre}
\eeq
with $r_g=\frac{3}{8}p^{\varrho_\mathrm{max}}_\mathrm{eriod}=2GM_\mathrm{tot}$. The Lematre style coordinate $\tau,\varrho$ is related with the usual Schwarzschild one through transformations of the following form
\bea{}
&&\hspace{-5mm}d\tau=dt+\big(\frac{r_g}{r}\big)^\frac{1}{2}\big(1-\frac{r_g}{r}\big)^{-1}dr
\\
&&\hspace{-5mm}d\varrho=dt+\big(\frac{r}{r_g}\big)^\frac{1}{2}\big(1-\frac{r_g}{r}\big)^{-1}dr
\\
&&\hspace{-5mm}\big(1-\frac{3\tau}{2r_g}\big)^\frac{2}{3}=\frac{r}{r_g}
\eea 
under which the metric function \eqref{SchwarzschildOutsideLemaitre} can be written into the usual Schwarzschild one 
\beq{}
ds^2_\mathrm{out}=-hdt^2+h^{-1}dr^2+r^2d\Omega^2,h=1-\frac{2GM_\mathrm{tot}}{r}
\label{SchwarzschildOutsideSchwarzschild}
\eeq

\begin{figure}[h]
\begin{center}
\includegraphics[width=0.49\textwidth]{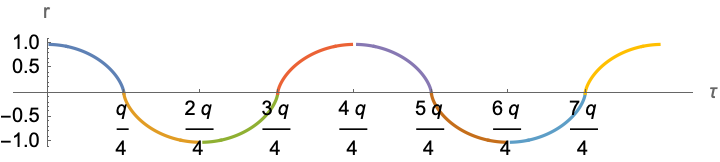}
\includegraphics[width=0.49\textwidth]{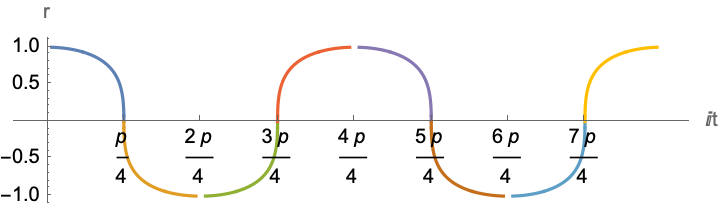}\end{center}
\caption{The oscillatory $r(\tau)$ and $r(it)$ curve of a representation particle on the surface of a spherical symmetric dust ball freely moving under self-gravitations. $it$ is the continuation of the Schwarzschild time $t$ defined by infinitely far away observers. We set $GM=1$ and $b=1$, so our representation particle is released from a less than horizon size configure and the oscillation period equals $q=4\pi$, $p=(8\sqrt{2}-10)\pi$. In the case $b\rightarrow2GM=r_h$, $p\rightarrow\infty$. While as $b>2GM$, that is the ball is released from larger than horizon size configure, $t_{r\rightarrow r_h+\epsilon}\rightarrow\infty$, but $\tau_{r\rightarrow r_h+\epsilon}$ is finite.}
\label{figOscillatoryScaleFactor} 
\end{figure}

Consider a representation particle co-moving with the surface of the matter occupation region, we can write down its equation of motion as follows,
\bea{}
&&\hspace{-5mm}\Big\{\begin{array}{l}
\ddot{t}+\Gamma^{t}_{tr}\dot{t}\dot{r}=0
\Rightarrow h\dot{t}=\mathrm{const}\equiv\sqrt{1-\frac{2GM}{b}}
\\
h\dot{t}^2-h^{-1}\dot{r}^2=1
,~h=1-\frac{2GM}{r}
\end{array}
\label{GRoscillationEq}
\eea
where for expression succinctness we have chosen writing $M_\mathrm{tot}=M$ and will assume that $r\leqslant b<2GM$, so the matter core is released freely from an inside horizon configuration. In this case equations \eqref{GRoscillationEq} can be integrated exactly to yield the 1/4th-period expression
\bea{}
&&\hspace{-5mm}r(\tau):\tau\!=\!\frac{b^{3/2}\arcsin\!\big[(1-\frac{r}{b})^\frac{1}{2}\big]}{\sqrt{2GM}}
\!+\!\frac{\sqrt{br(b\!-\!r)}}{\sqrt{2GM}}
\\
&&\hspace{-5mm}r(t):it=\frac{4GM\big(\arcsin\sqrt{\frac{b-r}{b-\frac{br}{2GM}}}\big)}{\sqrt{1-\frac{b}{2GM}}}\!-\!\sqrt{(b\!-\!r)r}
\\
&&\hspace{5mm}\!
-(4GM\!+\!b)\arcsin[(1-\frac{r}{b})^\frac{1}{2}]
\nonumber
\eea
The remaining 3/4th-period expression can be obtained by continuation or extensions very similar to those in equation \eqref{aExtension}.
We plot in FIG.\ref{figOscillatoryScaleFactor} the fully extended periodic behavior of $r(\tau)$ and $r(it)$ directly. From the figure, we easily see that  on each 1/4th periodical connection point, $r(\cdots)$ is continuous and infinitely smooth. Since for this representation particle $r(\cdots)=a[\tau,\varrho_\mathrm{max}]\varrho_\mathrm{max}$, $r(\cdots)$'s continuity and smoothness implies $a[\tau,\varrho_\mathrm{maxx}]$'s very directly.
\begin{figure}[h]
\includegraphics[totalheight=50mm]{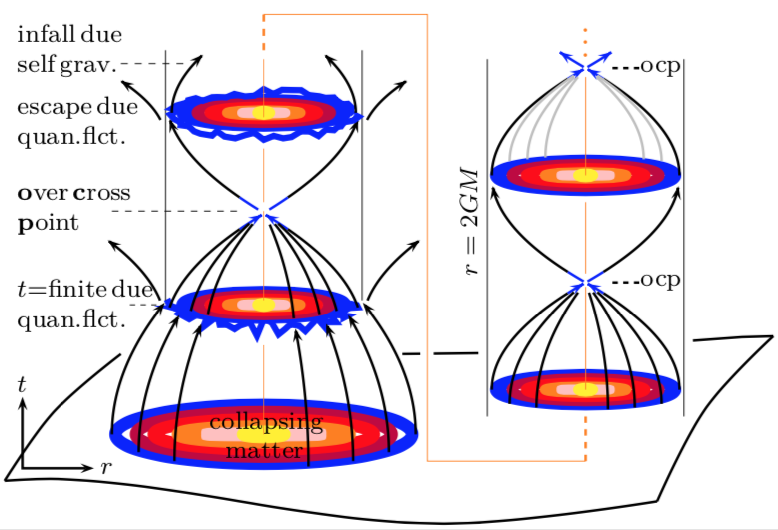}
\caption{Due to the uncertainty principle, even a far away observer of fixed position needs only a finite time to see a dust ball's contraction into the horizon surface determined by the ball mass. Because of non-interaction occurs among different parts of the matter contents, as they aggregate to the central point, they will over cross each other under the forward amplitude domination mechanism of quantum scattering and then periodically oscillate there. No-interaction assumption is a precise enough approximation when gravitation dominates all other interactions. Because of energy conservation, during the early few rounds of collapse and central-point-over-crossing-after expansion, every time the matters getting close to the horizon, quantum fluctuation will allow part of them fall across and escape, while others fall back into the horizon. But in latter time, they will form stable oscillation pattern inside the horizon, except the very weak spontaneous, i.e. Hawking radiation effects. This inside horizon oscillation could preserve initial mass profiles of the collapsing ball as microstates of the final black hole.}
\label{figCollapsingProcessDetail}
\end{figure}

If we have chosen $b>2GM$ so that the dust ball is released from an initially larger than horizon configuration, then we will see that the period of $r(t)$ is infinite, but that of $r(\tau)$ is still finite. This is nothing but the well known fact that a far away fixed-position observer needs to wait infinite time to see the black hole's formation. In other words, formations of the  event horizon occur outside the domain of time definition for such observers. This is obviously an obstacle for standard S-matrix description of a complete cycle of black hole formation and evaporation. In 't Hooft's work \cite{tHooft2021,tHooft2019}, this obstacle exhibits as the focused degrees of freedom becomes infinitely hard as the cauchy surface inhabiting them evolves towards the horizon. To avoid this obstacle, 't Hooft chooses to introduce techniques of antipodal identification and firewall transformation in the extended Schwarzschild space-time so that he can constrain his discussions in regions I and III of the Penrose-Carter diagram completely, see FIG.\ref{figBrickWallModel} for illustration. However, we would like to make a different choice. 

Our choice is, when the microscopic state definition of the black hole is concerned, we expand our discussion in regions II and IV. Just as we pointed out previously, from pure classic aspects, the inside horizon oscillation happens beyond the domain of time definition of outside observers. It belongs to physics of those observers' parallel universe or statistic ensemble world. Motion details of those microscopic states are observable only for those living inside the horizon, i.e. Maxwell's demon like observers. The outside observers has to adopt statistic mechanic viewpoint with those microscopic state. However, as long as the radiation process is concerned, our discussion comes back to regions I and III, referring FIG.\ref{figPCdiagramSchwz}. Rationalities of this choice could be attributed to quantum features of the horizon itself, see FIG.\ref{figCollapsingProcessDetail} and captions there for explanations. There are observations in the literature that (i) infalling of a mass shell into the horizon of a pre-existing Schwarzschild background could be observed in the domain of time definition of far away observers, see references \cite{SNZhang2009,SNZhang2010} for concrete example and \cite{Frolov1998} for more general speculations and (ii)  radiations originating from the collapse of a mass shell can occur and finish before the shell getting into the horizon, see references \cite{Stojkovic2007,Stojkovic2008a,Stojkovic2008b}. Observation (ii) is in fact a somewhat avatar of 't Hooft's firewall transformation result. All these works took the horizon surface as a purely classic geometric surface of zero thickness and precisely specifiable shape and size. However, the real world horizon, due to matters to fall across and facilitate its formation undergo quantum fluctuation ubiquitously, its formation can be done in finite time even from the viewpoint of observers far away from the system.

\subsection{Physics through Penrose-Carter Diagram}

\begin{figure}[h]
\rule{5mm}{0pt}\includegraphics[totalheight=47mm]{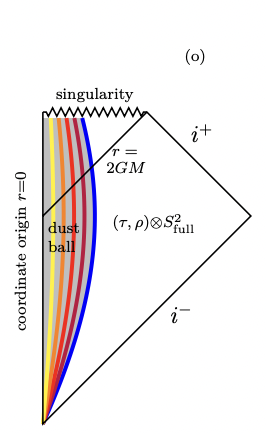}
\includegraphics[totalheight=47mm]{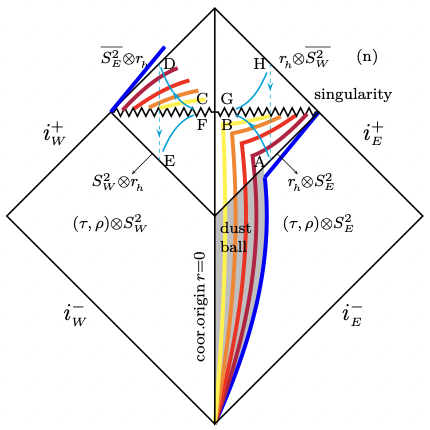}
\\
\includegraphics[totalheight=42mm]{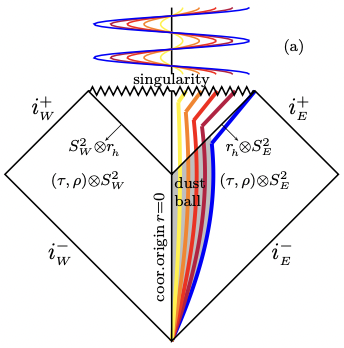}
\includegraphics[totalheight=42mm]{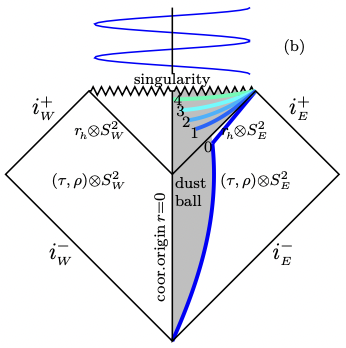}
\\
\includegraphics[totalheight=42mm]{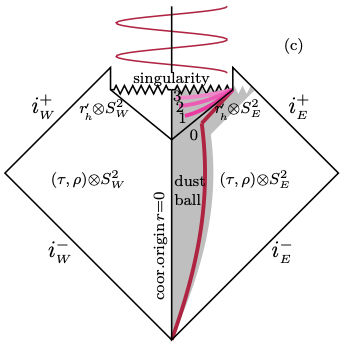}
\includegraphics[totalheight=42mm]{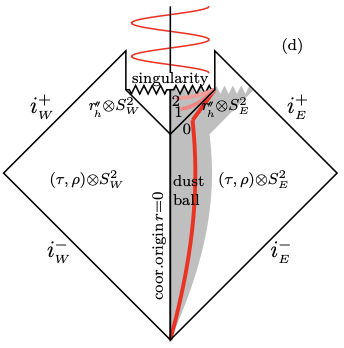}
\caption{Subfigure (o) is the Penrose-Carter diagram representation in conventional textbooks for a dust ball's collapsing and causing of black hole formation. Curves of different color correspond initially different equal $r$-shells consisting of the dust ball. The singularity's appearance is the ending of all story. Subfigure (n) is our Penrose-Carter diagram representation for the same process. For good looking only east panel of the dust ball is displayed explicitly. The dusts are allowed to oscillate inside the horizon. To exhibit this oscillation conveniently, the east and west panel of all 2-sphere of equal-time space are plotted independently. A complete oscillation period consists of motions from A to B and then -C-D-E-F-G-H-A in (n), C is the antipodal point of B, E is identified with D and et al. (a) is an alternative for (n). (b), (c) and (d) denote three initially equal-r shells' evolving and hitting of the singularity. For different shells, sizes of the spacetime region beyond the singularity surface thus being cut out in conventional textbooks are different.}
\label{figPCdiagramCollapse}
\end{figure}

In conventional textbooks, such as those of HE \cite{LargeScaleStructure} and MTW \cite{MTWtextbook}, since the inside horizon oscillation of matters is not considered, the Penrose-Carter diagram for gravitation collapse is usually plotted simpliy as part(o) of FIG.\ref{figPCdiagramCollapse}. However, things are different when the inside horizon motion of matters are considered. We plot in part(n) and (a) two representations of the relevant physics in some detail. Three new points are worth of emphasizing here. 

(i) In all subfigures except (o), the east and west semi-panel of the whole equal-time space are plotted separately and independently in the figure. 

(ii) The part of spacetime lying above the singular surface and being cut away in the usual PC diagram (o) for gravitational collapses are needed to be put back there to get a complete representation of oscillations inside the horizon (n). Spatial-coordinates of such being putted back region are identified with those of antipodal ones below the singularity surface.

(iii) For shells of different initial radius, sizes of spacetime being cut away in conventional PC diagram for gravitation collapses due to their lying beyond the singularity surface are different and are determined by masses wrapped by the corresponding spheric shell. The initial mass profile $M(\varrho)$ encodes all information of an exactly spherically symmetric classic black hole in general relativity. We will discuss its quantization in next section.

\begin{figure}[h]
\parbox{42mm}{\includegraphics[totalheight=30mm]{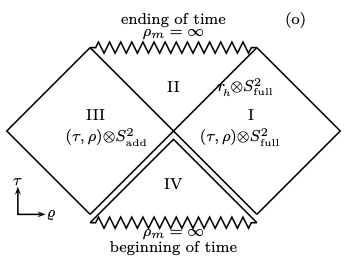}\\
\includegraphics[totalheight=30mm]{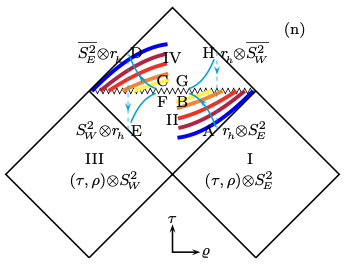}}
\parbox{42mm}{\includegraphics[totalheight=64mm]{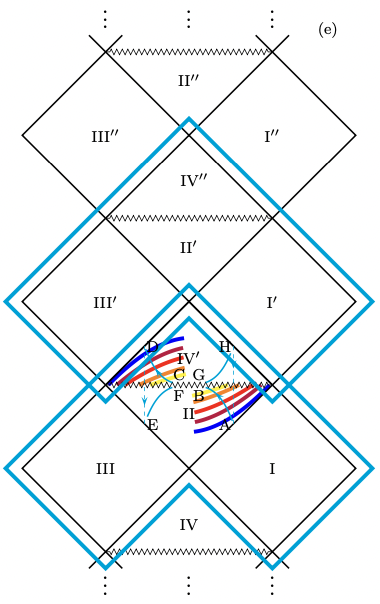}}
\caption{The left-upper, i.e. (o)sub is the usual PC diagram of extended Schwarzschild spacetime, I and III are two different asymptotically flat region, they share common inside-horizon region II and IV. II is interpreted as the black  hole inside region, IV is interpreted as the white hole inside region. The left-downer, i.e. (n)sub is our PC diagram for black holes with exact Schwarzschild outside metric but oscillatory matter cores. A full oscillation period consists space-time points like A-B-C-D-E-F-G-H-A, where C is the antipodal of B, E is identified as E and et al. The right panel, i.e. (e)sub is the infinitely extended version of the left-downer one. The union of two repeated units with the inside horizon region II+IV$^\prime$ \& II$^\prime$+IV$^{\prime\prime}$ identified and with the matters removed away will be equivalent with the usual eternal Schwarzschild spacetime of (o)sub.}
\label{figPCdiagramSchwzDetail}
\end{figure}

Now, let us shift the discussion to eternal black holes, whose inside-outside regions are both described by metrics of the form \eqref{SchwarzschildOutsideSchwarzschild}. In pure classic general relativity, such black holes are characterized by their total mass or central singularity - in fact an equal time surface -  completely, see FIG.\ref{figPCdiagramSchwzDetail}, left-upper part for PC-diagram representation. To assign any structure to their horizon surface or to their asymptotically infinite region will break their spherical symmetry inevitably. All such schemes to provide explanation for the black hole's microscopic state may be considered effective account of the underlying phyics instead of real cases. However, if we join an outside Schwarzschild metric with an inside horizon one supported by oscillatory matter cores such as those given by \eqref{genOSmetric}-\eqref{dustEMTensor}, then we could give interpretations for the microscopic state of black holes very naturally, they are nothing but the consisting matter's distribution pattern and oscillation modes, see FIG.\ref{figPCdiagramSchwzDetail}, left-downer part for PC-diagram representation. From the figure, we easily see that all matters consisting of the black hole or falling into its horizon will hit on the central singularity or equal-time singular surface inevitably, thus perfectly consistent with Penrose-Hawking's singularity theorem \cite{Penrose1965,Penrose1969,hawking1976,geroch1979,LargeScaleStructure}. But different from the usual Schwarzschild hole on the left-upper part, matters in our black holes will execute oscillations across the central point inside the horizon. This oscillation forms a resolution of the Schwarzschild singularity, in the sense that the conventional static and eternal singularity becomes a periodically forming and dissolving one. Singularities symboling the terminal of anything including time itsself becomes a normal, easily passing by equal time surface.

A very key observation of (o)sub in FIG.\ref{figPCdiagramSchwzDetail} is that, its region III denotes another asymptotically flat region independent of I. The two are connected with each other through Einstein-Rosen bridge \cite{EPR1935b,FullerWheeler1962,Friedmann1993,Gallowayetal1999}. Physics occurring in region III is fully entangled \cite{EPR1935a,EREPR} with those in I and this fact plays important roles in modern understanding \cite{Raamsdonk2010,Raamsdonk2012,Raamsdonk2013,Raamsdonk2016,Raamsdonk2018} of spacetime origin basing on entanglements. We can represent such extensions of spacetime by adding an additional copy of the physical one and letting the extra copy share common internal with the physical one, see the (e)sub of FIG.\ref{figPCdiagramSchwzDetail} for concrete doing. Somewhat different from 't Hooft, we hold preservations with the viewpoint that existence of two such fully entangled copy of spacetime is a redundant representation of physical reality, which means that researchers miss basic principles in choosing general coordinate to denote physical reality. See reference \cite{tHooft2019} for observations like ``In applying general coordinate transformations for quantized fields on a curved space-time background, to use them as a valid model for a physical quantum system, one must demand that the following constraint hold: the mapping must be one-to-one and differentiable. Every space-time point with Schwarzschild coordinates $(t, r, \theta, \phi)$ now maps onto exactly one point with Kruskal-Szekres ones $(\tau,\rho,\theta^\prime,\phi^\prime)$, without the emergence of cusp singularities''.

Obviously, without such constraints, we could add arbitrarily more extra copies such as I$^{\prime\prime}$+II$^{\prime\prime}$+III$^{\prime\prime}$+IV$^{\prime\prime}$ and $\cdots$ to the beginning one I+II+III+IV, through coordinate transformations of the Einstein-Rosen type \cite{EPR1935b}
\beq{}
(\tau,\rho,\theta,\phi)\rightarrow(\tau,\rho^{\prime},\theta,\phi),\rho^{\prime}=\rho^\frac{1}{2n}
\eeq
and let all these spacetime have common inside horizon region thus be connected by very complex wormholes. Note that physics occurring in all these additionally added spacetime in fact happens outside the domain of time definition of observers living in I+III, it is very natural to assign ensemble interpretation for these spacetime, that is, each of them corresponds to a member of a spacetime ensemble. All these members have common inside horizon region but inhabit different outside observers. Due to the quarantine policy arising from classic horizon, all these observers cannot definitely know what the microscopic state of their black hole is. This viewpoint can also be reversed, i.e., all these added spacetime have common outside horizon observer but different microscopic state of inside matter contents. This is exactly what we mean when the conception of microscopic state of black holes is used. We will provide quantum definitions for this concept in section \ref{sectionQuantizeInsideHorizonMotion}. But before that, let us consider a second exact solutions family for the inside horizon motion of matters in the following.

\subsection{2+1 Dimensional AdS-Schwarzschild}

Still use the Lemaitre style coordinates inside the matter occupation region, our exact solution family to the 2+1D Einstein equation sourced by negative cosmological constant and an oscillatory dust core has the following form
\beq{}
ds^2_\mathrm{in}=-d\tau^2\!+\!a^2(\tau)\big[\frac{d\varrho^2}{1\!+\!\varrho^2\ell^{\m2}\!-\!2GM(\varrho)}\!+\!\varrho^2d\phi^2\big]
\label{AdSmetricOscillation}
\eeq
\beq{}
a[\tau]=\cos[{\ell^{\m1}\tau}],~0\leqslant\varrho\leqslant\varrho_\mathrm{max}~M[\varrho_\mathrm{max}]=M_\mathrm{tot}
\label{AdSmetricScaleFactor}
\eeq 
whose energy momentum tensor reads
\beq{}
T_{\mu\nu}=\mathrm{diagonal}.\{\rho(\tau,\varrho)=\frac{M'\varrho^{-1}}{8\pi\,a[\tau]^2},~0,~0\}
\eeq
where $M(\varrho)$ is the co-moving mass function of the dust matter. It can be understood as the mass distribution of the system at an arbitrary initial time. On classic levels, the function form of $M(\varrho)$ is completely undeterminable, except being monotonically increasing and getting maximal at some given coordinate $\varrho=\varrho_\mathrm{max}$.

Outside the matter occupation region, using coordinate transformation of the following form
\beq{}
a(\tau)\varrho\equiv r\Rightarrow d\varrho^2=a^{-2}(dr-a^{-1}\dot{a}{r}d\tau)^2
\eeq
\bea{}
-d\tau^2+\frac{-2a^{-1}\dot{a}{r}d{r}d\tau\!+\!a^{-2}\dot{a}^2d\tau^2}{1+\varrho^2\ell^{-2}-2GM_\mathrm{tot}}
\equiv h(r)dt^2
\\
=-(1+\frac{{r}^2}{\ell^2}-2GM_\mathrm{tot})d{t}^2
\eea
the inside metric \eqref{AdSmetricOscillation}-\eqref{AdSmetricScaleFactor} can be easily seen join to those of the standard AdS-Schwaraschild black hole with infinite smoothness,
\beq{}
 ds^2_\mathrm{out}=-h(r)d{t}^2+h^{-1}(r)dr^2+r^2d\phi^2
\eeq
This is obviously the $J=0$ case of Banados-Teitelboim-Zanelli black hole \cite{BTZbh1992,Banados1996,Witten0706}, whose Bekenstein-Hawking temperature reads
\beq{}
k_{\scriptscriptstyle B}T=h'[r_h]=\frac{\sqrt{2GM-1}}{2\pi\ell}
\label{ads2p1HawkingTempterature}
\eeq

\begin{figure}[ht]
\includegraphics[totalheight=45mm]{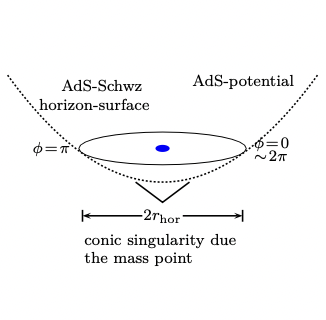}
\includegraphics[totalheight=45mm]{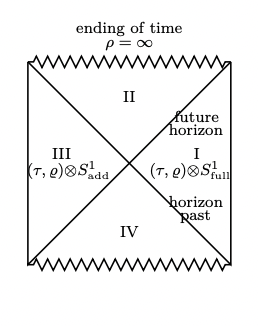}
\\
\includegraphics[totalheight=45mm]{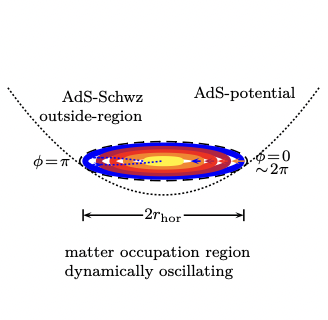}
\includegraphics[totalheight=45mm]{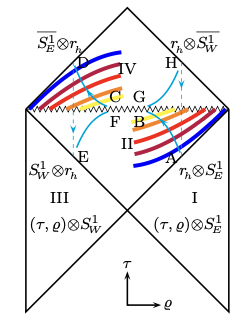}
\caption{Upper-left is a mass point in an otherwise exactly AdS space-time, conic singularity occurs on the central point. Upper-right, the PC diagram representation of the AdS-Schwarzschild spacetime due to the mass point on the left figure. Region III is another asymptotically AdS spacetime independent on, but connect through some Einstein-Rosen bridge like structure with region I. Lower-left, a dust disk with oscillatory inner structure in otherwise exactly 2+1D AdS spacetime and its Penrose-Carter diagram. A complete oscillation period of a dust point is A-B-C-D-E-E-F-G-H-A, C is the antipodal of B, E is identified with D and et al. The family of colored curves on the lower right figure correspond with the equal $r$-circle on the lower left.}
\label{figAdSschwarzschild}
\end{figure}

We compare in FIG.\ref{figAdSschwarzschild} spacetime structures of the mass-point type AdS-Schwarzschld black hole and that of an oscillatory dust core produced one with exact outside AdS-Schwarzschild geometry. In the former case, the conic singularity occurs on the center of the dust core. It is in fact a future equal-time hyper-surface unavoidable for any particle falling down and across the event horizon. While in the latter case, different from the former, conic singularity is only a transient phenomena. Dusts consisting of the ball, in facto two dimensional disk just oscillate across the central point. It should be emphasized that since we represent the east and west semi-circle of all equal $r$ space independently in the lower right figure, we need to take the union of region I and III of it to get the whole AdS-spacetime outside. In another word, we need to pile two copies of the displayed region I+II+III+IV in the lower right to get the extended spacetime represented by the upper right figure. But the same as in 3+1D Schwarzschild black hole case, physics occurs in the additionally piled up copy happens in fact outside the domain of time definition used by observers living in the original one. Such piling up of spacetime can be done arbitrarily many times and interpreted as constructions of a somewhat spacetime ensemble.

2+1D vacuum gravitation is well known of no dynamics, in the sense that it contains no propagating degrees of freedom. All dynamics in it manifest themselves through behaviors of the conic singularity \cite{Carlip1995,Strominger1996,Strominger1997,Banados1997,Carlip1997,Banados1998,Carlip2005,JBaez2007,Carlip1998,Witten0706}. In the AdS case, the cosmological constant provides a parabola potential well. When a cluster of dust gas is introduced, the system oscillates harmonically due to the potential well. The conic singularity's resolving in this case is very easily understandable. If we introduce no conic singularity a priori, then such singularities will not happen as the result of collapse of regular matters such as a cluster of continuous fluid. If we include small conic singularity atoms a priori, then no bigger extension of such atoms will be formed as the results of small atoms' collision due to pure gravitational interaction. Because, under the condition that all other interactions except gravitation is excluded, all these small atoms has only one kind of motion, oscillation due to the parabola potential well provided by the cosmological constant.

\subsection{Inside Horizon Oscillation, Memory Effects and Soft Hair}

To here, careful readers may be well suggested that we will use the consisting matters' inside horizon distribution and oscillation to explain microscopic state of black holes. A question arises here, will this physical picture contradict with other interpretations such as the soft hair theory of Hawking, Perry and Strominger \cite{gravMemoryChristodoulou1991,HPS201601,HPS201611,emMemorySabrina2017}. Just as we will discuss in this subsection, they are connected with each other through the memory effects thus are consistent with each other perfectly well. To see this more intuitively, let us take the volume mode oscillation of a spherically symmetric charged shell and its retarded potential as examples to illustrate this connection. 

\begin{figure}
\includegraphics[totalheight=45mm]{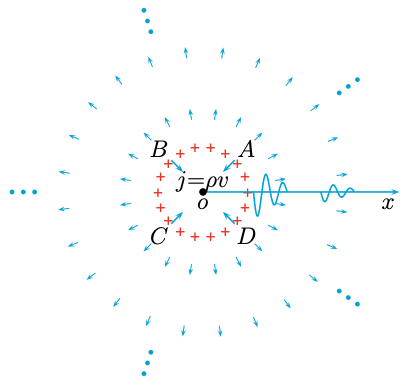}
\caption{The volume mode oscillation of spherical symmetric charged shell does not radiate electromagnetic wave. But the relevant potential function $A_\mu(t,\vec{x})$ will carry information of the shell's motion and propagate outward with the speed of light. This is the key idea of memory effects. In the case of gravitational collapses, the same physics occur. $A$, $B$, $C$ and $D$ are four points on the spherical shell, $\vec{x}$ is the coordinate of a point we focus on.}
\label{figRetPotentialMemory}
\end{figure}

By the word volume mode oscillation we mean the size oscillation of spherically symmetric charged shell. Referring to FIG.\ref{figRetPotentialMemory}, without loss of generality, we will let our focus point $\vec{x}$ lie on the x-axes of the coordinate system and use $x_s$ and $x$ to denote the shell radius and distance of the focus point $\vec{x}$ to the central of the system. With these notations, we can write the retarded potential of the system as follows,
 \beq{}
 \phi(t,\vec{x}){=}\!\!\!\sum_{\scriptscriptstyle\mathrm{semi}}^{\scriptscriptstyle\mathrm{sphere}}\!\!\!\left(\!\frac{2\rho_{\scriptscriptstyle\!A}d\mathbf{x}'}{|\vec{x}{-}\vec{x}'_{\scriptscriptstyle\!A}|}
{+}\frac{2\rho_{\scriptscriptstyle\!C} d\mathbf{x}'}{|\vec{x}{-}\vec{x}'_{\scriptscriptstyle\!c}|}\!\right)
 {=}\frac{q}{x}{+}\frac{a(t{-}\frac{x-x_s}{c})}{x^2}
 \label{retardPotentialAphi2}
 \eeq
  \beq{}
 \vec{A}(t,\vec{x}){=}\!\!\!\sum_{\scriptscriptstyle\mathrm{semi}}^{\scriptscriptstyle\mathrm{sphere}}\!\!\!\left(\!\frac{2\vec{j}_{\scriptscriptstyle A}^{\leftarrow}\!d\mathbf{x}'}{|\vec{x}{-}\vec{x}'_{\scriptscriptstyle\!A}|}
{-}\frac{2\vec{j}_{\scriptscriptstyle C}^{\rightarrow}\!d\mathbf{x}'}{|\vec{x}{-}\vec{x}'_{\scriptscriptstyle\!c}|}\!\right)
 {=}\frac{a(t{-}\frac{x-x_s}{c})\hat{x}}{x^2c}
 \label{retardPotentialAA2}
 \eeq
where $a(t{-}\frac{x-x_s}{c})$ is a somewhat retardation function whose concrete form depends on the motion feature of the shell and matters little for our analysis, $\rho_A$, $\rho_C$, $\vec{j}_A$, $\vec{j}_C$ are charge and current density of an arbitrary pair of antipodal points on the sphere, with $\vec{j}_{\scriptscriptstyle A}^{\leftarrow}$ and $\vec{j}_{\scriptscriptstyle C}^{\rightarrow}$ being just the horizontal component of $\vec{j}_A$ and $\vec{j}_C$ respectively. Obviously contributions of these charge and current elements from the oscillation shell to the potential function $\phi(t,\vec{x})$ and $\vec{A}(t,\vec{x})$ does not cancel each other, due to retardation effects of the propagation, although the whole charge shell oscillates spherical symmetrically. If we focus on potentials on the light-like surface surrounding the charge sphere, then we will get in fact the information of initial size and ingoing or outgoing velocities of the shell. That is, retardation potentials of system on the light like hyper surface remember what happens to the charged shell at initial epochs.

When more general charged ball and generalizations to gravitational collapses are concerned, the analysis is almost the same as the charged spherical shell above and involves no extra complexes, either technically or conceptually. Initial distribution and ingoing speed of matters to collapse and to form the black hole is remembered by the retarded gravitational potentials which propagate outward with the speed of light, and interpreted by Hawking Perry and Strominger as the soft hair of black holes. Obviously, our interpretations of the black hole microstate through the consisting matters' inside horizon distribution and oscillation is nothing but another description of the same physics as soft hair theory. It can be easily see that our interpretation adopts more intuitive physical pictures.

\section{Inside Horizon Oscillation Quantized}
\label{sectionQuantizeInsideHorizonMotion}

Now we are well prepared to discuss the quantization of consisting matter's distribution and oscillation inside the horizon of a black hole. We will start with the 3+1D Schwarzschild black holes and its inner structure.

\subsection{3+1D Schwarzschild Matter Contents Quantization}

We will consider the whole dust matter contents comprises of and moves inside the horizon of a Schwarzschild black hole as many concentric spherical shells, each with mass $m_i$ and time dependent size $r$. For each of such shells, spacetime metric affecting their motion can be written as the Schwarzschild form $ds^2=-h_idt^2+h_i^{-1}dr^2+\cdots$.  By the standard general relativity, the radius of each such shell evolves with time according to equations of the following form
\bea{}
&&\hspace{1mm}
h_i\dot{t}^2-h^{-1}_i\dot{r}^2=1
,~h_i=1-\frac{2GM_i}{r}
\label{eomFourSpeedNorm}
\\
&&\hspace{1mm}
\ddot{t}+\Gamma^{(i)t}_{tr}\dot{t}\dot{r}=0
\Rightarrow h_i\dot{t}=\gamma_i=\mathrm{const}
\label{eomGeodesic0}
\\
&&\hspace{1mm}\Rightarrow\rule{0pt}{4mm}\dot{r}^2-\gamma^2_i+h_i=0,
\gamma_i^2<0
\label{eomShellClassic}
\eea
This is nothing but the equation of geodesic line of representative particles co-moving with the dust shell $i$. The parameter $\gamma_i^2$ in eq\eqref{eomShellClassic} will be negative when a shell $i$ lies inside the horizon determined by the mass carried by all shells inside it, or although positive but always upper bounded when the shell starts oscillation from outside the horizon determined by the mass carried by all dusts internal but still inside the horizon determined by the total mass of the ball. In one word, equation \eqref{eomShellClassic} has oscillatory solutions just as we show in subsection \ref{subsec31exactsolution}, see FIG.\ref{figOscillatoryScaleFactor} there for referrings. While in FIG.\ref{figPCdiagramCollapse} we provide Penrose-Carter diagrammatic representation for the whole dynamic space-time, according to which the mass shell firstly falls onto the central point, then expands from that point and then falls back from the other side, thus oscillates around there periodically. This falling across and oscillation around the central point can be attributed to the forward amplitude domination when the dust particles collide together under inter-gravitations. Remember that we neglect all other interactions except gravitation.

\begin{figure}[h]
\begin{center}
\includegraphics[width=0.47\textwidth]{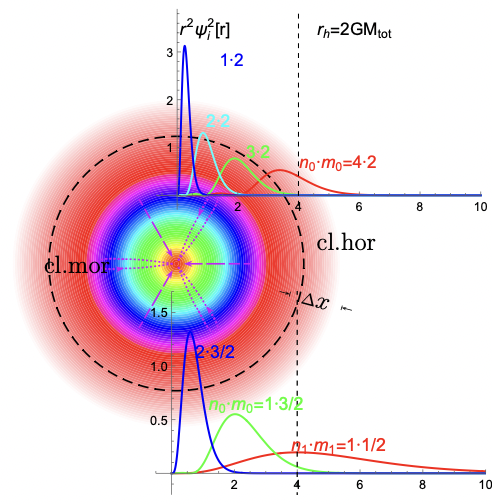}
\end{center}
\caption{Examples of probability density of the consistuent dust shell of a Schwarzschild black hole of mass $M_\mathrm{tot}=2M_\mathrm{pl}$. We let $\hbar=1$ and measure all dimensional quantities by $M_\mathrm{pl}$ or its inverse. The black holes described by the upper wave function consisting of just one shell of mass $m_{i(=0)}=2M_\mathrm{pl}$, so 4 square integrable microstate wave functions are allowed. The black holes described by the lower wave function consisting of two concentric shells, due to the constraining of quantization condition \eqref{enQuantizConditionSchwz}, the possible shell mass decomposition could be $\{m_{i(=0)},m_{i(=1)}\}=\{\frac{3}{2},\frac{1}{2}\}$, $\{\frac{3-\epsilon}{2},\frac{1+\epsilon}{2}\}$, $\cdots$, $\{1,1\}$ which is infinite and in-countable . The probability density of wave functions $\psi_{n0=1,m0=\frac{3}{2}}$,$\psi_{n0=2,m0=\frac{3}{2}}$, $\psi_{m1=\frac{1}{2}}$ are plotted explicitly.
}\label{figWFminkowski}\end{figure}

On quantizing equations of motion \eqref{eomShellClassic} for an arbitrary dust shell, we introduce a wave function $\psi_i(r)$ to describe the probability amplitude the shell be found of size $r$ and rewrite its radial momentum $m\dot{r}$ into operators $i\hbar\partial_r$ by the standard canonical method. Then acting the operatorized version of the equation on the wave function $\psi_i(r)$, we have
\beq{}
\big[-\frac{\hbar^2}{2m_i}\partial_r^2-\frac{GM_im_i}{r}-E_i\big]\psi_i(r)=0,~0\leqslant r<\infty
\eeq
where $E_i\equiv\frac{m_i}{2}(\gamma_i^2-1)$ is nothing but redefinition of the upper bounded integration constant $\gamma_i^2<1$ from equations \eqref{eomGeodesic0} and \eqref{eomShellClassic}. Now by the simple square integrability condition of $\psi_i$  or physical requirement that the probability the shell be found in the whole space is finite, we will get
\bea{}
&&\hspace{-5mm}\frac{\gamma^2_i\!-\!1}{2}m_i\equiv E_i=-\frac{(GM_im_i)^2m_i}{n_i^2\hbar^2}
,~n_i=1,2,\cdots
\label{enQuantizConditionSchwz}
\\
&&\hspace{-5mm}\psi_i=N_ie^{-x}xL_{n_i-1}^1(2x),x\equiv mr(1\mm\gamma_i^2)^\frac{1}{2}/\hbar
\label{eigenWavefunctionSchwz}
\eea
where $L_{n_i{-}1}^1$ is the associated Laguerr polynomial of order $(n_i{-}1,1)$, while $N_i$ the corresponding normalization constant. Equation \eqref{enQuantizConditionSchwz}, although manifests as a quantization condition, it does not predict discrete mass or energy eigenvalues for the shell.  $m_i$ and $\gamma_i$($M_i$ effectively) behave just like a pair of variables conjugate to each other. None of them is determined by \eqref{enQuantizConditionSchwz}. This is very different from J. D. Bekenstein's horizon area quantization scheme, which predicts discrete values for allowed masses of the black hole.

However, if we consider the black hole consisting of just one shell of mass $m_{i(=0)}=2M_\mathrm{pl}$, then the shell has only 4 square integrable microscopic state wave functions. Their corresponding probability density $r^2|\psi_i|^2$s are plotted in the upper panel of FIG.\ref{figWFminkowski}. $4\pi r^2|\psi_i|^2dr$ has the meaning of the corresponding mass shell be found of size $r\sim r+dr$. If we consider the black hole consisting of two concentric shells, then due to the quantization condition \eqref{enQuantizConditionSchwz}, the possible shell mass decomposition could be $\{m_{i(=0)},m_{i(=1)}\}=\{\frac{3}{2},\frac{1}{2}\}$, $\{\frac{3-\epsilon}{2},\frac{1+\epsilon}{2}\}$($0{<}\epsilon{<}1$), $\cdots$, $\{1,1\}$ which is infinite and in-countable. However, possible quantum numbers $\{n_{i(=0)},n_{i(=1)}\}$ defined by the quantization condition \eqref{enQuantizConditionSchwz} can only be $\{1\mathrm{or}2,1\}=\{\mathrm{Int}[0,\!\frac{9}{4}],\frac{1}{2}{\cdot}2\}$ and $\{1,1\mathrm{or}2\}$, which is obviously countable and finite. We plot in the lower panel of FIG.\ref{figWFminkowski} the probability density of wave function $\psi_{n0=1,m0=\frac{3}{2}}$,$\psi_{n0=2,m0=\frac{3}{2}}$, $\psi_{m1=\frac{1}{2}}$. We can further consider the possibility the black hole consisting of more layers. However, none of them would satisfy constraints of \eqref{enQuantizConditionSchwz}.

From FIG.\ref{figWFminkowski} we easily see that, all quantum wave functions have long fat nonzero tails outside the classic horizon region defined by the total mass of the dust ball. This means that, at quantum levels, we loss concepts of the conventional horizon surface with exact shape and position. We cannot determine a shell is inside or outside a purely geometric horizon surface defined by the formula $r_h=2GM_\mathrm{tot}$. On some sense, we get a truly quantum version of fuzzy ball picture of black holes. As comparison, the fuzzy ball picture of string theory \cite{MT0103,LuninMathur0202,LuninMaldacenaMaoz0212,GMathur0412,mathur0502,mathur0510,Jejjala0504,KST0704,Mathur0706} follows from a first quantized string living in classic spacetime with extra dimension. It is the classic motion of classic strings that blurs the horizon surfaces there. Of course, further quantization of such string is possible. But it is a very challenge task in the framework of first quantization of string theory.

Obviously, according conditions \eqref{enQuantizConditionSchwz} we can define finite and countable direct product quantum number $n_0{\otimes}n_1{\otimes}\cdots{n_k}$. However, just as we pointed out under equation \eqref{eigenWavefunctionSchwz}, for any such given quantum number set, we have infinite and non-countable ways to decompose the matter contents of black holes into concentric shells and setting their initial condition. This means that equation \eqref{enQuantizConditionSchwz} does not lead to discrete masses for black hole allowed and line shape for its radiation/emission spectrum as the naive idea of J. D. Bekenstein and V. F. Mukhanov does \cite{Bekenstein1974a,Mukhanov1986,Bekenstein1995,Bekenstein1997a,Bekenstein1997b}.
On the contrary, we note from equation \eqref{enQuantizConditionSchwz} that, for all shells consisting of the black hole, the minimal value of mass allowed  is
\beq{}
m_\mathrm{min}=\frac{n^{i}_\mathrm{outmost}\hbar}{GM^{i}_\mathrm{outmost}}\big(\frac{1{-}\gamma_i^2}{2}\big)^{\!\frac{1}{2}\mathrm{lower}}_{bound}=\frac{1}{GM_\mathrm{tot}}
\eeq
This is the mass of shell with most fat nonzero wave function tail living outside the classic horizon thus most probably be radiated away during the black hole radiation, see FIG.\ref{figWFminkowski}. It is obviously of the same order as hawking temperature.

\subsection{Microstate Definition and Counting}

Equation \eqref{enQuantizConditionSchwz} means that the way we look the dust ball as concentric shells and set their initial releasing condition $h_i\dot{t}_{r=r_\mathrm{release}}=\gamma_i$ are not arbitrary. This equation can be written as constraining condition of the following form
 \beq{}
 \frac{GM_im_i}{[(1-\gamma_i^2)/2]^{\frac{1}{2}}}=n_i\hbar, n_i\in\mathbb{Z}^+,\gamma_i^2<0
 \label{decQuantizConditionSchwz}
 \eeq
 where $M_i$ is the total mass of shells inside the $i$th shell and $i$th shell itself, $M_i=\sum_{\hat\imath=0}^im_{\hat\imath}$. When the dust matter contents are considered as $k$ shells, the wave function of the whole system can be written as the direct product of all consisting shells
 \beq{}
 \Psi[\{m_i(r)\}]_{i_\mathrm{max}}^{{{\scriptscriptstyle=}k}}=\psi_{\scriptscriptstyle0}(r)\otimes\psi_{\scriptscriptstyle1}(r)\otimes\cdots\psi_{k}(r)
 \label{directProductWavefunction}
 \eeq
 For latter convenience, we will call the shell mass combination $\{m_i\}$ a shell partition of $M_\mathrm{tot}$, the corresponding $\{n_i\}$ an excitation.  We provide in TABLE \ref{shellDecompositionMpl2} an example of some possible ways of shell partition and the corresponding exitation for the dust matter contents of an $M=2M_\mathrm{pl}$ Schwarzschild black hole. 
 
  \begin{table}
$$\begin{matrix}
\{G^\frac{1}{2}m_i\}&\{2\}&\{\frac{3}{2},\frac{1}{2}\}&\{\frac{5}{4},\frac{3}{4}\}&\{1,1\}
\\
\{G^\frac{1}{2}M_i\}&\{2\}&\{\frac{3}{2},2\}&\{\frac{5}{4},2\}&\{1,2\}
\vspace{2pt}\\
\{GM_im_i\}&\{4\}&\{\frac{9}{4},1\}&\{\frac{25}{16},\frac{3}{2}\}&\{1,2\}
\vspace{1pt}\\
\{\!\nu_i{\equiv}\big(\!\frac{1-\gamma_i^2}{2}\!\big)\!^\frac{1}{2}\!\}\!&\!\{\frac{4}{1},\frac{4}{2},\frac{4}{3},\frac{4}{4}\}\!&\!\{[\frac{9}{4},\frac{9}{8}],1\}\!&\!\{\frac{25}{16},\frac{3}{2}\}\!&\!\{1,[\frac{2}{1},\frac{2}{2}]\}
\vspace{2pt}\\
\{z_i{\equiv}\frac{GM_im_i}{\nu_i}\}\!&\!\{1,2,3,4\}\!&\!\{[1,2],1\}\!&\!\{1,1\}\!&\!\{1,[1,2]\}
\\
\#{\mathrm{p}{\otimes}\mathrm{e}}&4&2&1&2
\end{matrix}$$
\caption{Some possible ways of shell decomposition and initial releasing condition setting for the dust matter contents of an $M=2M_\mathrm{pl}$ Schwarzschild black hole. By distinguishability definition of \eqref{distinguishabilityRule}, the decomposition lies on the fourth column of this table is indistinguishable from both its predecessors and successor. Similarly all other decompositions such as $\{m_i\}=\{\frac{5}{4}\pm\epsilon,\frac{3}{4}\mp\epsilon\}$($\epsilon<\frac{1}{4}$) are distinguishable from $\{\frac{3}{2},\frac{1}{2}\}$, also from $\{1,1\}$. While other decompositions such as $\{\frac{3}{2}+\epsilon,\frac{1}{2}-\epsilon\}$($\epsilon<\frac{1}{4}$) are non allowable because their outmost shell break up quantization condition \eqref{decQuantizConditionSchwz}. Decompositions of the form $\{1-\epsilon,1+\epsilon\}$($\epsilon<\frac{1}{4}$) are also non allowable because their inner most shell break up condition \eqref{decQuantizConditionSchwz}. }
\label{shellDecompositionMpl2}
\end{table}

From the table we see that only the excitation number $n_0{\otimes}n_1{\otimes}\cdots{n_k}$\eqref{directProductWavefunction} is not enough for uniquely specifying the microstate of matter contents of a black hole. For example, cases may occur in which two different shell partition scheme share a common excitation number $n_0{\otimes}n_1{\otimes}\cdots{n_k}$. To lift this degeneracy, we will introduce the concept of partition precision as follows. 
 \bea{}
  &&\hspace{-8mm}two~partitions~of~equal~number~of~shells~\{m_i^1\}\&
  \label{distinguishabilityRule}
 \\
 &&\hspace{-8mm}\{m_i^2\}~are~indistinguishable~on~precision~\varepsilon~if
 \nonumber
 \\
 &&\hspace{-8mm}{\scriptstyle\Delta}m_i\equiv|m_i^1-m_i^2|\leqslant\min\{\frac{\varepsilon}{GM_{i1}},\frac{\varepsilon}{GM_{2i}}\}~for~all~i;
 \nonumber
 \\
 &&\hspace{-8mm} {two~partitions~of~unequal~shell~numbers}
 \nonumber\\
 &&\hspace{-8mm} {are~always~distinguishable}
 \nonumber
 \eea
As long as this partition precision  requirement is exerted, the number of microstates, or the ways looking the matter contents of a black hole as concentric shells and setting their initial conditions will be uniquely determined. For the $M=2M_\mathrm{pl}$ Schwarzschild black hole, TABLE \ref{shellDecompositionMpl2} tells us that on precision $\varepsilon=1$, the number of microstates is 8.

For small black holes, we can use method above to list and get their full microstate table. However, as $M_\mathrm{tot}$ increases, this method of brute force listing and counting quickly becomes exponentially complicating and time consuming. For larger mass values, we used strategies of the following idea to get fast estimations of the microstate number. Step1, the maximal number of shells a mass $M_\mathrm{tot}$ dust ball could be decomposed into happens in such a partition in which each shell has masses $m_i$ so that $GM_im_i=1$. This number is approximately $\frac{1}{2}M^2_\mathrm{tot}\equiv k$ which follows analytically from the relation
\bea{}
&&\hspace{8mm}GM_i(M_i-M_{i-1})=1
\label{shellMassQuantization}
\\
&&\hspace{-5mm}\Rightarrow M_i\stackrel{k\rightarrow\infty}{\approx}\sqrt{2i/G},~i_\mathrm{max}\approx\frac{GM^2_\mathrm{tot}}{2}\equiv k
\eea
Step2, other ways of decomposition could be implemented by combination of shells in this finest decomposition, for example
\bea{}
&&\hspace{-5mm}m_0,m_1,m_2,m_3,\cdots m_{i^\mm},m_{i},m_{i^+},\cdots m_{k}
\\
&&\hspace{-5mm}\Rightarrow m'_0,m'_1,\cdots m'_i,\cdots m'_\ell
\label{redecomposition}
\eea
This is an order-concerned partition of integer $k$, whose number is of $2^k$ exactly \cite{wikiComposition}. The number of layers
$\ell$ here may be much much less than $k$, may also be very very close to $k$. But a typical or average case is
\bea{}
&&\hspace{-5mm}m'_0=m_0,m'_1=m_1+m_2,m'_2=m_3+m_4+m_5
\\
&&\hspace{-5mm}\cdots,m'_i=\sum_{j=i/2}^{3i/2}m_j,~m'_\ell=\sum_{j=k-\ell}^km_j,~\ell\approx\sqrt{2k}
\eea
Step3, for this typical or average partition, the corresponding excitation number has the form
\beq{}
1\otimes\cdots\big[k{-}\sqrt{2k}{-}\sqrt{2(k{-}\sqrt{2k})}\big]\otimes(k{-}\sqrt{2k})\otimes k
\eeq
with the number of excitation ways reads
\bea{}
&&\hspace{-5mm}1\cdot\cdots\big[k{-}\sqrt{2k}{-}\sqrt{2(k{-}\sqrt{2k})}\big]{\cdot}(k{-}\sqrt{2k}){\cdot}k\equiv F(k)
\label{Fkdefinition}
\\
&&\hspace{-5mm}F(k)\stackrel{k\rightarrow\infty}{\approx}e^k
\eea
Step4, the total number of ways partitioning \& exciting the matter contents of a dust ball, i.e. the microstate of a black hole with such inner structures is
\beq{}
\#{\bf p}\mathrm{artition}\otimes{\bf e}\mathrm{xcitation}=2^{k}F(k)
\label{largeMtotEstimation}
\eeq
Substituting $k=\frac{GM_\mathrm{tot}^2}{2}$ into this estimation, we get the area law formula as expected.

\begin{figure}[h]
\includegraphics[scale=0.35]{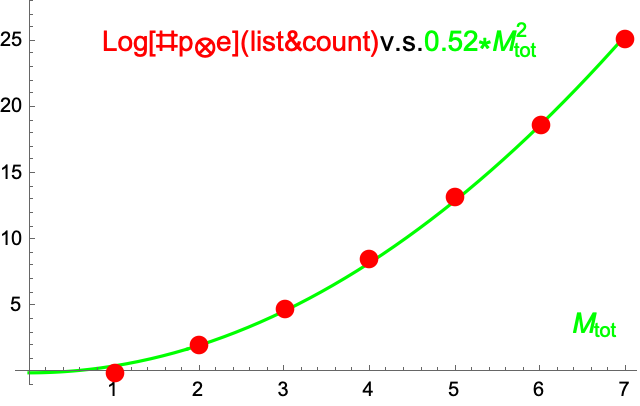}
\includegraphics[scale=0.35]{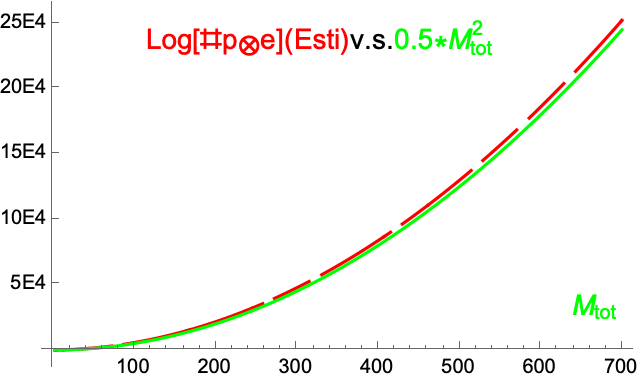}
\caption{The log number of ways looking the matter contents of a black hole as concentric shells and setting their initial moving condition for several $M_\mathrm{tot}$ values. The scattered points on the left panel come from exact listing and counting, while red discontiuous curve on the right comes from our estimations \eqref{shellMassQuantization}-\eqref{largeMtotEstimation}. Both are on precision  ${\varepsilon}=1$ and consistent with the area law expectation, denoted with continuous green curves on both panels.}
\label{figLogNpe}
\end{figure}

We plot in FIG.\ref{figLogNpe} the logarithmic number of our microstate counting for several typical values of $M_\mathrm{tot}$ as well as large $M_\mathrm{tot}$ estimation \eqref{largeMtotEstimation}.  Constrained by our computer power, our brute force listing and counting method provides only results for $M_\mathrm{tot}\leqslant7$.  From the figure, we easily to see that our microstates definition and counting yields results consistent with the area law formula of black hole entropy
\beq{}
k_{\scriptscriptstyle\!B}\mathrm{Log}[\#\,p{\otimes}e]{=}c(\varepsilon)\frac{4\pi(2GM)^2}{4G},~c(1){=}\frac{0.5{\sim}0.52}{4\pi}
\eeq
To get results exactly coincident with the Bekenstein-Hawking formula, we only need to tune the shell decomposition precision parameter $\varepsilon$ so that $c(\varepsilon)=1$. But it should be admitted that we currently find no physical reasons for such a tuning.

\subsection{Three Kinds of Partition Question}

There are three kinds of partition questions involved in this work, unordered partition of integers $P[N]$ \cite{Ramanujan1918}, ordered partition of integers $P_o[N]$ \cite{wikiComposition} and partition of positive real numbers under special quantization condition $P_q[M]$. Our interest is the limit behavior of relevant number of partition ways when the number being partitioned goes to infinite, $p[N\rightarrow\infty]$, $p_o[N\rightarrow\infty]$ and $p_q[N\rightarrow\infty]$.

Among these questions, the unordered partition is the most famous one. Ramanujan and Hardy was the earliest researchers of this question and have results
\beq{}
p(N\rightarrow\infty)\sim\frac{1}{4n\sqrt{3}}\exp\big(\pi\sqrt{\frac{2n}{3}}\big)
\eeq
The ordered partition of integers is also called composition of integers. It can be easily proven \cite{wikiComposition} that each positive integer $N$ has $2^{N-1}$ distinct compositions. So
\beq{}
p_o(N\rightarrow\infty)=2^{N-1}
\eeq
The third one, i.e. partition of general positive real numbers under special quantizing condition is the most rarely studied question in the literature. However, just as we discussed in the previous subsection, the microscopic state counting in our inner structured black hole models is just such a question. It can be expressed as the following pure mathematic theorem

 {\bf Theorem:} Let $M$ be an arbitrary positive real number, $\{m_i\}$ be a partition of it, i.e. $\sum_im_i=M$, $\{\nu_i\}$ be the corresponding excitation number, with all $\nu_i\geqslant1$, then for large $M$, the number of all distinguishable {\em partition} \& {\em excitation} ways is
\bea{}
&&\hspace{-7mm}\#\otimes\hspace{-5mm}{{\mathrm{partition}\{\!m_i\!\}}\atop{\mathrm{excitation}}\{\!\nu_i\!\}}\!\!\Big[M{=}\!\sum_i\!m_i,\frac{M_im_i}{\nu_i({\geqslant}1)}{=}\mathbb{Z}^+\Big]{=}\exp[c^{\varepsilon\!}M^2]
\label{PEformula}
\eea
where $M_i\equiv\sum_{i'}^{\leqslant i} m_{i'}$, $\mathbb{Z}^+$ denotes the positive integer number set and $c^{\varepsilon}$ is a partition precision dependent coefficient. Two partition ways $\{m_i\}$ and $\{m'_i\}$ will be considered indistinguishable on precision ${\varepsilon}$ if abstract of their difference $\{m_i\}{-}\{m'_i\}$ are one by one less than ${\varepsilon}/M_i$. 

{\bf Example:} Take $M=2$ as an example, its nine possible PE ways are as follows,
\beq{}
\begin{matrix}
\{m_i\}&\{2\}&\{\frac{3}{2},\frac{1}{2}\}&\{\frac{5}{4},\frac{3}{4}\}&\{1,1\}
\\
\{M_i\}&\{2\}&\{\frac{3}{2},2\}&\{\frac{5}{4},2\}&\{1,2\}
\\
\{M_im_i\}&\{4\}&\{\frac{9}{4},1\}&\{\frac{25}{16},\frac{3}{2}\}&\{1,2\}
\\
\{\nu_i\}\!&\!\{\frac{4}{1},\frac{4}{2},\frac{4}{3},\frac{4}{4}\}\!&\!\{[\frac{9}{4},\frac{9}{8}],1\}\!&\!\{\frac{25}{16},\frac{3}{2}\}\!&\!\{1,[\frac{2}{1},\frac{2}{2}]\}
\\
\{z_i\}\!&\!\{1,2,3,4\}\!&\!\{[1,2],1\}\!&\!\{1,1\}\!&\!\{1,[1,2]\}
\\
\#{\mathrm{p}{\otimes}\mathrm{e}}&4&2&1&2
\end{matrix}
\eeq
On precision ${\scriptstyle\Delta}=1$, partitions like those in the fourth column $\{m_i\}=\{\frac{5}{4},\frac{3}{4}\}$, or more generally $\{m_i\}=\{\frac{5}{4}\pm\epsilon,\frac{3}{4}\mp\epsilon\}$ with $\epsilon<\frac{1}{4}$ are all indistinguishable from its former and latter columns. While partitions like $\{1{-}\epsilon({<}1),\cdots\}$ or $\{\cdots\!,\frac{1}{2}{-}\epsilon({<}\frac{1}{2})\}$ are all non allowable because of their violation of the definition condition $\frac{M_im_i}{\nu_i(\geqslant1)}=Z^+$ on the innermost or outmost layer.

Currently, we do not how to prove this theorem exactly. However, it is indeed a question independent of any physical contexts. If be proved pure mathematically, it will become of course a support for our model of black hole inner structures.

\subsection{2+1D AdS-Schwarzschild Black Holes}

For the 2+1D Anti de Sitter Schwarzschild black hole, the story is ver similar, only with the mathematics becoming simpler. Following the same way in 3+1D Schwarzschild case, we continue to look the dust ball\footnote{Precisely saying, in 2+1 dimensional spacetime, the dust ball should be called dust disk and the shell should be called ring. But for expression convenience, we will use shell and ball uniformly.} as many concentric shells, each of which moves under controls
\bea{}
&&\hspace{-5mm}\ddot{t}+\Gamma^{(i)t}_{tr}\dot{t}\dot{r}=0
\Rightarrow h_i\dot{t}=\gamma_i=\mathrm{const}
\\
&&\hspace{-5mm}
h_i\dot{t}^2-h^{-1}_i\dot{r}^2=1
,~h_i=1+\frac{r^2}{\ell^2}-2GM_i
\\
&&\hspace{-5mm}\dot{r}^2-\gamma^2_i+h_i=0
\label{eomRingClassic}
\eea
which follows from the geodesic motion of a freely falling object accompanying the freely collapsing dust shell, with $-h_idt^2+h_i^{-1}dr^2+r^2d\phi^2\equiv ds^2$ being the effective geometry felt by the $i$-shell during the motion, $M_i$ is the mass of all shells inside $i$ and $i$-shell itself. When a shell is released from inside the horizon $r_\mathrm{release}<r_h$ determined by $M_i$, the integration constant $\gamma_i^2<0$. When it is released from outside the horizon, then $\dot{r}=0$, $\dot{t}=1$, $\gamma_i=h_{i}(r)_{r=r_\mathrm{release}}>0$. But as long as the shell is released from the horizon around, $\gamma_i$ is upper bounded. At classic levels, $\gamma_i^2<0$ will assure the shell's motion happens always inside the horizon. 

\begin{figure}[ht]
\begin{center}
\includegraphics[width=0.47\textwidth]{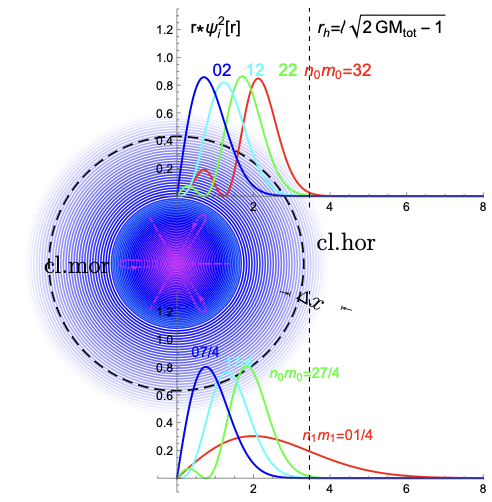}
\end{center}
\caption{Two examples of shell wave function for an AdS$_{2+1}$-Schwarzschild black hole of mass $M_\mathrm{tot}=2\ell^{-1}_\mathrm{ads}$. We let $G=\hbar=1$ and measure all dimensional quantities by $\ell_\mathrm{ads}$ or its inverse. If we consider the black hole consisting of just one shell of mass $m_{i(=0)}=2\ell^{-1}_\mathrm{ads}$, then the shell has 4 square integrable microstate wave functions, upper wave functions. If we consider the black hole consisting of two concentric shells, lower wave functions, then constrained by the quantizing condition \eqref{enQuantizCondition}, the possible shell mass decomposition could be $\{m^{\cdot\ell_\mathrm{ads}}_{i(=0)},m^{\cdot\ell_\mathrm{ads}}_{i(=1)}\}=\{\frac{7}{4},\frac{1}{4}\}$, $\{\frac{7-\epsilon}{4},\frac{1+\epsilon}{4}\}$, $\cdots$, $\{\frac{1}{\sqrt{2}},2{-}\frac{1}{\sqrt{2}}\}$ which is infinite and in-countable . However, possible quantum numbers $\{n_{i(=0)},n_{i(=1)}\}$ can only be $\{0/1/2,0\}=\{\mathrm{Int}[0,\!\frac{49}{16}-\frac{1}{2}],\frac{1}{4}\cdot2-\frac{1}{2}\}$, $\{1,1\}$ and  $\{0,0/1/2\}=\{\frac{1}{\sqrt{2}}\frac{1}{\sqrt{2}}{-}\frac{1}{2},\mathrm{Int}[0,(2{-}\frac{1}{\sqrt{2}}){\cdot}2{-}\frac{1}{2}]\}$. We can further consider the case the black hole consisting of more layers. However, constrained by the quantizing condition \eqref{enQuantizCondition}, the number of distinguishable shell quantum states in all cases are finite and countable. All shell wave function displayed in the figure has long tail of nonzero values outside the horizon. This makes the physic horizon of black holes a thick blurring or fuzzing region instead a cut-clear geometric surface.
}\label{figWFads}\end{figure}

At quantum levels, introduce wave function $\psi_i(r)$ to denote the probability amplitude the shell be found of radius $r$, write the radial momentum $m\dot{r}$ into operators $i\hbar\frac{\partial}{\partial r}$ and make the classic constraint \eqref{eomRingClassic} an operator and act on the wave function, we will find that $\psi_i(r)$ satisfy equations of the form
\beq{}
[(i\hbar m_i^{\m1}\partial_r)^2+1+\frac{r^2}{\ell^2}-2GM_i-\gamma^2_i]\psi_i(r)=0.
\label{eqSchrodingerAdS}
\eeq
Obviously, at this levels, the radial coordinate varies in the whole $0\leqslant r<\infty$ range. $\gamma_i^2<0$ does not assure $\psi_i(r)$'s being zero uniformly outside the horizon. This will make the meaning of mass parameter $M_i$ appearing in equation \eqref{eqSchrodingerAdS} become subtle. Consider two shells $i'<i$ with masses $m_{i'}$ and $m_{i}$, at classic levels, we can safely say that shell $i$ lives in the gravitational field thus is affected by the mass of $i'$ but not the contrary. But on quantum levels, shell $i$ has probabilities be found inside shell $i'$, so it may exert gravitations to shell $i'$. To avoid this subtlety, we will define $M_i$ as the mass summation of all shell $i'$s whose $[r|\psi_{i'}(r)|^2]_\mathrm{max}$ occur more close to the central point than $[r|\psi_{i}(r)|^2]_\mathrm{max}$ does.  Approximately, $M_i\approx\sum_{i'\leqslant i}m_{i'}$.

Excluding this subtle temporarily, eq\eqref{eqSchrodingerAdS} is nothing but the eigen-state Schr\"odinger equation of harmonic oscillators. The square integrability of the wavefunction requires that the following combination of mass and $\gamma$ parameter of the shell be quantized
\beq{}
\psi_i[r]=N_ie^{-\frac{m_ir^2}{2\ell}}\mathrm{HermiteH}[\frac{E_i\ell}{\hbar}-\frac{1}{2},\frac{m_i^{\!\frac{1}{2}}r}{\ell^\frac{1}{2}}]
\label{wfHarmOscillator}
\eeq
\bea{}
\big(GM_i+\frac{\gamma^2_i-1}{2}\big)m_i
\equiv{E_i}=(n_i+\frac{1}{2}){\hbar\omega}
\label{enQuantizCondition}\\
\omega\equiv\ell^{-1},~n_i=0,1,2\cdots.
\nonumber 
\eea 
$n_i$ here is upper bounded because both the total mass of the dust-disk and the maximal value of $\gamma_i^2$ are bounded. Recalling that, at classical level, to assure the outmost shell's move inside the horizon, $\gamma^2_{out.most}<0$ is required; while for those non-outmost shells, $\gamma_i^2$ are non-necessarily negative but are all upper bounded. Taking all shells together, we have
\beq{}
\sum_im_i=M_\mathrm{tot}, \psi_1\otimes\psi_2\otimes\cdots\psi_{imax}=\psi_\mathrm{tot}.
\label{massWFuncDirec}
\eeq
We displayed in Figure \ref{figWFads} examples of wave function of consisting dust shells of an AdS$_{2+1}$-Schwarzschild black hole of mass $M_\mathrm{tot}=2\ell^{-1}_\mathrm{ads}$. On the upper panel, the dust ball is considered consisting of just one shell, so each curve in the figure corresponds to a wave function of the system. While on the downer panel, the dust ball is considered consisting of two shells, the wave function of the system is direct product of the red line and the remaining three color lines. 

Similar with 3+1D Schwarzschild case, from the figure we easily see that, all shell wave functions are nonzero outside geometric surface $r=2GM$. This makes the physic horizon of and AdS2+1 Schwarzschild black hole also thickly blurring or fuzzy region instead border-clear geometric surface. Obviously, the lower bound of masses of the out-most shell allowed to have is lower bounded
\beq{}
m_\mathrm{min}=(GM_\mathrm{tot}+\frac{\gamma^2_{out.most}-1}{2})^{-1}\ell^{-1}
\eeq
The wave function of this shell has the most fat nonzero tail outside the horizon surface thus the biggest probability be radiated away from the whole disk if spontaneous radiation happens. Its consistence [up to an $\mathcal{O}(1)$ numeric factor] with the Hawking temperature \eqref{ads2p1HawkingTempterature} implies that
\beq{}
GM_\mathrm{tot}+\frac{\gamma^2_\mathrm{out.most}-1}{2}\approx\sqrt{GM}
\eeq
This could be attributed to the conspiracy between $\{m_i\}$ and $\gamma_i\sim\{n_i\}$'s value assigning, otherwise things like $M_i\neq\sum_{i'\leqslant i}m_{i'}$ would happen\footnote{Similar subtlety may also happen in the 3+1D Schwarzschild case, but that would affect exact form of $F(k)$ defined in equation \eqref{Fkdefinition} but not the exponential feature \eqref{largeMtotEstimation}.}. See discussions under equation \eqref{eqSchrodingerAdS}. 

The quantizing condition \eqref{enQuantizCondition} implies that the ways for us to decompose (more precisely, to look or to consider) the disk into concentric shells and setting their release parameter $\gamma_i$s are not arbitrary. Among various listing and counting ways, a special one could be implemented as follows, choosing each $\gamma_i$ in such a way that $\big(GM_i+\frac{\gamma^2_i-1}{2}\big)$ equals to 1 uniformly, in this case all $\frac{m_i}{\hbar\ell^{-1}}$ are forced to take integers. In another word, the excitation $\gamma_i$ in this case is no long a totally independent parameter for the microstate definition and their number counting becomes simply the conventional integer partition of $M_\mathrm{tot}$. By the partition-number formula of Ramanujan we have 
\bea{}
&&\hspace{-1mm}W\equiv\#\mathrm{microstate(}AdS_{2+1}\mathrm{dustdisk)}=
\label{Wformula}\\
&&\hspace{-5mm}\exp\big\{2\pi\sqrt{\frac{1}{6}\frac{M_\mathrm{tot}}{\hbar\ell^{-1}}}\big\}
=\exp\big\{\frac{2\pi\sqrt{2GM_\mathrm{tot}\ell^2}}{4G}\big(\frac{3}{4}\frac{\ell}{G}\big)^{-\frac{1}{2}}\big\}
\nonumber
\eea
Obviously, $k_{\scriptscriptstyle B}\log W$ is almost the Bekenstein-Hawking formula $S_{BH}=A/4G$ except for a numeric factor $\big(\frac{3}{4}\frac{\ell}{G}\big)^{-{1/2}}$. Ambiguities brought about by the subtlety involved in $M_i$'s definition and other possible ways of shell-decomposition and $\gamma_i$s' assigning will not change the basic fact that we have only finite number of ways to look the dust disk as concentric shells and setting their initial releasing condition as they collapse inside the horizon and begin to oscillation. Instead, precise consideration of those factors may bring us the factor $\big(\frac{3}{4}\frac{\ell}{G}\big)^{-{1/2}}$ back as the exact Bekenstein-Hawking formula requires. 

\section{A working example}
\label{secSimilarityDefinition}

To help readers more precisely understand physic pictures for our spontaneous radiation of black holes and their inner structure model, we provide in this section a working example of spontaneous radiation of an initially $W=6$ degenerating black hole. We will use equation \eqref{SchrodingerEq} and initial conditions \eqref{iniConSchrodingerEq} to follow up evolutions of such an initial black hole. This means that we need to introduce 1440 coefficients $c_{u^n}^{\vec{\omega}}$ to characterize quantum state of the system and set values for 184 coupling constant $g_{u^nv^\ell}$, according equation \eqref{monopoleCouplingStrength}-\eqref{similarityB} to account for transitions between all possible channels. 

When we get all $C_{16}^2=120$ similarity factors, we substitute them into equation \eqref{monopoleCouplingStrength} to get the coupling constant. As long as these preparation are ready, we can integrate equation \eqref{SchrodingerEq}-\eqref{iniConSchrodingerEq} to get all wave function coefficients $c_{u^n}^{\vec{\omega}}$. With the resulting $c_{u^n}^{\vec{\omega}}$, we can construct all needed density matrix $\rho^0_{ij}$, $\rho^b_{ij}$, $\rho^{m}_{ij}$, $\rho^{s}_{ij}$, $\rho^1_{ij}$. We will let $b\sim$big correspond $w=4$ black holes and illustrate that, tracing out states of the radiation products indeed leads to a non-diagonal density matrix for $\rho^b_{ij}$, although we are working in black hole inner structure eigenstate basis. Sizes and entropies of the black hole will be calculated as follows 
\bea{}
&&\hspace{-7mm}r_h(t)=|c_6^\phi|^2r_6+|c_5^{\vec{1}}|^2r_5+|c_4^{\vec{2}}|^2r_4+\cdots+|c_1^{\vec{5}}|^2r_1
\\
&&\hspace{-7mm}r_6=2GM_6,r_5=2GM_5,\cdots,r_1=2GM_\mathrm{pl}
\eea
\bea{}
&&\hspace{-7mm}s(t)=|c_5^{\vec{1}}|^2s_5+|c_4^{\vec{2}}|^2s_4+|c_3^{\vec{3}}|^2s_3+|c_2^{\vec{4}}|^2s_2+|c_1^{\vec{5}}|^2s_1
\\
&&\hspace{-7mm}s_6=s_0=0,s_5=5_1=1,s_4=s_2=2,s_3=3
\eea
Variations of this two quantity have been displayed in FIG.\ref{figWsixRadiationExample}.

\subsection{Microstate Wave Function}
Obviously, our first object should be the wave function of all possible intermediate state black holes. Among these black holes, the minimal mass one has degeneracy $W=1$, shell partition and excitation schemes of the form
\bea{}
W=1,M=1,\{m_i\}=\{1\},\{M_i\}=\{1\}
\\
\{M_im_i\}{=}\{1\},\{\nu_i\}{=}\{1\},\{z_i\}{=}\{1\},\#{\mathrm{p}{\otimes}\mathrm{e}}{=}1
\eea
Its wave function is a simple single factor one
\beq{}
\Psi_{1^1}[M(r)]=e^{-x}xL^1_0(2x),x=Mr\nu/\hbar=r
\eeq
with the average radius calculated easily
\beq{}
r^1_{11}=N_{11}\int_0^\infty\!\!\Psi_{1^1}\cdot r\cdot\Psi_{1^1}\cdot4\pi r^2dr=\frac{5}{2}
\eeq
where $N_{11}$ is a normalization factor for $\Psi_{1^1}$. It should be noted that $M=1$ is only lower bound of mass of black hole with $W=1$. Our inner structure model for black holes does not predict discrete spectrums for the mass of black hole holes.

For the $W=2$ intermediate state, we can easily determine that
\bea{}
W=2,M=\sqrt{2},\{m_i\}=\{\sqrt{2}\},\{M_i\}=\{\sqrt{2}\}
\\
\{M_im_i\}{=}\{2\},\{\nu_i\}{=}\{\frac{2}{1},\frac{2}{2}\},\{z_i\}{=}\{1,2\},\#{\mathrm{p}{\otimes}\mathrm{e}}{=}2
\eea
\bea{}
&&\hspace{-5mm}\Psi_{2^1}[M(r)]=e^{-x}xL^1_0(2x),x=Mr\nu/\hbar=2\sqrt{2}r
\\
&&\hspace{-5mm}\Psi_{2^2}[M(r)]=e^{-x}xL^1_1(2x),x=Mr\nu/\hbar=\sqrt{2}r
\eea
\bea{}
r^1_{21}=N_{21}\int_0^\infty\!\!\Psi_{2^1}\cdot r\cdot\Psi_{2^1}\cdot4\pi r^2dr=\frac{5}{4\sqrt{2}}
\\
r^1_{22}=N_{22}\int_0^\infty\!\!\Psi_{2^2}\cdot r\cdot\Psi_{2^2}\cdot4\pi r^2dr=\frac{55}{14\sqrt{2}}
\eea
where $N_{21}$ and $N_{22}$ are corresponding normalization factors.
Similar with the $W=1$ case, $M=\sqrt{2}$ is also only lower bound of mass of black holes with $W=2$. When $\sqrt{2}<M<M_{W=3}$, we can always fine tuning the value of $\nu_i$ so that $z_i\equiv\frac{GM_im_i}{\nu_i}=\{1,2\}$. From hereafter, we will not point out explicitly that the value of all mass parameters $M$ determined in the following $W=3,4,5,6$ cases are only lower bound.

Beginning from $W=3$, we have to note that the matter contents of a black hole could have two or more layers of shell structure
\bea{}
&&\hspace{8mm}W=3,M=\frac{\sqrt{5}+1}{2}\approx1.618\\
&&\hspace{-5mm}\begin{matrix}
\{m_i\}\!&\!\{1{+}\xi\}\!&\!\{1,\xi\}
\\
\{M_i\}\!&\!\{(1{+}\xi)\}\!&\!\{1,1{+}\xi\}\!&\!\xi(1{+}\xi){=}1
\\
\{M_im_i\}\!&\!\{(1{+}\xi)^2{<}3\}\!&\!\{1,\xi(1{+}\xi)\}\!&\!\Downarrow
\\
\{\nu_i\}\!\!&\!\!\{\frac{(1{+}\xi)^2\!}{1},\!\frac{(1{+}\xi)^2\!}{2}\}\!\!&\!\!\{1,1\}\!&\!\xi{=}
\\
\{z_i\}\!\!&\!\!\{1,2\}\!\!&\!\!\{1,1\}&(\sqrt{5}{-}1)/2
\\
\#{\mathrm{p}{\otimes}\mathrm{e}}\!&\!2\!&\!1
\end{matrix}
\eea
\bea{}
&&\hspace{-5mm}\Psi_{3^1}[M(r)]=e^{-x}xL^1_0(2x),x=\frac{Mr\nu}{\hbar}=(\sqrt{5}{+}2)r
\\
&&\hspace{-5mm}\Psi_{3^2}[M(r)]=e^{-x}xL^1_1(2x),x=Mr\nu/\hbar=\frac{\sqrt{5}{+}2}{2}r
\\
&&\hspace{-5mm}\Psi_{3^3}[M(r)]=e^{-x_1}x_1L_0^1(2x_1)\otimes e^{-x_2}x_2L_0^1(2x_2)
\\
&&\hspace{-5mm}x_1=m_1r\nu_1/\hbar=r,x_2=m_2r\nu_2/\hbar=\frac{\sqrt{5}-1}{2}r
\eea
\bea{}
&&\hspace{-5mm}r^1_{31}=N_{31}\int_0^\infty\!\!|\Psi_{3^1}|^2\cdot r\cdot4\pi r^2dr
=\frac{\scriptstyle2415{-}1080\sqrt{5}}{\scriptstyle6(17\sqrt{5}{-}38)}
\\
&&\hspace{-5mm}r^1_{32}=N_{32}\int_0^\infty\!\!|\Psi_{3^2}|^2\cdot r\cdot4\pi r^2dr
=\frac{\scriptstyle55(161-72 \sqrt{5})}{\scriptstyle7(17\sqrt{5}{-}38)}
\\
&&\hspace{-5mm}r^1_{33}{\otimes}r^2_{33}=N_{33}\int_0^\infty\!\!\Psi_{3^3}\cdot r\cdot\Psi_{3^3}\cdot4\pi r^2dr
\\
&&\hspace{-5mm}=\{\frac{\scriptstyle5}{\scriptstyle2},\frac{\scriptstyle5 (7{+}3\sqrt{5})}{\scriptstyle4(2{+}\sqrt{5})}\}
\eea
The radius of each shell in the two shell structure must be calculated independently and directly multiplying together. This direct product format is necessary for our calculation of similarities between different microstates.

For the $W=4$ case, we have
\bea{}
&&\hspace{-2mm}W=4,M=\sqrt{3}\approx1.732,\epsilon{<}2{-}\sqrt{3}{<}\frac{1}{2}\\
&&\hspace{-5mm}\begin{matrix}
\{m_i\}&\{\sqrt{3}\}&\{\frac{2}{\sqrt{3}},\frac{1}{\sqrt{3}}\}&\{\frac{2{-}\epsilon}{\sqrt{3}},\frac{1{+}\epsilon}{\sqrt{3}}\}
\\
\{M_i\}&\{\sqrt{3}\}&\{\frac{2}{\sqrt{3}},\sqrt{3}\}&\{\frac{2{-}\epsilon}{\sqrt{3}},\sqrt{3}\}
\\
\{M_im_i\}\!\!&\!\!\{3\}&\{\frac{4}{3},1\}&\{\frac{(2{-}\epsilon)^2\!}{3},1{+}\epsilon\}
\\
\{\nu_i\}\!&\!\{\frac{3}{1},\frac{3}{2},\frac{3}{3}\}\!&\!\{\frac{4}{3},1\}&\{\frac{(2{-}\epsilon)^2\!}{3},1{+}\epsilon\}
\\
\{z_i\}\!&\!\{1,2,3\}\!&\!\{1,1\}\!&\!\{1,1\}
\\
\#{\mathrm{p}{\otimes}\mathrm{e}}&3&1&\!\!0\mathrm{\,acc\,to\,}M_i^{\mm\frac{1}{2}}\!\!/2
\end{matrix}
\eea
\bea{}
&&\hspace{-5mm}\Psi_{4^1}[M(r)]=e^{-x}xL^1_0(2x),x=Mr\nu/\hbar=3\sqrt{3}r
\\
&&\hspace{-5mm}\Psi_{4^2}[M(r)]=e^{-x}xL^1_1(2x),x=Mr\nu/\hbar=\frac{3\sqrt{3}}{2}r
\\
&&\hspace{-5mm}\Psi_{4^3}[M(r)]=e^{-x}xL^1_2(2x),x=Mr\nu/\hbar=\frac{3\sqrt{3}}{3}r
\\
&&\hspace{-5mm}\Psi_{4^4}[M(r)]=e^{-x_1}x_1L_0^1(2x_1)\otimes e^{-x_2}x_2L_0^1(2x_2)
\\
&&\hspace{-5mm}x_1=m_1r\nu_1/\hbar=\frac{8r}{3\sqrt{3}},x_2=m_2r\nu_2/\hbar=\frac{r}{\sqrt{3}}
\eea
\bea{}
&&\hspace{-5mm}r^1_{41}=N_{41}\int_0^\infty\!\!\Psi_4^1\cdot r\cdot\Psi_4^1\cdot4\pi r^2dr
=\frac{\scriptstyle5}{\scriptstyle6\sqrt{3}}
\\
&&\hspace{-5mm}r^1_{42}=N_{42}\int_0^\infty\!\!\Psi_4^2\cdot r\cdot\Psi_4^2\cdot4\pi r^2dr
=\frac{\scriptstyle55}{\scriptstyle21\sqrt{3}}
\\
&&\hspace{-5mm}r^1_{43}=N_{43}\int_0^\infty\!\!\Psi_4^3\cdot r\cdot\Psi_4^3\cdot4\pi r^2dr
=\frac{\scriptstyle25}{\scriptstyle6\sqrt{3}}
\\
&&\hspace{-5mm}r^1_{44}{\otimes}r^2_{44}=\{\langle r_1\rangle,\langle r_2\rangle\}_4^4
=\{\frac{\scriptstyle15\sqrt{3}}{\scriptstyle16},\frac{\scriptstyle5\sqrt{3}}{\scriptstyle2}\}
\eea

For the $W=5$ case, we have
\bea{}
&&\hspace{3mm}W=5,M=\sqrt{3+\xi}\approx1.823,\epsilon={1/2}
\\
&&\hspace{-5mm}\begin{matrix}
\{m_i\}&\{\sqrt{3{+}\xi}\}\!&\!\{\frac{2{+}\xi}{\sqrt{3{+}\xi}}\!,\!\frac{1}{\sqrt{3{+}\xi}}\}\!&\!\{1,\frac{1{+}\epsilon}{\sqrt{3+\xi}}\}
\\
\{M_i\}&\{\sqrt{3}{+}\xi\}\!&\!\{\frac{2{+}\xi}{\sqrt{3{+}\xi}}\!,\!{\scriptstyle\sqrt{3{+}\xi}}\}\!\!&\!\{1,{\scriptstyle\sqrt{3{+}\xi}}\}
\\
\{M_im_i\}\!\!&\!\!\{3{+}\xi\}\!&\!\{\frac{(2{+}\xi)^2}{3{+}\xi}{<}2,1\}\!&\!\{1,1{+}\epsilon\}
\\
\{\nu_i\}\!\!&\!\!\{\frac{3{+}\xi}{1}\!,\!\frac{3{+}\xi}{2}\!,\!
\frac{3{+}\xi}{3}\}\!&\!\{\frac{(2{+}\xi)^2\!}{3{+}\xi}\!,\!1\}&\{1,1{+}\epsilon\}
\\
\{z_i\}\!&\!\{1,2,3\}\!&\!\{1,1\}&\{1,1\}
\\
\#{\mathrm{p}{\otimes}\mathrm{e}}&3&1&1
\end{matrix}
\\
&&\hspace{3mm}1{+}\frac{1{+}\epsilon}{\sqrt{3{+}\xi}}=\sqrt{3{+}\xi}
\Rightarrow \xi=\frac{\sqrt{7}}{2}{-}1
\eea
\bea{}
&&\hspace{-5mm}\Psi^1_5[M(r)]=e^{-x}xL^1_0(2x),x=\frac{Mr\nu}{\hbar}=\frac{(1{+}\sqrt{7})^3r\!\!}{8}
\\
&&\hspace{-5mm}\Psi^2_5[M(r)]=e^{-x}xL^1_1(2x),x=\frac{Mr\nu}{\hbar}=\frac{(1{+}\sqrt{7})^3r\!\!}{16}
\\
&&\hspace{-5mm}\Psi^3_5[M(r)]=e^{-x}xL^1_2(2x),x=\frac{Mr\nu}{\hbar}=\frac{(1{+}\sqrt{7})^3r\!\!}{24}
\\
&&\hspace{-5mm}\Psi^4_5[M(r)]=e^{-x_1}x_1L_0^1(2x_1)\otimes e^{-x_2}x_2L_0^1(2x_2)
\\
&&\hspace{-5mm}x_1=\frac{m_1r\nu_1}{\hbar}=\frac{\scriptstyle(\sqrt{7}{+}2)^3r}{\scriptstyle(\sqrt{7}{+}1)^3}
,x_2=\frac{m_2r\nu_2}{\hbar}=\frac{2r}{\scriptstyle\sqrt{7}{+}1}
\\
&&\hspace{-5mm}\Psi^5_5[M(r)]=e^{-x_1}x_1L_0^1(2x_1)\otimes e^{-x_2}x_2L_0^1(2x_2)
\\
&&\hspace{-5mm}x_1=m_1r\nu_1/\hbar=r,x_2=m_2r\nu_2/\hbar=\frac{\scriptstyle9r/2}{\scriptstyle\sqrt{7}{+}1}
\eea
\bea{}
&&\hspace{-5mm}r_{51}^1=N_{51}\!\int_0^\infty\!\!\Psi_5^1\cdot r\cdot\Psi_5^1\cdot4\pi r^2dr
=\frac{\scriptstyle5(5\sqrt{7}{-}11)}{\scriptstyle27}
\\
&&\hspace{-5mm}r_{52}^1=N_{52}\!\int_0^\infty\!\!\Psi_5^2\cdot r\cdot\Psi_5^2\cdot4\pi r^2dr
=\frac{\scriptstyle110(5\sqrt{7}{-}11)}{\scriptstyle189}
\\
&&\hspace{-5mm}r_{53}^1=N_{53}\!\int_0^\infty\!\!\Psi_5^3\cdot r\cdot\Psi_5^3\cdot4\pi r^2dr
=\frac{\scriptstyle85(5\sqrt{7}{-}11)}{\scriptstyle69}
\\
&&\hspace{-5mm}r_{54}^1{\otimes}r_{54}^2
=\{\frac{\scriptstyle5(115{-}41\sqrt{7})}{\scriptstyle27},\frac{\scriptstyle1{+}\sqrt{7}}{\scriptstyle4}\}
\\
&&\hspace{-5mm}r_{55}^1{\otimes}r_{55}^2
=\{\frac{\scriptstyle5}{\scriptstyle2},\frac{\scriptstyle5(1{+}\sqrt{7})}{\scriptstyle9}\}
\eea

The final is the $W=6$ case, although only $\Psi_6^1$ will be used for our concrete example, we choose to write down all 6 wave functions explicitly here.
\bea{}
&&\hspace{3mm}W=6,M=\sqrt{3{+}x}\approx1.932,\epsilon=1/2
\\
&&\hspace{-5mm}\begin{matrix}
\{m_i\}&\{\sqrt{3{+}x}\}&\{\sqrt{2},\frac{1}{\sqrt{3{+}x}}\}\!&\!\{\frac{2{-}\epsilon{+}x}{\sqrt{3{+}x}}\!,\!\frac{1{+}\epsilon}{\sqrt{3{+}x}}\}
\\
\{M_i\}&\{\sqrt{3{+}x}\}&\{\sqrt{2},{\scriptstyle\sqrt{3{+}x}}\}\!\!&\!\!\{\frac{2{-}\epsilon{+}x}{\sqrt{3{+}x}}\!,\!{\scriptstyle\sqrt{3{+}x}}\}
\\
\{M_im_i\}&\{3{+}x\}&\{2,1\}\!&\!\!\!\{\!\frac{(2{-}\epsilon{+}x)^2\!\!}{3{+}x}\!,\!1{+}\epsilon\!\}
\\
\{\nu_i\}\!&\!\{[\frac{3{+}x}{1,2,3}]\}\!&\!\{[\frac{2}{1},\frac{2}{2}],1\}\!\!&\!\!\{\!\frac{(2{-}\epsilon{+}x)^2\!\!}{3{+}x}\!,\!1{+}\epsilon\!\}
\\
\{z_i\}\!&\!\{1,2,3\}\!&\!\{[1,2],1\}\!&\!\{1,1\}
\\
\#{\mathrm{p}{\otimes}\mathrm{e}}&3&2&1
\end{matrix}
\\&&\hspace{3mm}
\sqrt{2}{+}\frac{1}{\sqrt{3{+}x}}{=}\sqrt{3{+}x}\Rightarrow x{=}\sqrt{3}-1
\eea
\bea{}
&&\hspace{-5mm}\Psi^1_6[M(r)]=e^{-x}xL^1_0(2x),x=Mr\nu/\hbar=\frac{\scriptstyle(\sqrt{3}{+}1)^3r\!}{\scriptstyle\sqrt{8}}
\\
&&\hspace{-5mm}\Psi^2_6[M(r)]=e^{-x}xL^1_1(2x),x=Mr\nu/\hbar=\frac{\scriptstyle(\sqrt{3}{+}1)^3r\!}{\scriptstyle2\sqrt{8}}
\\
&&\hspace{-5mm}\Psi^3_6[M(r)]=e^{-x}xL^1_1(2x),x=Mr\nu/\hbar=\frac{\scriptstyle(\sqrt{3}{+}1)^3r\!}{\scriptstyle3\sqrt{8}}
\\
&&\hspace{-5mm}\Psi^4_6[M(r)]=e^{-x_1}x_1L_0^1(2x_1)\otimes e^{-x_2}x_2L_0^1(2x_2)
\\
&&\hspace{-5mm}x_1=m_1r\nu_1/\hbar=2\sqrt{2}r,x_2=m_2r\nu_2/\hbar=\frac{\scriptstyle\sqrt{2}r}{\scriptstyle\sqrt{3}{+}1}
\\
&&\hspace{-5mm}\Psi^5_6[M(r)]=e^{-x_1}x_1L_0^1(2x_1)\otimes e^{-x_2}x_2L_1^1(2x_2)
\\
&&\hspace{-5mm}x_1=m_1r\nu_1/\hbar=\sqrt{2}r,x_2=m_2r\nu_2/\hbar=\frac{\scriptstyle\sqrt{2}r}{\scriptstyle\sqrt{3}{+}1}
\\
&&\hspace{-5mm}\Psi^6_6[M(r)]=e^{-x_1}x_1L_0^1(2x_1)\otimes e^{-x_2}x_2L_0^1(2x_2)
\\
&&\hspace{-5mm}x_1=m_1r\nu_1=\frac{\scriptstyle85\sqrt{2}{-}39\sqrt{6}}{\scriptstyle16}r,x_2=m_2r\nu_2=\frac{\scriptstyle9\sqrt{2}r}{\scriptstyle4(\sqrt{3}{+}1)}
\eea
\bea{}
&&\hspace{-5mm}r^1_{61}=N_{61}\!\int_0^\infty\!\!\Psi_6^1\cdot r\cdot\Psi_6^1\cdot4\pi r^2dr
=\frac{\scriptstyle5}{\scriptstyle2\sqrt{16{+}15\sqrt{3}}}
\\
&&\hspace{-5mm}r^1_{62}=N_{62}\!\int_0^\infty\!\!\Psi_6^2\cdot r\cdot\Psi_6^2\cdot4\pi r^2dr
=\frac{\scriptstyle55}{\scriptstyle7\sqrt{16{+}15\sqrt{3}}}
\\
&&\hspace{-5mm}r^1_{63}=N_{63}\!\int_0^\infty\!\!\Psi_6^3\cdot r\cdot\Psi_6^3\cdot4\pi r^2dr
=\frac{\scriptstyle765}{\scriptstyle46\sqrt{16{+}15\sqrt{3}}}
\\
&&\hspace{-5mm}r^1_{64}{\otimes}r^2_{64}=
=\{\frac{\scriptstyle5}{\scriptstyle4\sqrt{2}},\frac{\scriptstyle5(1{+}\sqrt{3})}{\scriptstyle2\sqrt{2}}\}
\\
&&\hspace{-5mm}r^1_{65}{\otimes}r^2_{65}=
=\{\frac{\scriptstyle5}{\scriptstyle2\sqrt{2}},\frac{\scriptstyle55(1{+}\sqrt{3})}{\scriptstyle14\sqrt{2}}\}
\\
&&\hspace{-5mm}r^1_{66}{\otimes}r^2_{66}=
=\{\frac{\scriptstyle20\sqrt{2}}{\scriptstyle85{-}39\sqrt{3}},\frac{\scriptstyle5(\sqrt{2}{+}\sqrt{6})}{\scriptstyle9}\}
\eea

\subsection{Similarity Factors and Transition Amplitude}
\begin{figure}[h]
\includegraphics[totalheight=54mm]{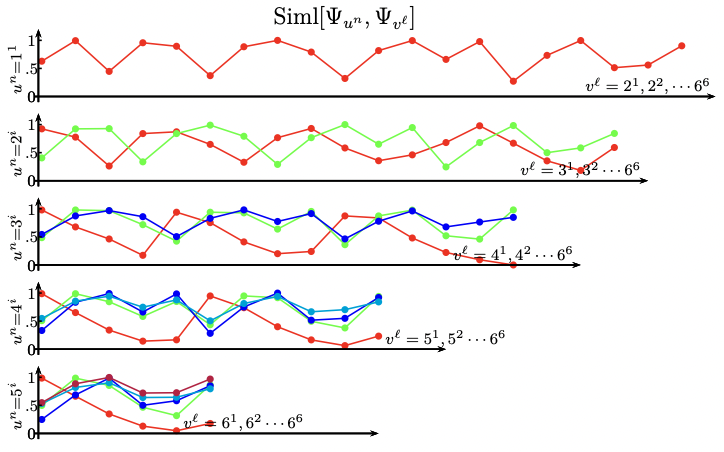}
\caption{Similarity factors between all different inner structure wave function defined by \eqref{similarityA}-\eqref{similarityB}.}
\label{figSimilarity}
\end{figure}

By definitions \eqref{similarityA}-\eqref{similarityB} and \eqref{monopoleCouplingStrength}, we can calculate all the relevant transition amplitude or similarity factors as follows
\beq{}
\mathrm{Siml}[\Psi_{1^1},\Psi_{2^1}]=\frac{2r^1_{11}\cdot r^1_{21}}{(r^1_{11})^2+(r^1_{21})^2}=\frac{4\sqrt{2}}{9}
\eeq
\beq{}
\mathrm{Siml}[\Psi_{1^1},\Psi_{2^2}]=\frac{2r^1_{11}\cdot r^1_{22}}{(r^1_{11})^2+(r^1_{22})^2}=\frac{154\sqrt{2}}{219}
\eeq
\beq{}
\mathrm{Siml}[\Psi_{1^1},\Psi_{3^1}]=\frac{2r^1_{11}\cdot r^1_{31}}{(r^1_{11})^2+(r^1_{31})^2}=\frac{1}{\sqrt{5}}
\eeq
\beq{}
\mathrm{Siml}[\Psi_{1^1},\Psi_{3^2}]=\frac{308 (870 + 533\sqrt{5})}{663545}
\eeq
\bea{}
&&\hspace{-7mm}\mathrm{Siml}[\Psi_{1^1},\Psi_{3^3}]=\Big[\frac{2r^1_{11}\cdot r^1_{33}}{(r^1_{11})^2{+}(r^1_{33})^2}\cdot\frac{2r^1_{11}\cdot r^2_{33}}{(r^1_{11})^2{+}(r^2_{33})^2}\Big]^\frac{1}{2}
\\
&&\hspace{-7mm}=\frac{\sqrt{2}}{5^{1/4}}
\nonumber
\eea
Other similarity factors could be calculated similarly. Our results are compiled in FIG.\ref{figSimilarity} explicitly. From the figure, we easily see that, in most cases, $\mathrm{Siml}[\Psi_{u^n},\Psi_{v^\ell}]=\mathcal{O}[1]$, there very few cases $\mathrm{Siml}[\Psi_{u^n},\Psi_{v^\ell}]\ll1$. So, in crude estimations, we can set all $g_{u^nv^\ell}=1$ for efficiency.

\subsection{Schr\"odinger Equation and Partial State Tracing out}

Substituting results above into transition amplitude \eqref{monopoleCouplingStrength} and differential equations \eqref{SchrodingerEq}-\eqref{iniConSchrodingerEq}. We can integrate and get 1440 time dependent coefficient functions $c_{u^n}^{\vec{\omega}}(t)$. They form wave functions of an evaporating black hole and its radiation products which initially is set on pure state $c_{u^n}^{\vec{\omega}}=\delta_{u6}\delta_{n1}\delta^{\vec{\omega}\phi}$, symbols $\phi$ here is used denoting zero radiation product state.  Our key results has been displayed in FIG.\ref{figWsixRadiationExample} already. We provide here evidences that tracing out state of the radiation products will make the density matrix of the black hole non-diagonal even when we are working on basis of inner structure eigenstate of the black hole during the whole process.

\begin{figure}[h]
\includegraphics[totalheight=25mm]{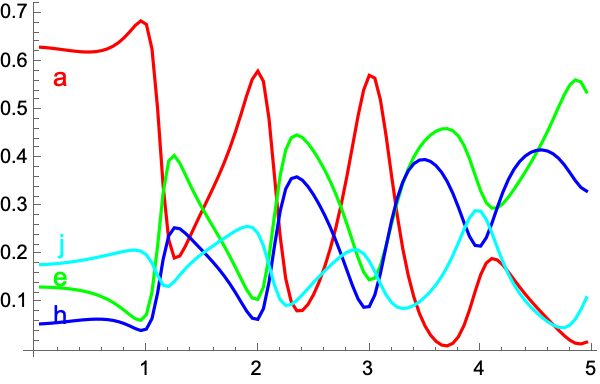}
\includegraphics[totalheight=25mm]{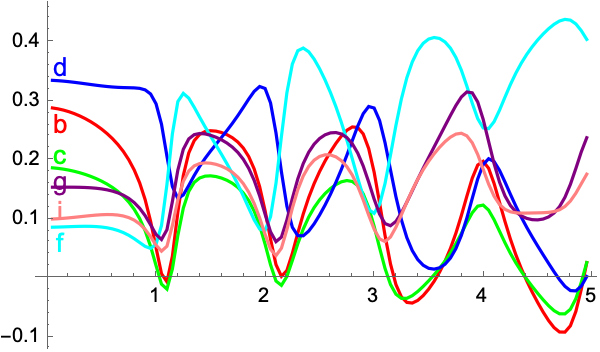}
\caption{Variations of the diagonal and off diagonal elements of density matrix \eqref{densityMatrixTracingout}, the horizontal axis is time.}
\label{figTracingout}
\end{figure}

We will take the subspace of $u^n\otimes\vec{\omega}=$ $4^{n=}_{1234}\otimes$ $\{\{2^1\},\{1^{j=}_{1\cdot\cdot5},1^1\}\}$ as example. This is a 24 dimensional Hilbert space, where $u^n$ is the possible black hole microstate, $\vec{\omega}$ is the possible radiation product state. If we write this vector explicitly, its form reads like
\bea{}
c_{u^n}^{\vec{\omega}}{=}\{c_{4^1}^{\{2^1\}},c_{4^1}^{\{1^11^1\}},c_{4^1}^{\{1^21^1\}},\cdots c_{4^1}^{\{1^51^1\}},c_{4^2}^{\{2^1\}},c_{4^2}^{\{1^11^1\}}
\\
,c_{4^2}^{\{1^21^1\}},\cdots c_{4^2}^{\{1^51^1\}},\cdots\cdots c_{4^4}^{\{1^51^1\}}\}^\mathrm{transpose}
\nonumber
\eea
The corresponding density matrix has forms (since initially our black holes has $w=6$, we will call its radiation residue $w=4$ big here)
\bea{}
\rho^b_{ij}=c_{u^n}^{\vec{\omega}}c_{u^n}^{\vec{\omega}\dagger}=
\left[\begin{matrix}
a_{6\times6}&b_{6\times6}&c_{6\times6}&d_{6\times6}
\\
b_{6\times6}&e_{6\times6}&f_{6\times6}&g_{6\times6}
\\
c_{6\times6}&f_{6\times6}&h_{6\times6}&i_{6\times6}
\\
d_{6\times6}&g_{6\times6}&i_{6\times6}&j_{6\times6}
\end{matrix}
\right]
\eea
subscript $i,j$ on $\rho_{ij}$ specify state of the black hole and radiation products. Tracing out state of the latter means that we take traces on each $6\times6$ submatrix and get a $4\times4$ matrix
\bea{}
\rho^{b}_{hh'}=
\left[\begin{matrix}
a&b&c&d
\\
b&e&f&g
\\
c&f&h&i
\\
d&g&i&j
\end{matrix}
\right]
\label{densityMatrixTracingout}
\eea
subscripts $h,h'$ on this matrix density specify state of the black hole only. Although during the whole process we are kept on basis of eigenstate of the black hole inner structure and radiation products' energy, tracing out microstate of the latter provides us a non-diagonal matrix $\rho^{b}_{hh'}$. We provide in FIG.\ref{figTracingout} their time dependence explicitly.

\section{Some discussions} 
This section is some discussion or critics on our inside horizon oscillation picture and its development.

\subsection{On the meaning of inside horizon oscillation}

\begin{figure}[h]
\includegraphics[totalheight=27mm]{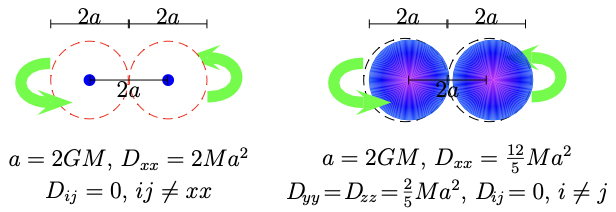}
\caption{A binary black holes' inspiral and merging process. The left hand side black holes point like one, quadrupole $D_{ij}$ of the sysetm have only one non-0 component. The right hand side black holes have oscillatory cores, the system have rather different quadrupoles. When they inspiral and merging, this two systems will experience rather different tidal deformation thus emitting rather different gravitational wave.}
\label{figBinary}
\end{figure}

Many people believe that, it is of nonsense to talk about a spherically symmetric black hole's inner matter distribution and motion modes due to the horizon's forbidding of such inner structure's being measurable by outside observers.  However, this argument applies only in single body context. When we let two spherically symmetric black holes (one with totally known inner structure and the other unknown) spin around and merge with each other, spherical symmetries of both black holes will be broken due to tidal deformation and the structure of the previously unknown black hole will be measurable from gravitational waves following from the two bodies' merging process. The basic reason here is that, black holes with different inner matter distribution and oscillation modes, for example two with singular central points and two without such singular centers, will exhibit different tidal deformation and lead to different gravitational waves as the merging proceeds \cite{dfzeng2020}. We display in FIG.\ref{figBinary} the quadrupole of two such static binary systems. Although the difference between the two static systems is a diagonal constant matrix which has no observable effects, when they spin around their own central vertical line, the right system will experience different tidal deformation and produce different gravitation waves relative to the left one. 

If the method of effective field theory is adopted to describe inspiral and mergering processes of such inner structured objects \cite{Porto2016,eftGoldberger2006,eftGoldberger2007}, we have to write
\bea{}
&&\hspace{-3mm}\exp\{iS_\mathrm{EH}[g^{\scriptscriptstyle\!L}_{\mu\nu}(x)+iS_\mathrm{eff}[x^\alpha_\mathrm{cm}(\sigma),g^{\scriptscriptstyle\!L}_{\mu\nu}(x)]]\}=
\\
&&\hspace{-5mm}\int\!\!\mathscr{D}\!g^{\scriptscriptstyle\!S}_{\mu\nu}\mathscr{D}\!\delta\!{x}^{\!\alpha}\!(\sigma_p\!)
\exp\{iS_\mathrm{EH}[g_{\mu\nu}\!(x){+}iS_\mathrm{eff}[x^\alpha_\mathrm{cm}\!(\sigma_p\!),g_{\mu\nu}(x)]]\}
\nonumber
\eea
where $g^{\scriptscriptstyle\!L}_{\mu\nu}$ and $g^{\scriptscriptstyle\!S}_{\mu\nu}$ are the long and short degrees of freedom of spacetime geometry;
$x^{\alpha}_\mathrm{cm}(\sigma)$($\alpha=1,2$) denote the center of mass coordinate of the black holes under consideration; $\sigma$ parameterizes their worldlines; ${\delta}x^\alpha_p=x^\alpha_p-x^\alpha_\mathrm{cm}$ represent displacement of constituents of the black hole relative to their own centers. Integrating out short the distance degrees of freedom, we will have 
\bea{}
S_\mathrm{eff}[x_\mathrm{cm},g^{\scriptscriptstyle\!L}_{\mu\nu}]=\!\!\int\!\!d^4x\delta[x-x_\mathrm{cm}(\sigma)]\Big\{
\\
-m\sqrt{g^{\scriptscriptstyle\!L}_{\mu\nu}\dot{x}^\mu_\mathrm{cm}(\sigma)\dot{x}^\nu_\mathrm{cm}(\sigma)}
\nonumber\\
+C_R\int R^{\scriptscriptstyle\!L}(g^{\scriptscriptstyle\!L}_{\mu\nu})\sqrt{g^{\scriptscriptstyle\!L}_{\mu\nu}\dot{x}^\mu_\mathrm{cm}(\sigma)\dot{x}^\nu_\mathrm{cm}(\sigma)}
\nonumber\\
+C_V\int R^{\scriptscriptstyle\!L}_{\mu\nu}(g^{\scriptscriptstyle\!L}_{\mu\nu})\frac{\dot{x}^\mu_\mathrm{cm}(\sigma)\dot{x}^\nu_\mathrm{cm}(\sigma)}{\sqrt{g^{\scriptscriptstyle\!L}_{\mu\nu}\dot{x}^\mu_\mathrm{cm}(\sigma)\dot{x}^\nu_\mathrm{cm}(\sigma)}}
+\cdots
\Big\}
\nonumber
\eea
Ellipsis on the last line denotes higher order terms in long wave length expansion of the world line action, $C_R$, $C_V$ and those omitted $C_{X}$ are series of parameters called Wilson coefficients. If the black hole owns no inner structure, all the relevant Wilson coefficients $C_{i}$ will be determined solely by the its mass $M$ and Newton couplings $G$, \cite{Bini2012,Chakrabarti2013a,Chakrabarti2013b,Bini2015}. However, if the black hole own non-trivial inner structure, such as we discussed in this work, then these Wilson coefficients will manifest correspondingly features reflecting the underlying structure. This will make the black hole behave more closely with neutron stars \cite{eftBinnington,eftDamour2009,Hinderer2008,Hinderer2010,Vines2011,Yagi2013,Donneva2013,Bernuzzi2014,Agathos2015,Chatziioannou2015}
  
In practice, data collected in the LIGO/VIRGO observation may not be sufficient to uncover precisely the inner structure of black holes being observed \cite{gw150914,gw170817a,gwScalarSoliton,cardoso2016,afshordi2016,cardoso2017}. But it may have potentials to tell us if black holes have singular centers or not. So to talk about the consisting matters' inside horizon oscillation is not a pure theoretical question, it is an experimental one.

In single body case, implications of matter's oscillation inside the horizon could be seen in some theoretical examination. For example, to recover Hawking radiation's unitarity and to get proper interpretation for the Bekenstein Hawking entropy's origin, 't Hooft \cite{tHooftISSP2016,tHooftISSP2018,tHooft1809,tHooft1804,tHooft1612,tHooftFP2018,tHooft1605,tHooft1601} pointed out that extra assumption, such as Antipodal Points Identification (API) about the horizon itself is almost unavoidable. API assumes that points in regions III and IV of the Penrose diagram of the Schwarzschild black hole should be identified with the antipodal ones in regions I and II. Obviously, in black hole with oscillatory inner structures such as those described by our exact solutions, API is not an artificial assumption, but a natural deduction of the math formalism. Further more, when quantum effects are considered, matters consisting of the black hole have very large probabilities be measured outside its horizon, thus making the horizon a blurring or fuzzing region instead a cut-clear surface. Observers outside the hole need not waiting infinite durations to see signals unveiling the matters' falling into this region and participating oscillations inside it. On pure general relativistic level, the fact that inside horizon oscillations happen beyond the domain of time definition of outside observer only implies that, we need to consider them as physics occurring in the parallel universe or statistic ensemble world. Details of such motions are measurable to observers living in the corresponding universe, i.e. Maxwell's demon like observers. To us, statistic mechanic viewpoint or many world explanation is enough.

To emphasize the role of matters consisting of the black hole instead of focusing on pure gravitational or geometric degrees of freedom in the information missing puzzle's resolving is also the key of works \cite{Stojkovic2007,Stojkovic2008a,Stojkovic2008b,Stojkovic2009,Stojkovic2013,Stojkovic2014,Stojkovic2015}. However, there is popular sayings that the number of degree of freedom following from the constituent matter would lead to a volume-law entropy instead of area-law type one. This is a very misleading doctrine. Since the number of particles constrained inside the horizon are proportional to the black hole mass $N\propto M$, thus in 3+1D Schwarzschild case are proportional to its horizon size $N\propto r_h=2GM$, simple calculation
\bea{}
&&\hspace{-2mm}Z=\int\!d\vec{x}_1d\vec{p}_1{\cdots}d\vec{x}_{\scriptscriptstyle\!N}d\vec{p}_{\scriptscriptstyle\!N} e^{-\beta H(\vec{x}_1,\vec{p}_1,\cdots)}
\\
&&\hspace{-5mm}S\sim(1-\beta\partial_\beta)\ln Z\sim N^{1-\epsilon}\sim r_h^{1-\epsilon},\epsilon>0
\eea
will give us neither volume-law, nor area-law, but diameter-law entropy at most. Then how do we get area law entropy in our work? The key point here is, the objects carrying degrees of freedom evaluated by the area-law entropy is not the usual particles consisting of the black hole, but {\em collective motion modes} of them. According to our quantizing condition \eqref{enQuantizConditionSchwz}, the characteristic mass of such modes is $m_\mathrm{i=out.most}\propto (GM_i)^{-1}=(GM_\mathrm{tot})^{-1}$. In a black hole of mass $M_\mathrm{tot}$, the number of such objects is $N_{\scriptscriptstyle\!CMM}\propto GM_\mathrm{tot}^2$, which is obviously proportional to the area of the black hole under consideration.

\subsection{Development of our ideas}

Finally, we wish to make a short review about our ideas on taking the inside horizon oscillation modes as the origin of Bekenstein-Hawking entropy. We firstly proposed this possibility in reference \cite{dfzeng2017}. In that work, we write the inside horizon metric in the following form
 \bea{}
 &&ds^2=-h^{-1}A(\tau,r)d\tau^2+h^{-1}dr^2+r^2d\Omega^2
 \label{InnerSoldynamic}
 \\
 &&A=\frac{\dot{m}^2}{{m'}^2}+h,~h=1-\frac{2m(\tau,r)}{r}
 \nonumber
 \eea
 and tries to build a quantization description for $m(\tau,r)$ basing on Wheeler-DeWitt equation \cite{deWitt1967,QCosmologyHalliwellLecture,QBHAllen1987,QBHFangLi1986,QBHLaflamme1987,QBHNambuSasaki1988,QBHNagai1989,QBHRodrigues1989}. But our equations (18) and (21) there contain errors. The correct form should be
\beq{}
\big[\frac{16r^2 \rho m'^{-1}}{h^2}{+}\frac{\ell_p^2\delta^2}{\delta m(r)^2}+\frac{4r^4\rho^2m'^{-2}}{h^{2}}\big]\Psi[m(r)]=0
\label{funcSchrodinger}
\eeq 
\begin{gather}
\forall r\in[0,r_h],~\big[\big(\frac{8mr}{\xi}r^2\!\rho
{+}\frac{16r^2}{\xi^2}r^4\!\rho^2\big)\big(1{-}\frac{2m}{r}\big)^{\m2}(\ln{\!r})^2
\nonumber
\\
\rule{5mm}{0pt}+\hbar^2\big(\ln{\!r\,}\partial_\xi-\partial^2_\xi\big)\big]\Psi(\xi)=0
,~m=\frac{r_h}{2}\big(\frac{r}{r_h}\big)^\xi
\label{normalEigenvalueproblem}
\end{gather}
After this correction, the quantitative feature of wave functions listed in the Fig.4 there would be revised. But the key conclusions do not change.

The functional Schr\"odinger equation \eqref{funcSchrodinger} is very difficult to solve. So in reference \cite{dfzeng2018a} we try an alternative way to consider the question. We look the matter contents of a spherically symmetric black hole as many concentric shells. By applying quantization scheme to each of these shells and taking their total mass being equal to the mass of black hole under consideration as a constraint, we build our first version of inside horizon oscillation model in this work. Basing on this model, we write down an explicitly hermitian hamiltonian to describe their radiation and argues that the spectrum is thermal.

In reference \cite{dfzeng2018b}, we further polish our ideas in reference \cite{dfzeng2018a} and provide more evidences for the degeneracy counting of  inside horizon oscillation modes of their matter contents leads in indeed area law formulas for the Bekenstein Hawking entropy. By the hermitian Hamiltonian describing the radiation of black holes, we provide numeric evidences that the information missing during the radiation of a black hole could be retrieved from its size variation curve. But both in\cite{dfzeng2018a} and in \cite{dfzeng2018b}, we find no ways to prove exactly how thermal spectrums could be produced from such an explicitly hermitian Hamiltonian.

In reference \cite{dfzeng2020}, we find exact dynamical solutions to the Einstein equation sourced by oscillatory and shell structured dust matter, both in asymptotically AdS2+1 and Minkowskian 3+1 spacetime. We discuss their quantization and implications for the Bekenstein Hawking formulas. We also speculate this inside horizon structure's being detectability.

Finally, in the current work, we compiles all these results together and build concrete theories for the spontaneous radiation of black holes. By applying Wigner-Wiesskopf approximation, we prove with rather strictness that an explicitly hermitian hamiltonians like those in \cite{dfzeng2018a} and \cite{dfzeng2018b} indeed yield thermal radiations spectrum as expected in Hawking radiation.

\section{Conclusion} 

We build an explicitly hermitian hamiltonian description for the spontaneous radiation of black holes. Our description is essentially a many-mode multi-degeneracy generalization of Jaynes-Cummings model, which we call Z-model according to which particles are radiated away from or absorbed into the black hole through monopole type couplings between the black hole and radiation particles. Focusing on single particle's radiation, by the well known method of Wigner-Wiesskopf approximation, we find that our model yields power spectrum completely the same as Hawking radiation requires. While as many particles' radiation is concerned, numeric method can be used to trace the full process of a black hole's radiation evolution, from which proper curve for the variation of hawking particles entropy follows rather naturally. 

Our model analysis point out that, to let hawking particles carry information and escape away from a black hole controlled by no hair theorem, two ingredients are necessary. The first is, hawking particles must be produced through direct/indirect couplings with the black hole microscopic state, instead of vacuum pair production around the classic no-hair horizon, as Hawking imagined originally. Mathur's small correction theorem provides us strong implication on this point. The second is, a two-levels superposition happening in the intermediate state of an evaporating black hole must be accounted for properly. That is, any such intermediate state are superpositions of both (i) microscopic states of black hole with equal mass but different inner structures and (ii) states of black hole with all possible masses less than or equal to the initial one. We explore in this language working logics of the replica wormhole method and point out that, it can be interpreted as an effective account of the two-levels superposition essence of the intermediate state of evaporating black holes.

We then discuss the most important basis for the Z-model of spontaneous radiation of black holes, that is, their atomic like inner structures embodied in the Bekenstein-Hawking entropy. We provide exact inside horizon oscillatory solutions family to the Einstein equation sourced by spherically symmetric, shell structured dust ball, both in Minkowskian and Anti-de Sitter asymptotic.  After quantization, the horizon of such inside horizon oscillatory solutions becomes highly blurred and the whole black hole becomes large amount of fuzzy balls, the matter contents of each have remarkably big probability being found outside the horizon. More importantly, we find that spectrums of these fuzzy balls' quantum wave functional are discrete and countable, with the degeneracy almost perfectly consistent with the area law formula of Bekenstein-Hawking entropy, both in 3+1D Schwarzschild black hole and in 2+1D AdS-Schwarzschild case. Comparing with string theory fuzzy balls, our Schwarzschild fuzzy ball has two sides of advantage at least. The first is, both of its classic metric and quantum wave functions can be written out explicitly. The second is, its construction involves no hyper-concepts of as extra dimensions or super symmetry.

We response to some critics to our ideas published previously and review our works on this topic in the next-to-final section.  As is well known, necessities of the quantum gravitation is not excited experimentally, but logically. The information missing puzzle is just such an excitation source. So, frankly speaking, we do not want to see any satisfactory resolutions to it if they do not bring us new research area at the same time. Quantum information and quantum geometry accompanying with the replica wormhole method are of course such new areas. We wish inner structure of black holes be also such an area, both theoretically and observationally.

\section*{Acknowledgements}
This work is supported by NSFC grant no. 11875082 \url{https://search.crossref.org/funding}..

\end{document}